\documentclass[nonacm,sigconf]{acmart}

\usepackage{xspace}
\usepackage{pifont}
\usepackage[inline]{enumitem}
\usepackage{cleveref}
\usepackage{subcaption}

\DeclareMathOperator*{\argmin}{arg\,min}

\newcommand{\bx}{\mathbf{x}}

\newcommand{\bfn}{\mathbf{f}}
\newcommand{\bu}{\mathbf{u}}
\newcommand{\by}{\mathbf{y}}

\AtBeginDocument{%
  }

\setcopyright{acmlicensed}
\copyrightyear{2025}
\acmYear{2025}
\acmDOI{XXXXXXX.XXXXXXX}

\acmConference[SIGMOD '25]{Proceedings of the 2025 International Conference on Management
of Data}{June 22--27,
  2025}{Berlin, Germany}
\acmISBN{978-TBD}

\newcommand{\sparagraph}[1]{\vspace{1mm}\noindent \textbf{#1}} 
\newcommand{\sysname}[0]{BayesQO\xspace}
\newcommand{\bo}[0]{BO\xspace}
\newcommand{\random}[0]{Random\xspace}

\newcommand{\circleOne}{\ding{182}\xspace}
\newcommand{\circleTwo}{\ding{183}\xspace}
\newcommand{\circleThree}{\ding{184}\xspace}
\newcommand{\circleFour}{\ding{185}\xspace}
\newcommand{\circleFive}{\ding{186}\xspace}
\newcommand{\circleSix}{\ding{187}\xspace}
\newcommand{\circleSeven}{\ding{188}\xspace}
\newcommand{\circleEight}{\ding{189}\xspace}
\newcommand{\circleNine}{\ding{190}\xspace}

\newcommand{\edit}[1]{#1}

\begin{document}
\setlength{\textfloatsep}{0pt}


\title{Learned Offline Query Planning via Bayesian Optimization}

\author{Jeffrey Tao}
\orcid{0000-0001-6407-1316}
\affiliation{
  \institution{University of Pennsylvania}
}
\email{jefftao@seas.upenn.edu}

\author{Natalie Maus}
\orcid{0000-0002-6616-8506}
\affiliation{
  \institution{University of Pennsylvania}
}
\email{nmaus@seas.upenn.edu}

\author{Haydn Jones}
\orcid{0000-0002-1006-4126}
\affiliation{
  \institution{University of Pennsylvania}
}
\email{haydnj@seas.upenn.edu}

\author{Yimeng Zeng}
\orcid{0009-0001-9676-0893}
\affiliation{
  \institution{University of Pennsylvania}
}
\email{yimengz@seas.upenn.edu}

\author{Jacob R. Gardner}
\orcid{0000-0003-1897-8384}
\affiliation{
  \institution{University of Pennsylvania}
}
\email{jacobrg@seas.upenn.edu}

\author{Ryan Marcus}
\orcid{0000-0002-1279-1124}
\affiliation{
  \institution{University of Pennsylvania}
}
\email{rcmarcus@seas.upenn.edu}

\renewcommand{\shortauthors}{Tao et al.}

\begin{abstract}
Analytics database workloads often contain queries that are executed repeatedly. Existing optimization techniques generally prioritize keeping optimization cost low, normally well below the time it takes to execute a single instance of a query. If a given query is going to be executed thousands of times, could it be worth investing significantly more optimization time? In contrast to traditional online query optimizers, we propose an offline query optimizer that searches a wide variety of plans and incorporates query execution as a primitive. Our offline query optimizer combines variational auto-encoders with Bayesian optimization to find optimized plans for a given query. We compare our technique to the optimal plans possible with PostgreSQL and recent RL-based systems over several datasets, and show that our technique finds faster query plans.

\end{abstract}



\maketitle

\section{Introduction}

Query optimization is a long-standing problem in the database community~\cite{system_r,howgood,qo_unsolved}. Recent advancements in \emph{learned} query optimization (LQO)~\cite{balsa,lero,roq,loger,hybrid_lqo,lstm_jo,qo_state_rep,rejoin,hero_qo,concurrent_lqo,leon_qo} have shown significant promise, often delivering  2-10x improvements in query runtime. However, deploying LQO is complicated due to two main challenges: (1) query regressions (``my query was fast yesterday, why is it slow today?'') and (2) the tight integration of machine learning components into the core query processing pipeline (which are generally engineered with different levels of reliability in mind).

Despite various efforts to address these challenges~\cite{bao,fastgres}, most real-world deployments of LQO (such as at Meta~\cite{autosteer}, Microsoft~\cite{bao_scope, bao_scope2}, and Alibaba~\cite{eraser_lqo,pilotscope}) have separated learned query optimization two components, an offline component and an online component. The \emph{offline component} tests new query plans and caches those plans that perform better than the plans produced by the traditional optimizer. The \emph{online component} then checks this cache for a plan; if a cached plan is not found, it calls the traditional optimizer instead.

This compromise --- spending additional resources offline to find good query plans for specific queries --- solves issues with query regressions and avoids the need to put ML primitives into the query processing pipeline, but it is also motivated by the nature of analytic workloads. In many analytic systems, the majority of compute resources are spent executing repetitive report generation~\cite{superopt_vision} or dashboarding queries~\cite{redshift_pred_cache}, with some queries being executed hundreds of times per day~\cite{stage} or hundreds of thousands of times per year~\cite{superopt_vision}. \edit{Recent studies of Amazon Redshift showed that, for the median database, 60\% of all queries executed were repeated queries (verbatim)~\cite{wu_stage_redshift} and that roughly 10\% of all Redshift clusters have their entire workload consisting of queries that repeated within the last day ~\cite{redshift_workload}.} \footnote{\edit{Some unknown proportion of these verbatim repeats may involve staging tables or views, for which the underlying  query may be changing.}} Substantial repetition means that even a small improvement in query latency can be amplified many times over, making it worthwhile to invest additional optimization resources.

We call the goal of these offline components the \textbf{offline query optimization problem}: to find the best query plan using as few offline resources (i.e., offline query time) as possible. Unlike traditional query optimizers, which generally seek to be so fast that optimization time amortizes to zero compared to execution time, an offline query optimizer is expected to take many times longer than a single query execution. 

Learned query optimizers using the ``offline/online'' compromise are implicitly performing offline query optimization. Current systems must solve two fundamental problems: first, offline query optimizers must have a \emph{search strategy} to decide which plans to test. Second, offline query optimizers must have a \emph{timeout strategy} to deal with query plans that take too long to execute.

\begin{figure}
    \centering
    \includegraphics[width=\linewidth]{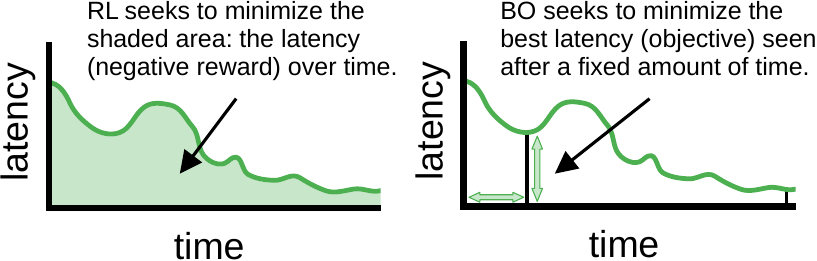}
    \caption{A comparison of reinforcement learning (RL, left) and Bayesian optimization (BO, right). BO is a better match for the offline query optimization problem because we do not care about ``regressions'' in the offline search phase; we only care about the quality of the best-discovered plan.}
    \label{fig:bayes_vs_rl}
\end{figure}

\sparagraph{Search strategy} Existing techniques either use a coarse-grained search of predefined alternative plans~\cite{autosteer,bao_scope2}, or adopt a fine-grained reinforcement learning procedure~\cite{pilotscope,eraser_lqo}. The biggest drawback of the coarse-grained approach is that the set of plans explored is limited, and significant improvements may be outside of the search space. The drawback of the RL approach is more subtle.

\emph{Reinforcement learning (RL) is fundamentally the ``wrong tool for the job'' of offline query optimization.} Specifically, the objective function of reinforcement learning is poorly aligned with offline query optimization. Consider the latency of a query over time as the query is optimized, as depicted on the left side of Figure~\ref{fig:bayes_vs_rl}. RL seeks to minimize the shaded area, representing latency (negative reward) over time. This corresponds to balancing exploration and exploitation~\cite{rl_book}: the optimizer is penalized each time it chooses a bad plan, so the optimizer must frequently choose ``lower risk'' plans that lower the area under the curve, but might not be as informative as ``higher risk'' plans. This is perfectly aligned with the goals of \emph{online} query optimization, however, in the context of \emph{offline} query optimization, what we really care about is finding the best query plan observed during the time-bounded optimization phase (that is, the minimum of the plotted function, not the area under the curve), as shown on the right side of Figure~\ref{fig:bayes_vs_rl}.

\sparagraph{Trouble with timeouts} An important, but often overlooked, dimension of learned query optimization is query timeouts: since some query plans are orders of magnitude worse than others~\cite{howgood}, any learned query optimizer (either online or offline) has a non-zero chance of hitting a poor-performing plan. Current approaches solve this fundamental dilemma in ad-hoc ways, often by ``timing out'' (termination prior to completion) proposed query plans after a fixed threshold. These timed-out values are problematic because (1) they represented a large amount of wasted time (a poor query plan was selected~\cite{balsa}) and (2) no new information was gained (since updating an RL model with the timed-out value would cause the model to strictly underestimate the cost of the timed-out query~\cite{neo}). Thus, a successful offline query optimizer must have a strategy for \emph{selecting timeout values} and for \emph{learning from timed-out queries}.

\sparagraph{Bayesian optimization (BO).} We propose an offline query optimizer, \sysname, that closely mirrors recent work in computational drug discovery~\cite{lolbo}: we use a learned encoder and decoder to translate query plans to and from vectors (called the \emph{latent space}), such that similar query plans are mapped to nearby vectors. Then, off-the-shelf and well-studied Bayesian optimization techniques manipulate the encoded vectors in the latent space, using query execution as a reward signal. Existing coarse-grained techniques that generate a fixed set of plan variants for each query can be used to initialize the process, ensuring that the best-found plan is at least as good as the best plan in the initialization set.  

We show that off-the-shelf BO techniques~\cite{bayes_survey} can be easily adapted to the offline query optimization objective (that is, finding the fastest possible plan in the least amount of time). Furthermore, we show that the framework of Bayesian optimization enables robust \emph{learning from timeouts} as well as \emph{selecting timeouts} on a learned, plan-by-plan basis. In other words, we can meaningfully represent ``query latency $> x$'' within the learned model as a \emph{censored observation}, and our model's confidence intervals provide a robust way of selecting timeouts to maximize information gain. Finally, we show how cross-query information can be incorporated into \sysname by fine-tuning a language model to provide database-specific initialization points for the BO search process.

In our experiments, we show that offline optimization can yield 10-100x performance improvements over prior learned query optimization for some queries, and we demonstrate that our approach can find modest improvements for nearly every query in several benchmarks. Since our system targets repetitive analytic queries, even modest gains can be significantly amplified in practical settings. Our contributions include:

\begin{enumerate}[leftmargin=*]
    \item We formalize the problem of offline query optimization,
    \item We implement \sysname, an offline query optimizer that applies Bayesian optimization techniques,
    \item We show  how recent developments in Bayesian optimization that accommodate high-dimensionality and censored observations can be integrated into \sysname,
    \item We show how fine-tuning a language model can be used to incorporate cross-query information into \sysname,
    \item We show that \sysname can outperform online learned query optimization techniques in an offline setting.
\end{enumerate}

\section{Related work}

Query optimization is a long-standing problem in the databases community. System R~\cite{system_r} proposed the heuristic query optimization scheme now used in most production databases~\cite{cascades}, consisting of a cost model, cardinality estimates, and a dynamic programming search.
Conventional query optimizers are designed according to the ``query optimization contract''~\cite{chaudhuri2009_rethinking_qo}, which expects query optimizers to produce plans quickly (within hundreds of milliseconds, as the actual execution of the plan might be very quick); this work proposed that in order to improve query optimizers, we should consider ``breaking'' this contract in a number of ways, such as allowing the optimizer to intrusively examine the base data (as opposed to keeping cheap histograms), spend a long time on optimization, or even adaptively change the query plan during execution. We believe our work falls into this ``breaking the contract'' category. Older work~\cite{fittest_qo} considered amortizing the cost of searching parts of the plan space across multiple executions, but did not consider offline execution. Query \emph{reoptimization}, perhaps the first ``contract breaker,'' is the task of proactively modifying or recreating a query plan during execution, based on information found during execution, with the overall goal of minimizing total latency~\cite{reopt,rio_reopt,inc_reopt}.

A related concept from the compilers literature is ``superoptimization''~\cite{superopt_coined}, in which a program compiler, which traditionally follows a similar ``contract'' as a query optimizer (i.e., fast compilation times), instead uses a large time budget to produce the best possible sequence of assembly instructions for a given program. Our work can be considered a sort of ``superoptimization for query plans''. GenesisDB~\cite{genesisdb} represents a similar effort, focusing on developing fast implementations over relational operators, instead of entire query plans (thus GenesisDB is mostly orthogonal to the work presented here). Kepler~\cite{kepler} uses a genetic algorithm and exhaustive execution to map the plan space for parameterized queries, which can be viewed as a type of superoptimization. SlabCity~\cite{slabcity} takes an approach similar to traditional superoptimization, by considering SQL-level semantic rewrites of queries to improve performance (e.g., query simplification). Finally, DataFarm~\cite{datafarm} \edit{and HitTheGym~\cite{hitthegym}} investigated how to best produce datasets for machine learning powered database components, including query optimizers.

In recent years, the databases community has been increasingly engaged in applying machine learning techniques to query optimization, including latency prediction~\cite{learning_latency,zeroshot_latency_model,qppnet,contender,jennie_sigmod11,stage}, cardinality estimation~\cite{deep_card_est2, flowloss, neurocard, quicksel, geom_card_est, gridar_lce, asm_lce,alece_lce, robopt}, and cost models~\cite{learn_cost}. Other works have attempted to either augment existing optimizers with learned components (e.g.,~\cite{bao,fastgres,workload_reopt,leon_qo,hybrid_lqo,opengauss}) or entirely replace query optimizers with reinforcement learning (e.g.,~\cite{neo,balsa,lero,pilotscope, eraser_lqo, roq, loger, qo_rank, q_transformer, mcts_qo, lstm_jo, skinnerdb}). Most of these works are focused on the online optimization setting: they must complete quickly while avoiding performance regressions relative to traditional heuristic-based optimizers. Most of these works also employ reinforcement learning, seeking to manage regret from exploring alternatives instead of exploiting the current known-best plan. In comparison, we apply Bayesian optimization to the superoptimization problem because we are principally concerned with finding the query plan with the best possible latency, and ignore suboptimal plans. In the superoptimization setting, bad plans are only bad insofar as executing them until the timeout consumes part of the optimization time budget.

\edit{Bayesian optimization is not the only sample-efficient learning technique. For example, NeuroCARD~\cite{neurocard} learns join distributions efficiently by uniformly sampling tuples from the full outer join of all tables in a schema. Reiner et al.~\cite{geom_card_est} show how domain knowledge can be incorporated into learned models to improve sample efficiency via geometric deep learning. LlamaTune~\cite{kanellis_llamatune} uses database documentation to accelerate DBMS knob-tuning.}

\edit{While our plan encoding was inspired by work in molecular dynamics (i.e, SELFIES~\cite{selfies} strings as used by Maus et al.~\cite{bayes_latent}), a representation with similar goals for query plans was presented by Reiner et al.~\cite{geom_card_est}. Our approaches mainly differ in what we are trying to represent: as our format only seeks to encode join orderings, it does not encode predicates. Furthermore, while Reiner et al. use invariances to give joins with the same cardinality the same representation, our encoding format may have multiple representations for the same join ordering. Further motivation for this design choice is given in ~\Cref{sec:string-format}. Other works have also looked at non-string representations of queries based on graphs~\cite{geom_card_est}, trees~\cite{neo}, and recurrences~\cite{tree_lstm}, typically by using neural network architectures which model these structures.}


The random search heuristic we presented in ~\Cref{sec:eval} can be considered a modified version of QuickPick~\cite{quickpick}. Instead of sampling random query plans and then using a cost model to evaluate their quality, we simply evaluate the quality of the random plans by actually executing them. Such a suggestion would seem ludicrous in the original context of~\cite{quickpick}, but for offline optimization, executing terrible query plans is \emph{not} off the table, if it eventually leads to a better plan! 

Since Bayesian optimization is a relatively old technique, it may be reasonable to ask ``why now?'' Recent innovations in the machine learning community have made it practical to apply Bayesian optimization to \emph{structured} (i.e. non-continuous) inputs with high dimensionality~\cite{lolbo,high_dim_bo,bayes_local}, which was previously impossible. The key innovation that enabled this advancement was attention transformer models~\cite{attention}, which allowed sequences to be efficiently and accurately mapped into vector spaces. In the databases literature, Bayesian optimization has been most frequently applied to tuning configuration knobs~\cite{database_hyperparameter_optimization, cgptuner, gptuner}. To our knowledge, this is the first work to apply BO directly to the optimization of individual queries.

\edit{Perhaps most similar to this work is LimeQO~\cite{limeqo}, a system that uses offline query execution to find the best query hint for each query in a workload. LimeQO can be viewed as a practical way of finding an optimal Bao~\cite{bao} model for a given workload. LimeQO is arguably much simpler than the present work, requiring only linear methods (e.g., no VAE or Bayesian optimization). However, LimeQO only considers a finite set of query hints to apply to each query in a workload, whereas we fully construct query plans. As a result, the present work can potentially find better plans. Additionally, LimeQO focuses on optimizing an entire workload of queries at once (i.e., considering which queries are best to explore next), whereas we focus on optimizing only a single user-specified queries.}
\section{System model \& problem definition}

We define the offline optimization problem for database query planning and show an approach to offline optimization based on Bayesian optimization. We implement this approach in \sysname.


\sparagraph{Challenges} An obvious naive strategy to perform offline optimization is to exhaustively enumerate the space of all possible query plans. This is obviously computationally intractable even for relatively simple queries on few tables: the number of join orderings alone grows super-factorially with the number of joined tables: for a query joining $n$ tables, considering only binary joins and ignoring physical operator selection, there are $n! \cdot C_n = \frac{2n!}{n!}$ distinct join orderings, where $C_n$ is the $n$-th Catalan number~\cite{joe_complexity}.

A refinement of this strategy might be to consider query planning with ``perfect cardinalities,'' since cardinality estimates are often hypothesized to be the main culprit for poor query plan performance~\cite{howgood}. Exhaustively computing cardinality estimates for even a simple query can take \emph{months}~\cite{flowloss}, and adaptively measuring only the cardinality estimates used by the query planner leads to the infamous ``fleeing from knowledge'' problem in which the optimizer repeatedly picks poor query plans due to underestimation from the independence assumption~\cite{fleeing_from_knowledge}.

Furthermore, the plan space contains numerous bad plans which on their own are intolerable to execute to completion as they are many orders of magnitude slower than the optimal. Thus, an offline optimization method must efficiently explore the space of query plans while avoiding executing these bad plans to completion.

It is also crucial that the optimization process use information gained from executing candidate plans to inform its exploration. A necessary property of an offline optimization method is that it can be run for longer in order to obtain better results.

\begin{figure}
    \centering
    \includegraphics[width=\linewidth]{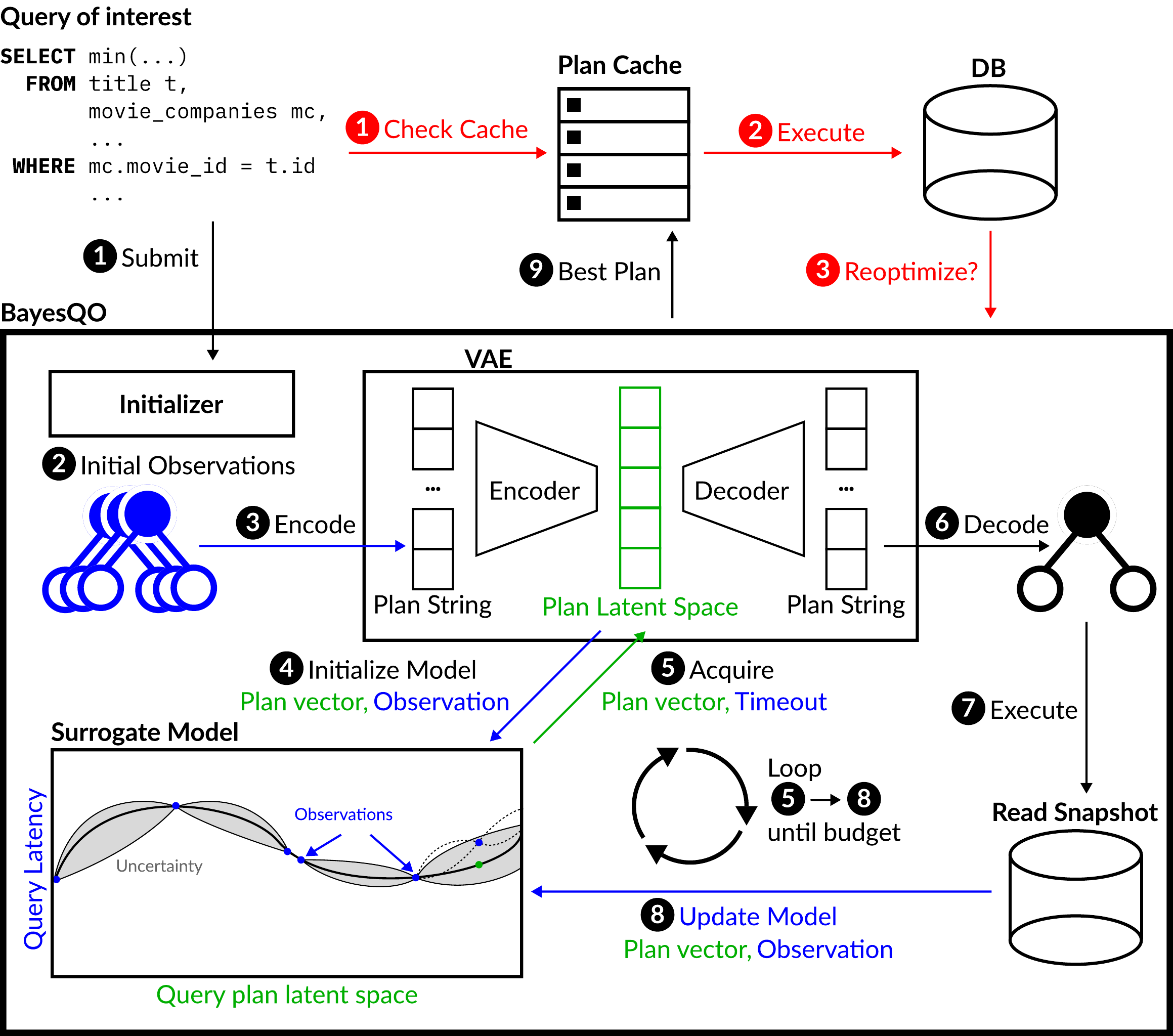}
    \caption{\sysname workflow}
    \label{fig:system_model}
\end{figure}

\sparagraph{System model} \Cref{fig:system_model} depicts the architecture of \sysname. First,
\circleOne a user identifies a query $Q$ that they wish to optimize offline. \sysname will then begin searching for a fast plan for $Q$. To do so, \sysname uses a \emph{Bayesian optimization} loop:
\circleTwo an initialization strategy (discussed in \cref{sec:initialization_strategies}) is used to produce a set of plans
\circleThree the plans are translated into strings (discussed in \cref{sec:string-format}) and embedded into a vector space using a learned model, 
\circleFour the embedded plans and their observed execution latencies are used to initialize the surrogate model, 
\circleFive an \emph{acquisition function} (discussed in \cref{sec:bo-background}) is used to select a point in the latent space, and then
\circleSix the latent space vector is decoded back to a query plan.
\circleSeven This plan is then given a timeout value $TO(P)$, and executed against a read-only snapshot of the database. The query either executes successfully with latency $L(P) < TO(P)$, or times out.
\circleEight the Bayesian optimization algorithm uses the observed latency of the new query plan to improve its understanding of the query space. Then, the process repeats steps \circleFive $\to$ \circleEight until a time budget is exhausted or the user is satisfied with the achieved latency.
\circleNine Finally, the best seen plan goes into a cache.

When the query is being executed online, \textcolor{red}{\circleOne} the user submits a runtime query to the system. \textcolor{red}{\circleTwo} If the plan is present in the cache because it was optimized offline, the cached plan is used. Otherwise, the DBMS' optimizer is used. \textcolor{red}{\circleThree} Looking at the runtime statistics of the executed plan, \sysname decides whether the query should be re-optimized.

\sparagraph{Problem definition} Our goal is to find a query plan $P$ with low latency for a query $Q$ using as few additional resources as possible (i.e., as quickly as possible).\footnote{Note that the constraint on time is required to make the problem non-trivial: if we have infinite computational resources, we can simply test every possible query plan and pick the fastest one.} Unlike traditional query optimizers, \sysname will continuously test new query plans until terminated (either by a time budget or directly by the user). We denote the sequence of queries produced by \sysname at a given time $t$ as $S_t = P_1, P_2, \dots, P_n$. Let $\mathbb{I}_i$ be an indicator such that when $\mathbb{I}_i = 0$, the plan $P_i$ completed successfully, and when $\mathbb{I}_i = 1$, the plan $P_i$ timed out. Thus, the cost a sequence $S_t$ is denoted as:

\begin{equation*}
    Cost(S_t) = \sum_i \mathbb{I}_i \times TO(P_i) + (1 + -\mathbb{I}_i) \times L(P_i)
\end{equation*}

\noindent ... and the best latency achieved within $S_t$ is denoted as:
\begin{equation*}
Latency(S_t) = \min_i 
\begin{cases}
    L(P_i)  & \mbox{if } \mathbb{I}_i = 0 \\
    \infty  & \mbox{if } \mathbb{I}_i = 1 \\ 
\end{cases}
\end{equation*}

\noindent Our goal is minimize latency while staying within a user-specified cost budget $B$:

\begin{equation}
\begin{aligned}
    \min_{S_t} \quad & Latency(S_t) \\
    \mbox{subject to } \quad & Cost(S_t) < B
\end{aligned}
\end{equation}

\sparagraph{Discussion} Traditional and previous learned query optimizers solve this problem for (very small) budgets $B$ (e.g. for Neo~\cite{neo}, $B < 500ms$). Here we consider $B$ large enough to actually test the query plan. Prior optimizers are designed to have the optimization time amortized out compared to execution, our approach assumes optimization time is many times higher than the query latency.

\sparagraph{Assumptions} Our system model assumes the following about the DBMS and workload:
\begin{enumerate}[leftmargin=*]
    \item There is a \emph{default query optimizer} that produces reasonable but not globally optimal query plans for any given query.
    \item Queries can be executed against \emph{read snapshots} of the database.
    \item The execution engine can \emph{accept physical plans/hints} that specify join orders and physical join operators.
    \edit{\item Joins within queries are PK-FK equijoins. \footnote{\edit{This is mostly a constraint of our current implementation; future work could  straightforwardly extend this technique to support non-key or non-equijoin queries, since our core technique only needs to know which tables are involved in which joins.}}}
\end{enumerate}

\sparagraph{Why a read snapshot?} \sysname needs to execute many plans over the course of optimizing a query, some of which may be bad plans with very large intermediate results. So as not to disrupt the database's currently-running workload, we assume the ability to execute queries against read snapshots.


\sparagraph{Example: a trivial offline optimizer} The easiest way to implement an offline optimizer in our framework is with random plan search, similar to Quickpick \cite{quickpick} but ignoring cost estimates. Given a query $Q$, first measure the latency of the query plan $P_d$ produced by the default query optimizer $L(P_d)$. Then, select a query plan $P_1$ for $Q$ at random, and execute it with timeout equal to the latency of the default plan $TO(P_1) = L(P_d)$. Ignore the feedback, and continue executing random plans up to the budget B. While the odds of finding a better plan than $P_d$ are poor, note that you never exceed the budget B and your final plan is always at least as good as the default optimizer. We experimentally evaluate this simple baseline in Section~\ref{sec:eval}.

\sparagraph{Cross-query learning} A hidden benefit to \sysname is that each time a query is optimized, a large number of query plans are executed. These execution traces can be used as training data for cross-query models. \sysname does this by using past execution traces to fine-tune an LLM to conditionally generate a query plan string (described in Section~\ref{sec:string-format}) for a given query. While this LLM obviously will not always produce a high-quality query plan, we show experimentally that these LLM-generated plans are reasonable starting points for future optimizations. Thus, \sysname creates a ``virtuous cycle:'' each time a query is optimized, additional training data is collected. That training data can be used to fine-tune an LLM. The LLM can then generate good initialization points for optimizing the next query, and so on. We describe our technique for cross-query learning in Section~\ref{sec:initialization_strategies}.

\section{Bayesian Optimization for Query Plans}
\label{sec:technique}

Given our goal to perform offline optimization of queries while minimizing the cost of executing extra queries against the database, and given that we do not know how to efficiently explore the space of possible query plans, Bayesian optimization is a promising approach. Bayesian optimization (BO) enables optimization of expensive-to-evaluate, black-box functions while requiring relatively few evaluations of the expensive function. However, Bayesian optimization techniques operate over continuous, real-valued domains, whereas query plans are discrete tree structures.

Prior work on Latent Space Bayesian Optimization (LSBO)
has allowed Bayesian optimization to be applied over other discrete, combinatorial spaces by using a deep autoencoder model (DAE) to transform a discrete, structured search space $\mathcal{X}$ into a continuous, numerical one $\mathcal{Z}$ \cite{Weighted_Retraining,ladder,gomez2018automatic,Huawei,eissman2018bayesian,kajino2019molecular, lolbo}.
For example, \citet{lolbo} applied LSBO to problems in drug discovery using a DAE trained on string representations of molecules. 
Inspired by this work, we define a string encoding format for query plans that can represent all possible query plans and in which each string decodes unambiguously to a particular query plan (\Cref{sec:string-format}). We train a variational autoencoder (VAE) on strings of this format using plans derived from a traditional query optimizer (\Cref{sec:technique-vae}). We perform BO in the latent space of this VAE (\Cref{sec:bo-background}), making novel contributions in the selection of timeouts (\Cref{sec:censored_observations}). Finally, we discuss different strategies for initializing the local BO process, as the quality of initialization points impacts BO performance (\Cref{sec:initialization_strategies}).

\subsection{Query Plan String Format}
\label{sec:string-format}
Our first step towards optimizing query plans with Bayesian optimization is to build a string language for query plans that we encode and decode into a vector space. We first explain the desired properties of this string language. Then, we explain the string language we designed to have these properties. While we believe our string languages makes some good tradeoffs, in theory any string language with the desired properties could be equally effective.

\sparagraph{Desiderata for string representations.} We identify two important attributes of our string language from prior work on molecular optimization. Maus et al.~\cite{lolbo} identify two essential properties that are key to success of a string language for Bayesian optimization (in the case of~\cite{lolbo}, the string language is called SELFIES \cite{selfies} and describes molecules). We translate those design constraints from the domain of~\cite{lolbo} to query optimization, and adopt these properties as requirements for our query plan language:
\begin{enumerate}[leftmargin=*]
    \item \textbf{Completeness.} Any valid query plan in $X$ must be representable as a string using the language. If this is not true, high-quality plans that are not representable will not be discoverable by our optimization algorithm.
    \item \textbf{Decoding validity.} Any sequence of characters in the language must correspond to a unique and valid query plan. If latent codes of the DAE decode to invalid query plans, optimization would need to be done under additional feasibility constraints, which adds unnecessary complexity to the optimization problem. Validity was a primary goal of the models in both \citet{JTVAE} and \citet{maus2022local}.
\end{enumerate}

\edit{The ideal string representation would also be \emph{injective}, meaning that each unique string maps to a unique plan~\cite{geom_card_est}. We were unable to find a string representation with all three of these properties, so we settle for a complete representation with decoding validity that is not injective (the same plan might be represented by multiple strings). Taking this tradeoff is motivated by prior work: Maus et al.~\cite{bayes_latent} showed that a string representation with only completeness and decoding validity (SELFIES~\cite{selfies}) outperformed an injective representation with completeness (SMILES~\cite{weininger_smiles}). We leave the investigation of alternative string representations to future work.}

\sparagraph{String language for query plans} We design an encoding format for binary-tree-structured query plan defining join orders and operators (henceforth, a ``join tree''). \edit{The join tree does not encode other aspects of the query such as selections, join and filter predictes, and aggregations. When the join tree is decoded to executable SQL, we translate the join tree to a hint string and prepend it to the original SQL text which contains these aspects of the query.} We observe that a join tree can be unambiguously reconstructed by knowing for each non-leaf node the left and right subtree that compose it, and that valid trees have left and right subtrees that are disjoint; we simply need an unambiguous way to identify these subtrees.

In our encoding format, each join subtree is expressed as a 3 symbol sequence (left, right, operator). Each physical join operator ($h$ash, $m$erge, $n$ested loops) is given a unique symbol. The fully specified query plan is simply the concatenation of these sequences.

The leaves of a join tree are always the tables being joined, so we define a unique symbol for each base table in the schema. However, we cannot define symbols for each possible join subtree, as doing so would be tantamount to defining a unique symbol for every possible query plan. Instead, we observe that after a table symbol is used once to specify a join subtree, it will never be used again, as any table will only ever appear as a leaf once. The same is true for each join subtree after it is specified to be the child of some larger subtree. As such, to the right of a particular join subtree sequence, the symbols composing the join instead refer to the larger subtree.

\edit{For multiple occurrences of the same table under different aliases within the same query, we rename such aliases to be numbered (e.g. \texttt{movies1}, \texttt{movies2}, ...), and we define a unique symbol in our language per table-number pair. This does require choosing a maximum number of possible aliases of a single table; in our experiments we select the maximum occurrences of a particular table within the benchmark queries.}

The leftmost occurrence of a table symbol in a plan string always references the base table itself, but subsequent occurrences represent the largest subtree that the table is part of. For example, in a join between three tables $A$, $B$, and $C$, $(A \bowtie_\text{hash} (B \bowtie_\text{merge} C))$, the valid encoding strings are $(B, C, \bowtie_\text{m}, A, B, \bowtie_\text{h})$ and $(B, C, \bowtie_\text{m}, A, C, \bowtie_\text{h})$.

To fulfill our second requirement, decoding validity, we use a simple trick. We maintain state about the partially-specified join tree as we decode the string from left to right. If the decoder encounters a symbol that is not syntactically valid (e.g. a table in place of a join operator) or semantically valid (e.g. a table that is not part of the join), it deterministically resolves it to a symbol that is valid by constructing a list of all valid symbols and using the invalid symbol's integer value as an index into the list.

\edit{We note that the choice of replacement symbol is arbitrary, but this scheme for ensuring decoding validity is preferable to more obvious ideas such as simply refusing and resampling when encountering invalid strings or decoding them to some default plan. Bayesian optimization will be performed over a vector space that decodes to potential plan strings. Rejecting strings would prevent the surrogate model learning about vast regions of the plan space. Decoding all invalid strings to some default plan would make vast regions of the space undifferentiated in performance. Our technique ensures that all strings decode to valid plans, that similar invalid strings are mapped to somewhat different valid query plans, and that these decodings are a valid function of the input string.}

\edit{
\sparagraph{Limitations} Our language does not represent subqueries and CTEs. When processing queries that contain such structures, they are left untouched, so the decoded query plan hint will not contain any reference to them or to tables only occurring within them.
}

\subsection{Encoding \& Decoding Query Plans} \label{sec:technique-vae}
While BO can be applied to various search spaces, it is most straightforward in a continuous, real-valued domain. This presents a challenge when optimizing query plans, which are inherently discrete tree structures. To address this, we construct a mapping between query plans and points in a continuous domain. Having defined our string representation in \Cref{sec:string-format}, we train a deep autoencoder (DAE) model on these encoded strings. This process generates a \emph{latent space} – a continuous, real-valued domain that serves as a proxy for the discrete space of query plans, enabling application of BO techniques. Intuitively, the goal of the DAE is to construct a latent space in which similar query plans are mapped to similar vectors. This way, a search routine that finds a particularly good plan can look at the ``neighbors'' of that plan in the latent space for similar plans. The notions of ``similar'' and ``neighbors'' are both highly approximate: no actual neighborhoods or similarity scores are computed, but instead this property is \emph{implicitly} created when training the DAE.

A DAE consists of an encoder $\Phi: X \to Z$ that maps from an input space $X$ to a latent space $Z$ (sometimes called a bottleneck~\cite{bottleneck}, as it is often lower dimensional than $X$ in order to force a degree of compression) and a decoder $\Gamma : Z \to X$ that maps from the latent space back to the input space. We use a type of DAE known as a variational autoencoder (VAE)~\cite{KingmaW13}, in which the encoder produces a distribution over latent points $\Phi(Z | X)$, and the decoder produces a distribution over $X$ given $Z$, $\Gamma(X | Z)$. The model is trained by maximizing the evidence lower bound (ELBO):
$$
\mathbb{E}_{\Phi(Z | X)}[\log \Gamma(X | Z) - \text{KL}(\Phi(Z | X) || p(Z) ]
$$

In training such a DAE, we construct a mapping $\Gamma : Z \to X$ that can produce string-encoded query plans given points in the latent space. The VAE regularization (the KL term, representing relative entropy) makes the search space smooth, facilitating more effective optimization. This is precisely what we need in order to perform BO: the surrogate model is defined over the latent space, and we evaluate the black-box function $f$ for points in the latent space by decoding the point to a string query plan through the DAE and executing them against the real database.

\sparagraph{Training data} In order to train the DAE, we compute a large set of encoded query plans ($\sim$1 million) from the database schema. Importantly, this process does not require any query execution, and can be done exclusively with only metadata from the DBMS. The idea is to create a suitably diverse set of ``reasonable'' query plans so that the DAE can learn a smooth probability distribution of the space of query plans. By ``reasonable'' here, we do not mean that these plans must all be optimal, just that they must be somewhat representative of the \emph{family} of optimal query plans---\edit{the purpose of this is to create a space of plans in which points that are close to each other have similar performance characteristics. However, the space still contains points for \emph{all possible} query plans.}


\edit{To generate this set of plans, we sample random PK-FK equijoin queries from the schema by constructing the ``alias-$k$ reference graph'' which contains $k$ nodes corresponding to each table and edges corresponding to all PK-FK references between tables. We choose $k$ equal to the highest number of aliases of the same table used in any query in the workload. From this alias-$k$ reference graph, we sample queries by selecting random connected subgraphs with varying numbers of vertices. Given a particular subgraph, we produce a query joining all table aliases with join predicates corresponding to all present edges.}

For each sampled query, we plan the query using the existing default query optimizer (e.g. PostgreSQL), encode the plan in our string encoding format, and add it to the VAE training set. In order to expand the diversity of plans used to train the VAE, we additionally produce encoded plans using hints~\cite{url-pg_hints} to the default query optimizer (e.g. disable nested loops, disable sequential scans).

Our training data generation process makes two key design choices: (1) sampling random queries from the database schema, and (2) generating query plans with the database's default optimizer. The first decision ensures that we have coverage for a wide variety of input queries. Our goal is to train the DAE once per schema, and then reuse the DAE for every query over the schema. The second decision ensures that the query plans we get are somewhat reasonable. For example, the underlying database optimizer is unlikely to pick a plan full of cross joins, and is likely to take advantage of index structures if applicable.

\subsection{Background on Bayesian Optimization} \label{sec:bo-background}
Given a query plan language and a trained DAE to translate query plan strings into vectors in a latent space (and back), we can now optimize queries inside of the latent space using Bayesian optimization (BO). Intuitively, BO in our application works by learning the relationship between the DAE's latent space and actual query plan latency. BO learns this relationship by repeatedly testing points sampled from the latent space. If the BO algorithm can get a good estimation of the relationship between the latent space and query latency, then good plans can be found. This section gives important background on the BO technique we use in this paper. Then, in Section~\ref{sec:censored_observations}, we explain some of the small changes we made to traditional algorithms to address query optimization specifically.

\sparagraph{Bayesian Optimization} This section provides a brief overview of Bayesian Optimization (BO). For readers unfamiliar with BO, we recommend the comprehensive book by \citet{garnett_bayesoptbook_2023}. Our methodology builds upon the approach developed by \citet{eriksson2019scalable}, \edit{with specific novel modifications tailored for optimizing query plans and execution latency in a DBMS.}

Bayesian Optimization is a method for optimizing black-box functions that are expensive to evaluate, aiming for \emph{sample efficiency}. Given an input space $\mathcal{X}$ and an unknown objective function $f: \mathcal{X} \rightarrow \mathbb{R}$, BO seeks to find an input $x^* \in \mathcal{X}$ that minimizes $f(x)$ in as few evaluations of $f$ as possible. This is particularly useful when each evaluation of $f(x)$ is costly---for example, when $f(x)$ involves executing a query plan in a DBMS to measure its runtime.

BO operates by constructing a probabilistic surrogate model of the objective function, which is iteratively refined as new data is acquired. The general optimization procedure follows these steps:

\begin{enumerate}[leftmargin=*]
\item \textbf{Initialization}: Build a surrogate model of the objective function $f$.
\item \textbf{Acquisition Function Optimization}: Use an acquisition function to select the next point $x_{\text{next}}$ to evaluate, balancing exploration and exploitation.
\item \textbf{Evaluation}: Compute the true function value $f(x_{\text{next}})$ by executing the query plan corresponding to $x_{\text{next}}$.
\item \textbf{Model Update}: Update the surrogate model with the new observation $(x_{\text{next}}, f(x_{\text{next}}))$, and repeat steps 2--4.
\end{enumerate}

To efficiently navigate the search space, BO leverages the surrogate model along with the acquisition function to select promising candidate plans while minimizing the number of expensive evaluations. In this work, we use \emph{Thompson Sampling}~\cite{thompson} as the acquisition function.

\sparagraph{Local BO} Standard BO methods can struggle with high-dimensional or discrete optimization problems, such as those encountered in query plan optimization, due to the curse of dimensionality and the combinatorial explosion of the search space. To address this, we incorporate methods from the local BO literature, specifically \emph{TuRBO}~\cite{eriksson2019scalable}. TuRBO maintains a hyper-rectangular ``trust region'' within the input space, which constrains the region from which points are sampled. By dynamically adjusting the size and location of these trust region based on the the optimization success / failure, TuRBO can balance global exploration with local exploitation, allowing for efficient optimization in high-dimensional spaces. \footnote{\edit{Though called ``local BO'', this is a \emph{global} optimization process that can produce results significantly different from the initialization points. Local BO methods are the most competitive methods in high-dimensional spaces, as established in Eriksson et al. ~\cite{eriksson2019scalable}.}}

\sparagraph{Right-Censored Observations}
During typical Bayesian optimization, when we make an observation at a point $\bx$, we obtain an associated objective value $f(\bx) = y$. 
For a right-censored observation, when we observe at the point $\bx$, we instead only learn that $y$ was greater than some threshold $\tau$. In our application, right-censored observations represent query timeouts: If a query $\bx$ is observed to execute for $\tau$ seconds before timing out, then we know that the true latency of $q$ is \emph{at least} $\tau$: $f(\bx) \geq \tau$. 

In the query optimization setting, using right-censored observations is particularly important. Obtaining true values for arbitrary plans in the space of possible plans can be infeasible, as bad plans may take days or even weeks. Thus, it is more efficient if we can \emph{time out} plans that perform poorly and update the surrogate model with knowledge that the running time of $x$ is ``at least as bad as $y$''. Intuitively, for regions of $X$ that contain truly awful  plans, for the purposes of finding optimal plans, it is not necessary to know exactly how bad a particular plan is---it suffices to know that plans like it should be avoided. 

BO in the presence of censored observations was first explored by Hutter et al. ~\cite{DBLP:journals/corr/HutterHL13}, where an EM-like algorithm was used to impute the value of censored responses.
They applied this method to algorithm configuration, terminating any runs that exceeded a constant factor of the shortest running time observed so far. Building on this, Eggensperger et al. ~\cite{eggensperger2020neural} trained a neural network surrogate on a likelihood based on the Tobit model to directly model right-censored observations:
\begin{equation}
\begin{aligned}
\label{eq: tobit}
p(\mathbf{y}|\mathbf{f}) &= \phi(\mathbf{z})^{1-\mathbf{I}}(1-\Phi(\mathbf{z}))^\mathbf{I} \\
\mathbf{z} &= \frac{\mathbf{f}-\mathbf{\mu}}{\mathbf{\sigma}^2}\\ 
I&= \begin{cases}
       0, & \text{if } \mathbf{y} \text{ is uncensored} \\
       1, & \text{if } \mathbf{y} \text{ is censored}
    \end{cases}
\end{aligned}
\end{equation}

\noindent where $\phi$ and $\Phi$ denote the Gaussian density and cumulative density function respectively. In~\cite{eggensperger2020neural}, timeout thresholds were chosen as a fixed percentile of existing observations.

\sparagraph{Approximate Gaussian Processes}  
Because the space of query plans is large, we anticipate needing to test a large number of query plans. As a result, we must select a surrogate model that (1) allows for \emph{probabilistic inference}, that is, gives a probability distribution at each point instead of a simple point estimate, and (2) can scale to a large number of observations. Thus, we select an \emph{approximate} Gaussian Process (GP) model.

Approximate GP models, such as the popular Scalable Variational Gaussian Process (SVGP)~\cite{svgp}, use inducing point methods in combination with variational inference to allow approximate GP inference on large data sets~\cite{hensman2013gaussian,titsias2009variational}. The standard evidence lower bound (ELBO) on the log-likelihood used to train a SVGP model is the following:
\begin{align}
\label{eq: svgp}
\log p(\by) & \geq  \mathbb{E}_{q(\bfn)}[\log p(\by \mid \bfn)] - \textrm{KL}(q(\bu)\,||\,p(\bu)) 
\end{align}
%

\subsubsection{Bayesian Optimization with Censored Observations} \label{sec:censored_observations}

While previous work on Bayesian optimization with censored observations (censored BO) did not use approximate SVGP~\cite{svgp} models, \edit{we contribute a straightforward extension} of SVGP~\cite{svgp} models to the censored BO setting. 
Starting from \Cref{eq: svgp} and using the Tobit likelihood given in~\Cref{eq: tobit}, we derive the new ELBO:
\begin{equation*}
\centering
\begin{aligned}
& \log p(\by) \geq  \mathbb{E}_{q(\bfn)}[\log p(\by \mid \bfn)] - \textrm{KL}(q(\bu)\,||\,p(\bu)) \\
			& = \mathbb{E}_{q(\bfn)}[\log \phi(\mathbf{Z})^{1-\mathbf{I}}(1-\Phi(\mathbf{Z}))^\mathbf{I}] - \textrm{KL}(q(\bu)\,||\,p(\bu)) \\
			& = \mathbb{E}_{q(\bfn)}[\log \phi(\mathbf{Z})^{1-\mathbf{I}} + \log(1-\Phi(\mathbf{Z}))^\mathbf{I}] - \textrm{KL}(q(\bu)\,||\,p(\bu)) \\
            & = \mathbb{E}_{q(\bfn)}[\log \phi(\mathbf{Z_u})] + \mathbb{E}_{q(\bfn)}[\log(1-\Phi(\mathbf{Z_c}))] - \textrm{KL}(q(\bu)\,||\,p(\bu)) 
\end{aligned}
\end{equation*}
Here, $\mathbf{Z}_{u}$ correspond to $\frac{\mathbf{f} - \mu}{\sigma^2}$ values for uncensored observations, and $\mathbf{Z}_{c}$ correspond to censored observations. The first term
can be computed analytically as in standard SVGP models. The second term, $\mathbb{E}_{q(\bfn)}[\log(1-\Phi(\mathbf{Z_c}))]$, can be computed using one dimensional numerical techniques like Gauss-Hermite quadrature.  

During optimization, we select a threshold $\tau$ for each executed query plan $\bx$, and cut off execution once the running time exceeds $\tau$. This results in right-censored observations. 
Selecting the timeout for any given observation is crucial: selecting too low of a timeout deprives BO of important knowledge about the space of plans, whereas selecting too high of a timeout wastes time executing bad plans. Previous work in BO uses a constant multiplier over the best observation seen so far~\cite{hutter2013_bocensored}, or a fixed percentile across all observations~\cite{eggensperger2020_censored}. Balsa~\cite{balsa} also uses a fixed multiplier  in order to bound the impact of executing bad plans. 
\edit{We use an uncertainty-based method for selecting timeouts that, compared to prior work, ensures that the surrogate model will be sufficiently confident that a particular point is suboptimal before timing out.}


Before evaluating a new candidate query plan $x_t$ during step $t$ of optimization, we dynamically set a new timeout threshold $\tau_t$. 
We select thresholds so that, after conditioning on the right-censored observation $(\bx_{t}, \tau_{t})$, we are \textit{confident} that the best query plan observed so far, $\bx^{*}_{t}$, is still a better design than the candidate plan $\bx_{t}$. 
Because we do not want to waste additional running time evaluating $f(\bx_{t})$, we ideally want the \textit{smallest} such $\tau_{t}$. 

\paragraph{Selecting $\tau_{t}$.} The above discussion leads to the following optimization problem, where we find the smallest threshold $\tau$ so that our incumbent is confidently better than $\bx_{t}$ \textit{after conditioning on $(\bx_{t}, \tau)$}:
\begin{align*}
    \tau_{t}^{*} &= \argmin \tau \\
    & \textrm{s.t.}\;\; y^{*}_{t} \leq \mu'_{t}(\tau) - \kappa \sigma'_{t}(\tau)
\end{align*}

On \edit{its} surface, this optimization problem is challenging, as evaluating our constraint for a given $\tau$ involves updating the Gaussian process surrogate model with that value $\tau$ as the observed timeout. This is similar to other acquisition functions in the Bayesian optimization literature that use fantasization to do lookahead, e.g., knowledge gradient~\cite{frazier2009knowledge}.

Because we use variational GPs, there are several inexpensive strategies that we can use to evaluate the constraint. For example, \citet{maddox2021conditioning} recently proposed an efficient routine for online updating sparse variational GPs, both with conjugate and non-conjugate likelihoods. Alternatively, a few additional iterations of SGD can be used to update the model in a less sophisticated way.

Finally, we note that the value of $\mu'_{t}(\tau) - \kappa \sigma'_{t}(\tau)$ should generally be monotonic in $\tau$---fantasizing that $\bx_{t}$ cut off with a larger threshold should strictly increase the gap between our belief about $y_{t}$ and $y^{*}_{t}$. Therefore, given a routine to cheaply evaluate the constraint, the constrained minimization problem over $\tau$ can be solved e.g. with binary search.

\subsection{Initialization Strategies} \label{sec:initialization_strategies}
The initial step of BO for a given query typically involves selecting points within the latent space using the acquisition function. As the surrogate is initialized with a random prior, this amounts to selecting random points within the latent space. Theoretically, given sufficient time for BO execution, this approach would yield optimal results. However, to improve the practicality of BO within high dimensional spaces, it is helpful to initialize the process with a small number of precomputed $(x, f(x))$ pairs representing high-quality plans. We explore multiple methods of generating these initialization points.



\sparagraph{Hinted plans (Bao)} We can leverage an existing traditional query optimizer that accepts hints, such as PostgreSQL, to generate the initialization points. We exhaust all of the combinations of join and scan hints (as in the hint sets used by Marcus et al.'s Bao~\cite{bao} optimizer) to produce 49 initialization points for each query. These 49 initialization points are \emph{guaranteed} to contain the best plan that could have possibly been chosen by Bao. We note that it is not important \emph{which} of the queries in the initialization set is optimal, merely that the initialization set contains some queries that represent a promising starting point for optimization. Thus, it is not necessary to ``prune'' hint sets from Bao, as is recommended in~\cite{bao}.

\sparagraph{The default optimizer plan} A simpler strategy would be to generate a single optimization point by using the DBMS' underlying optimizer. This approach has the advantage of simplicity, since the underlying DBMS almost surely has an optimizer. Unfortunately, we found that this approach does not work well in practice, mostly because initializing BO with a single initialization point seems to be suboptimal~\cite{lolbo}.

\sparagraph{LLM} Inspired by previous work demonstrating the effectiveness of large language models (LLMs) in optimizing program runtimes \cite{pie}, we explore the use of fine-tuned LLMs for generating initialization points. We collected trajectories from 606 \sysname runs, selecting the top-1 and top-5 query plans for each query to construct a fine-tuning dataset. Using this dataset, we fine-tuned GPT4o-mini for one epoch. For each new query, we use the fine-tuned model to sample 50 initialization points. This approach leverages the model's ability to learn patterns from previous optimization runs, potentially producing high-quality plans that outperform those generated by traditional query optimizers. Our evaluation (\cref{sec:eval-llm}) demonstrates that this LLM-based strategy can often produce the best query plan among all initialization strategies considered here.

\sparagraph{Extensibility} \sysname simply admits sets of initialization pairs $(x, f(x))$, so any strategy can be used to generate these pairs. As such, our approach can incorporate future improvements in traditional or learned query optimization techniques.


\subsection{Random plans} \label{sec:random}
Though not related to BO, we implement random plan search, which can be thought of as a completely exploration-based algorithm. The intuition behind this method is that joins are commutative but that cross-joins are generally bad for performance. This strategy samples random plans from the space of all plans that do not contain any cross joins.



\edit{Given a particular query over a set of table aliases, we construct the subgraph of the schema's alias-$k$ reference graph containing only the table aliases referenced in the query.}
From this query graph, we can construct a random join tree by constructing a spanning tree. Whenever an edge is added to the spanning tree, we add the join between the two newly connected components to the join tree. Physical join operators are selected uniformly randomly.

One potential benefit of utilizing this strategy is that its viability implies that it may be possible to perform offline optimization in the absence of a traditional query optimizer. As we show in \Cref{sec:eval}, this strategy can be used on its own to perform offline optimization.

\section{Experiments} \label{sec:eval}

We sought to answer the following research questions about \sysname:

\begin{enumerate}[label=\textbf{RQ\arabic*.}, leftmargin=*]
    \item How effectively does \sysname reduce query latency for a workload given a certain time budget? (\Cref{sec:eval-overall}, \Cref{sec:eval-case-studies}) \label{rq:optimization-tradeoff}
    \item How important are our modifications to Bayesian optimization to the performance of \sysname? (\Cref{sec:eval-timeout-ablation}) \label{rq:bo-ablation}
    \item How robust are plans generated by \sysname to data drift? Can previously optimized plans help jump-start reoptimization? (\Cref{sec:eval-drift}) \label{rq:drift}
    \item Can we train an LLM from offline optimization results to generalize to unseen queries? (\Cref{sec:eval-llm}) \label{rq:llm}
\end{enumerate}



\subsection{Setup}
We evaluate \sysname over four sets of queries:
\begin{enumerate}[leftmargin=*]
    \item \textbf{JOB}: The entire Join Order Benchmark (introduced by Leis et al. \cite{job}), which consists of 113 queries over the IMDB dataset.
    \item \textbf{CEB}: A subset of the Cardinality Estimation Benchmark (introduced by Negi et al. in ``Flow-Loss'' \cite{flowloss}), which consists of $\sim3000$ queries divided across 16 query templates over the IMDB dataset.  We select the top 100 and bottom 100 queries by improvement of the optimal hint set vs. the PostgreSQL default plan and the 100 queries with longest-running PostgreSQL default plans. There is some overlap between these categories, resulting in 234 total queries representing 13 templates.
    \item \textbf{Stack}: A subset of the StackOverflow benchmark (introduced by Marcus et al. in ``Bao'' \cite{bao}), which consists  of $\sim6000$ queries divided across 16 query templates. We selected the longest-running queries from each template in equal proportion (excluding templates consisting entirely of queries that took less than 1 second), producing a list of 200 queries.
    \edit{
    \item \textbf{DSB}: 3 generated queries from each of 30 templates, based on TPC-DS but enhanced with more complex data distributions and query templates (introduced by Ding et al. ~\cite{ding_dsb}). We use generated queries from the ``agg'' and ``spj'' template sets. Following Wu et al.~\cite{learned_shift}, we use a scale factor of 50.
    }
\end{enumerate}

Workload characteristics are summarized in \Cref{table:workloads}.

\begin{table}[]
    \centering
    \begin{tabular}{c|c|c|c}
        \textbf{Name} & \textbf{Size on Disk}  & \textbf{Queries} & \textbf{Median joins per query} \\
        \hline
         JOB & 8GB & 113 & 7 \\
         CEB & 8GB & 234 & 10 \\
         Stack & 64GB & 200 & 6 \\
         DSB & 89GB & 90 & 5 \\
    \end{tabular}
    \caption{Characteristics for the four evaluation workloads.}
    \label{table:workloads}
\end{table}

The offline query planning setting has not received much attention in the query optimization literature. To our knowledge, this work is one of the first to demonstrate an offline optimization technique. As such, we compare plans generated by \sysname to plans from PostgreSQL, Bao, Balsa, and the random non-cross-join plan generation technique described in \cref{sec:initialization_strategies}, which we refer to as \random. \edit{The \random strategy can be seen as representing the gains from performing offline optimization at all, and we use it to understand whether our technique brings further improvement when rethinking the query optimization contract~\cite{chaudhuri2009_rethinking_qo}.}

Instead of actually running Bao, we instead execute all hint sets (49 total, comprised of all combinations of join and scan hints) and take the hint set with the fastest runtime. This is the best plan that Bao could ever produce, since it focuses on steering PostgreSQL's existing optimizer using hints. We choose this baseline as representing the best that traditional heuristic-based query optimizers can do in the offline query optimization setting. Unless stated otherwise, all BO runs in the following section are initialized using these 49 hinted ``Bao'' plans and runtimes.

We choose Balsa~\cite{balsa} as a baseline representing reinforcement learning-based systems, which are also not originally designed for the offline query optimization setting. We use the default configuration for Balsa, except that we set $S = 1.5$ (the multiplier on query timeout values), which we found to universally improve results in our experimental setup. We note that the comparison to Balsa is not entirely fair: as a reinforcement learning-powered query optimizer, Balsa seeks to minimize \emph{regret}, whereas our BO-based approach seeks to minimize the \emph{best latency found}. This distinction is illustrated in Figure~\ref{fig:bayes_vs_rl}, but the most important experimental consequence is that Balsa will occasionally repeat a query plan that it considers to be ``good'' in order to maximize reward, which is obviously suboptimal in an offline optimization scenario. In these experimental results, we use a plan cache to avoid actually executing any exactly-duplicated query plans.

\random can be thought of as an offline optimization technique that purely explores the space of possible plans. It learns from feedback only insofar as it decreases the time spent on bad plans by settings its timeout to the runtime of the best plan seen (as, unlike with BO, there is no point in executing plans worse than the best-seen). The initial timeout greatly affects how many plans can be tried within a given time budget, as a lower timeout results in more plans tried within a particular time period, but we do not know a priori what the runtime of the fastest possible plan is. We initialize the \random plan generation process with the PostgreSQL default plan runtime as the initial timeout.

All queries are executed against PostgreSQL version 16.3. For all workloads, we disable JIT and set \texttt{join\_collapse\_limit} to 1. Physical operator hints are specified using \texttt{pg\_hint\_plan}. We configure PostgreSQL with 32GB shared buffers and 16MB work memory. For the Stack queries, we disabled GEQO as it was causing high variability in plan performance.

We create indexes on all join keys on the IMDB, Stack, \edit{and DSB} datasets. For Stack, some tables have compound join keys. We build all indexes such that for each set of join predicates between two tables present in all queries in the workload, an index exists containing all of the referenced columns on each side.

Our VAE is based on the transformer VAE architecture introduced in \cite{lolbo}. For each database, the set of training query plans were divided into an 80\%/20\% train-test split, over which the VAE was trained for $800,000$ steps. To determine an appropriate latent space size, VAEs were trained with varying latent dimensionality on the IMDB dataset to evaluate the trade-off between compression and reconstruction, results of which are shown in \Cref{table:vae_accuracy}. A latent space of $64$ dimensions was chosen as it represented a good balance between latent dimension and reconstruction accuracy. Each VAE was trained on 2 Ampere A6000s for $\sim24$ hours. On the GPU cloud that we were using, this cost roughly $\$24$ for each VAE.

\begin{table}
    \centering
    \begin{tabular}{c|c}
        \textbf{Latent Dimension} & \textbf{Reconstruction Accuracy} \\
        \hline
        128 & 97.93\% \\
        64 & 89.67\% \\
        32 & 58.71\% \\
        16 & 24.79\% \\
        8 & 8.49\% \\
        \end{tabular}
    \caption{VAE reconstruction accuracy on the validation set at different latent dimensions, higher is better.}
    \label{table:vae_accuracy}
\end{table}

\subsection{Plan Optimization} \label{sec:eval-overall}
\begin{figure*}
    \centering
    \includegraphics[width=\textwidth]{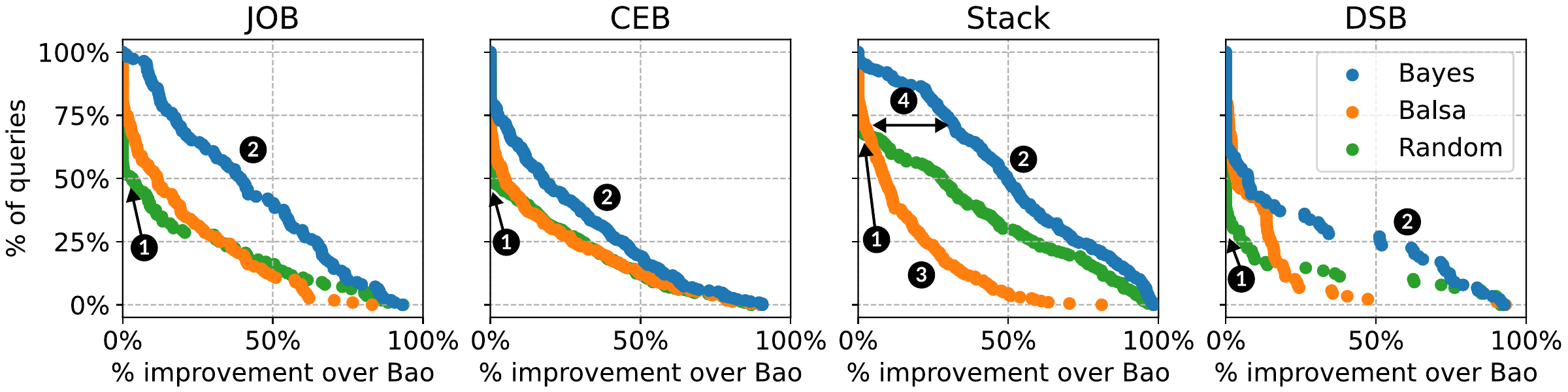}
    \caption{Best plans found at the end of optimization with each technique for each workload. Towards the top right corner is better. \circleOne On each workload, \random fails to find any improvement over the best Bao plan for 30-50\% of queries. \circleTwo \sysname always finds the most improvement compared to the other methods, with this difference being more pronounced on JOB and Stack than on CEB. On Stack, \sysname finds over 2\texttimes~ improvement on 50\% of queries. \circleThree Balsa underperforms when used as an offline optimizer on Stack due to longer query runtimes limiting exploration by the RL algorithm. \circleFour \sysname finds improvement on the \textasciitilde 25\% of Stack queries where the other two techniques find none.}
    \label{fig:workload-summaries}
\end{figure*}

In order to answer \labelcref{rq:optimization-tradeoff}, we executed each of our baseline optimization techniques for several hours for each query in each of our three workloads. \Cref{fig:workload-summaries} visualizes the results at the end of optimization for \bo, Balsa, and \random across our three workloads by comparing the cumulative distribution of queries that achieve at least a certain percentage improvement in plan runtime compared to Bao. ``\% improvement over Bao'' refers to the percentage reduction in plan runtime for a certain query compared to the runtime of the optimal Bao plan (e.g. a reduction in runtime from 1 second to 200 milliseconds would be an improvement of 80\%). The visual separation between the series for the different techniques trending towards the top right of each plot indicates gaps in performance in which one technique is finding plans for more queries with better performance than another technique.

We strove to make comparisons between techniques fair by giving each optimization technique the same optimization budget, since offline optimization techniques can hypothetically be executed for an indefinite amount of time to find potentially faster plans. We only considered time spent executing proposed plans against the database as consuming budget and exclude the overhead of executing each technique (the overhead for \sysname is analyzed in \Cref{sec:overhead}). Each optimization technique was executed for 4000 plan executions. We choose this because query runtimes span from tens of milliseconds to tens of seconds, and we did not want to optimize some queries with $1000\times$ more observations than others.

These plots make obvious the fact that \sysname is finding more plans with better performance compared to the baselines across all three workloads. JOB and Stack are highly differentiated, while all techniques perform similarly on CEB. \edit{DSB is notable in its proportion of queries for which no technique finds much improvement over Bao, but for those queries where improvement can be found, \sysname is moderately differentiated from the baselines.}

\subsection{Case Studies} \label{sec:eval-case-studies}
\begin{figure*}
    \centering
    \begin{subfigure}[t]{0.32\linewidth}
        \centering
        \includegraphics[width=\linewidth]{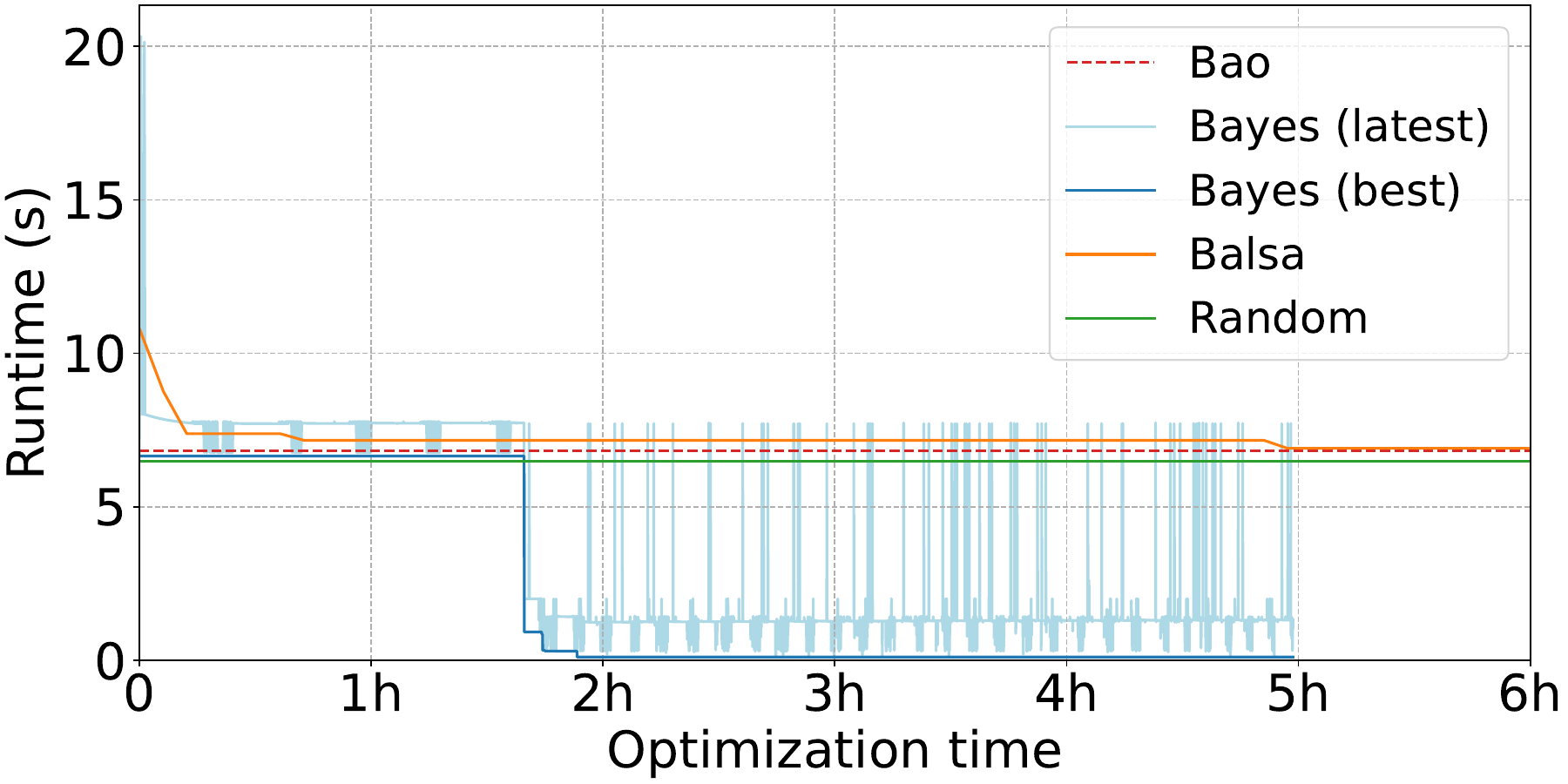}
        \caption{STACK\_Q2-025: \bo finds a better plan than any of the other techniques}
        \label{fig:case-study-best}
    \end{subfigure}
    \hfill
    \begin{subfigure}[t]{0.32\linewidth}
        \centering
        \includegraphics[width=\linewidth]{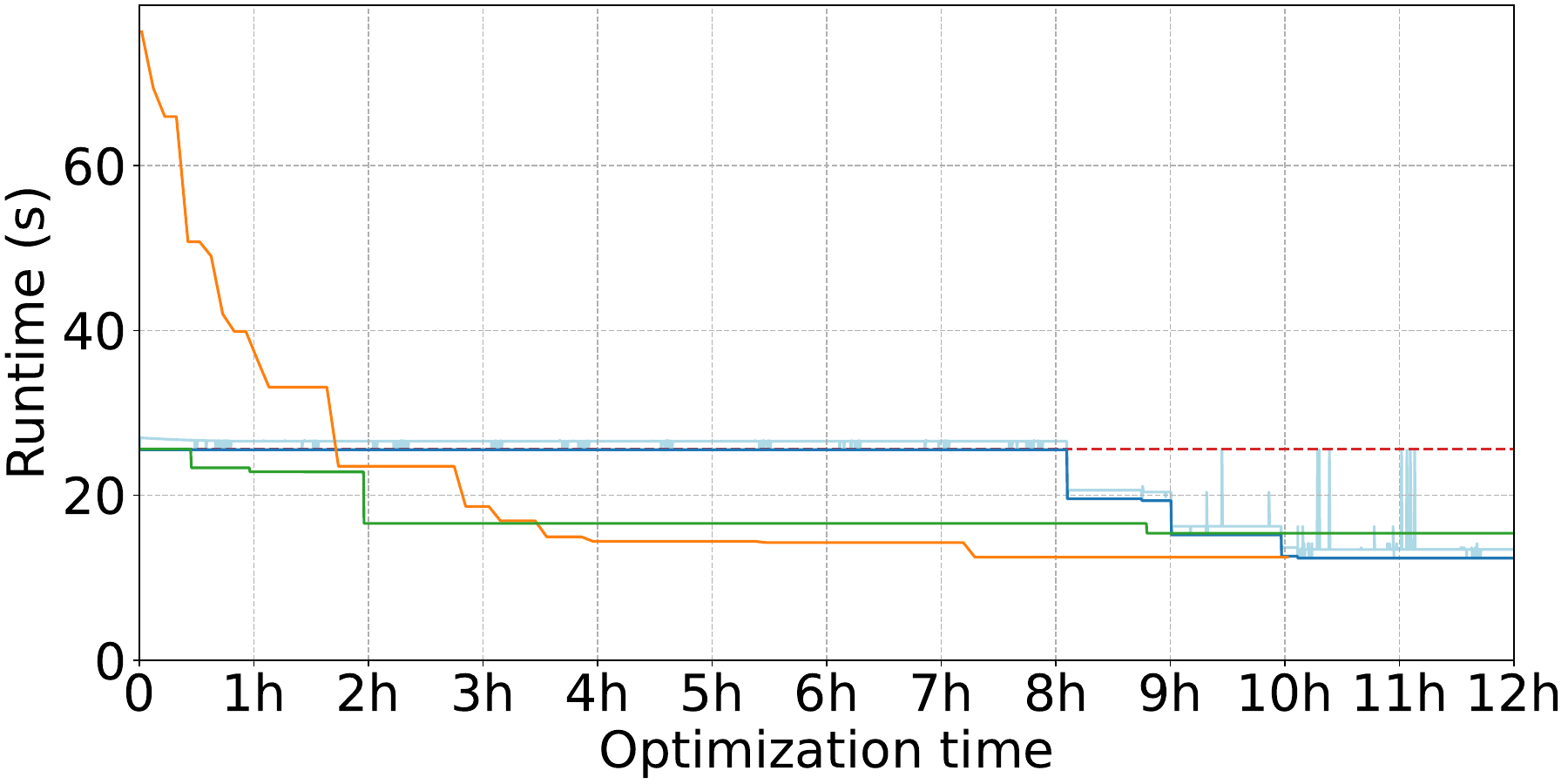}
        \caption{CEB\_11A102: All techniques find good plans, though \bo takes notably longer.}
        \label{fig:case-study-moderate}
    \end{subfigure}
    \hfill
    \begin{subfigure}[t]{0.32\linewidth}
        \centering
        \includegraphics[width=\linewidth]{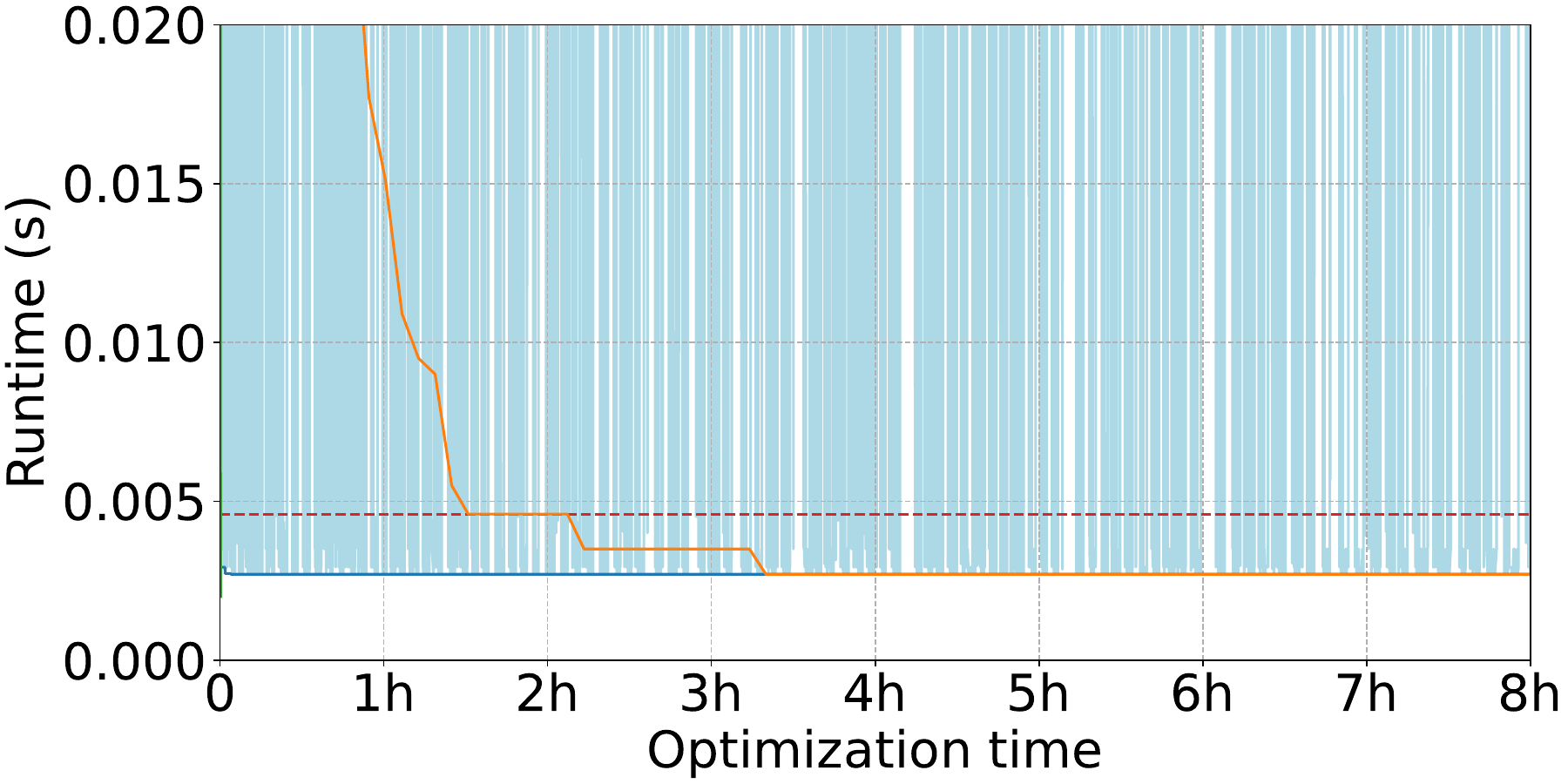}
        \caption{JOB\_1B: All techniques converge to the same (very small) runtime.}
        \label{fig:case-study-worst}
    \end{subfigure}
    \caption{Case studies showing optimization time vs. plan performance for individual queries. Lower is better.}
    \label{fig:case-studies}
\end{figure*}

We select three queries from the optimization workloads to illustrate the different optimization outcomes for \sysname. The optimization runs for these three queries are visualized in \Cref{fig:case-studies}. All three plots visualize the runtime of the best plan found so far over the course of an optimization run across all of the optimization techniques. Bao is visualized as a horizontal line showing the runtime of the best plan because it cannot improve after its hint sets are executed. The light blue ``Bayes (latest)'' line illustrates the runtime of the plan run most-recently by BO, hence its constant fluctuation as BO explores the space of possible plans. The $x$ axis in each of these plots captures cumulative execution time on the database and ignores time spent in the rest of the optimization algorithms such as plan proposal and model updates.

In \Cref{fig:case-study-best}, we highlight a case where the advantages of BO are most obvious. In the first hour, BO is exploring the space immediately around its initialization points, executing plans for slightly longer than the best Bao initialization due to uncertainty-based timeouts. Around 1.5 hours into the optimization run, it finds a plan within a trust region that has substantially better performance and rapidly exploits this new information by trying nearby plans. Note that in most of the queries made afterwards, timeouts are lowered to be closer to the new optimal as the BO surrogate model no longer needs to know if plans are much worse than the new optimal. We also note that neither \random nor Balsa manages to find a plan better than the Bao optimal.

In \Cref{fig:case-study-moderate}, all optimization strategies converge to the same best plan runtime after several hours of execution. BO takes notably longer than \random and Balsa to find this plan. That \random finds an optimized plan so quickly suggests that the space of possible plans contains many good plans that perform approximately this well, but they may be quite different from the initialization points given to BO. We also note that despite the fact that we give Balsa training examples including the Bao optimal plan, it begins its search with plans that are considerably worse before eventually passing the Bao optimal.

\Cref{fig:case-study-worst} shows a query for which all techniques converge relatively quickly to the same plan runtime and do not make any progress afterwards. We note that this optimized plan executes in 2ms, which is short enough to be indistinguishable from noise in our experimental setup. We observe that Balsa takes the longest time to arrive at this plan whereas the other techniques find it nearly instantly, which we take as further evidence of the unsuitability of RL-based algorithms for offline optimization. The plateaus in Balsa's progress occur when it is exploiting its best known plan in order to minimize regret instead of trying to find a faster plan.

\subsection{Timeout Ablation Study} \label{sec:eval-timeout-ablation}
\begin{figure}
    \centering
    \begin{subfigure}[t]{\linewidth}
        \centering
        \includegraphics[width=0.8\linewidth]{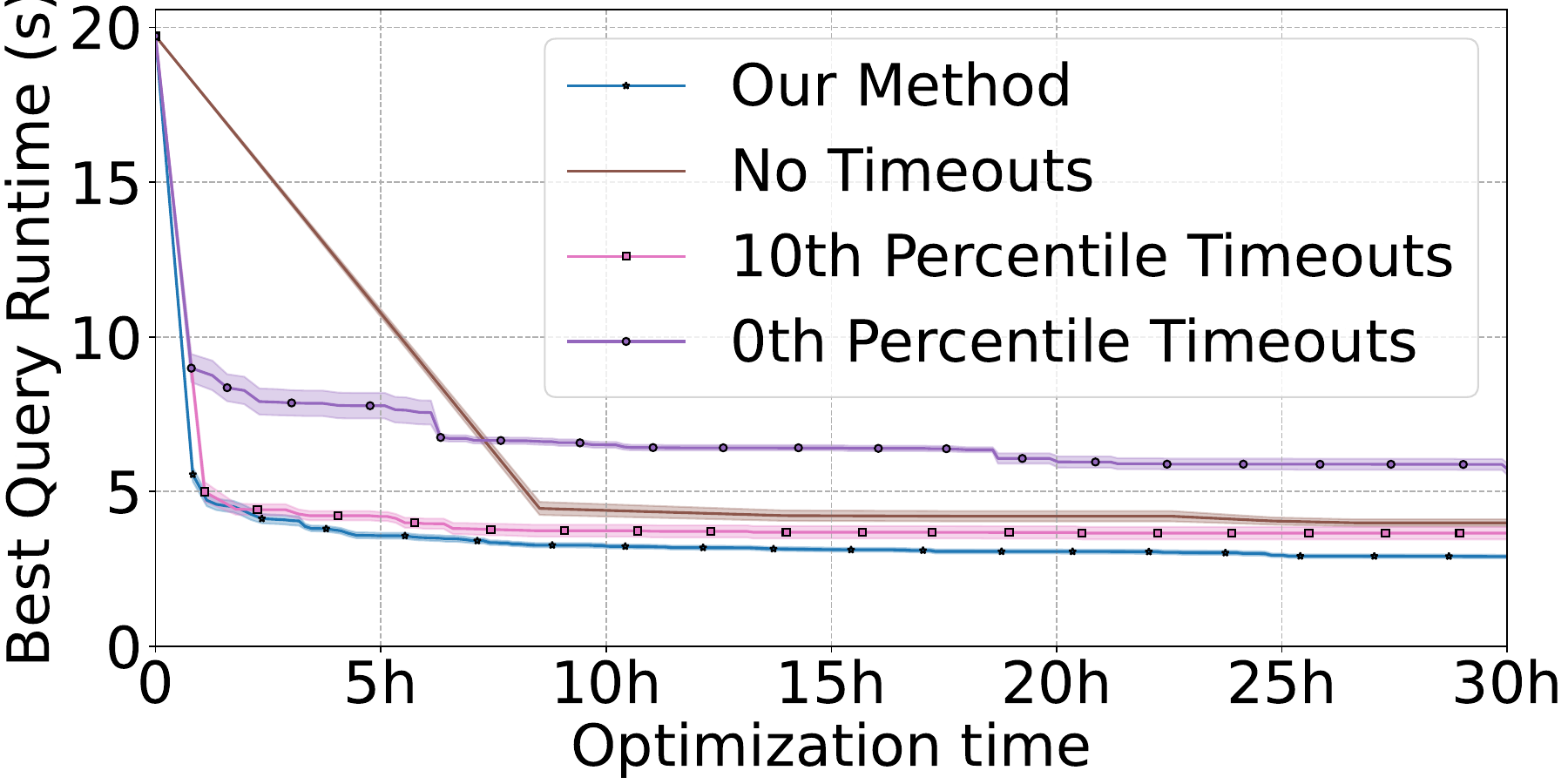}
        \caption{Optimization time vs. best plan runtime for different timeout schemes when running BO.}
        \label{fig:bo-ablation-timeout}
    \end{subfigure}
    \hfill
    \begin{subfigure}[t]{\linewidth}
        \centering
        \includegraphics[width=0.8\linewidth]{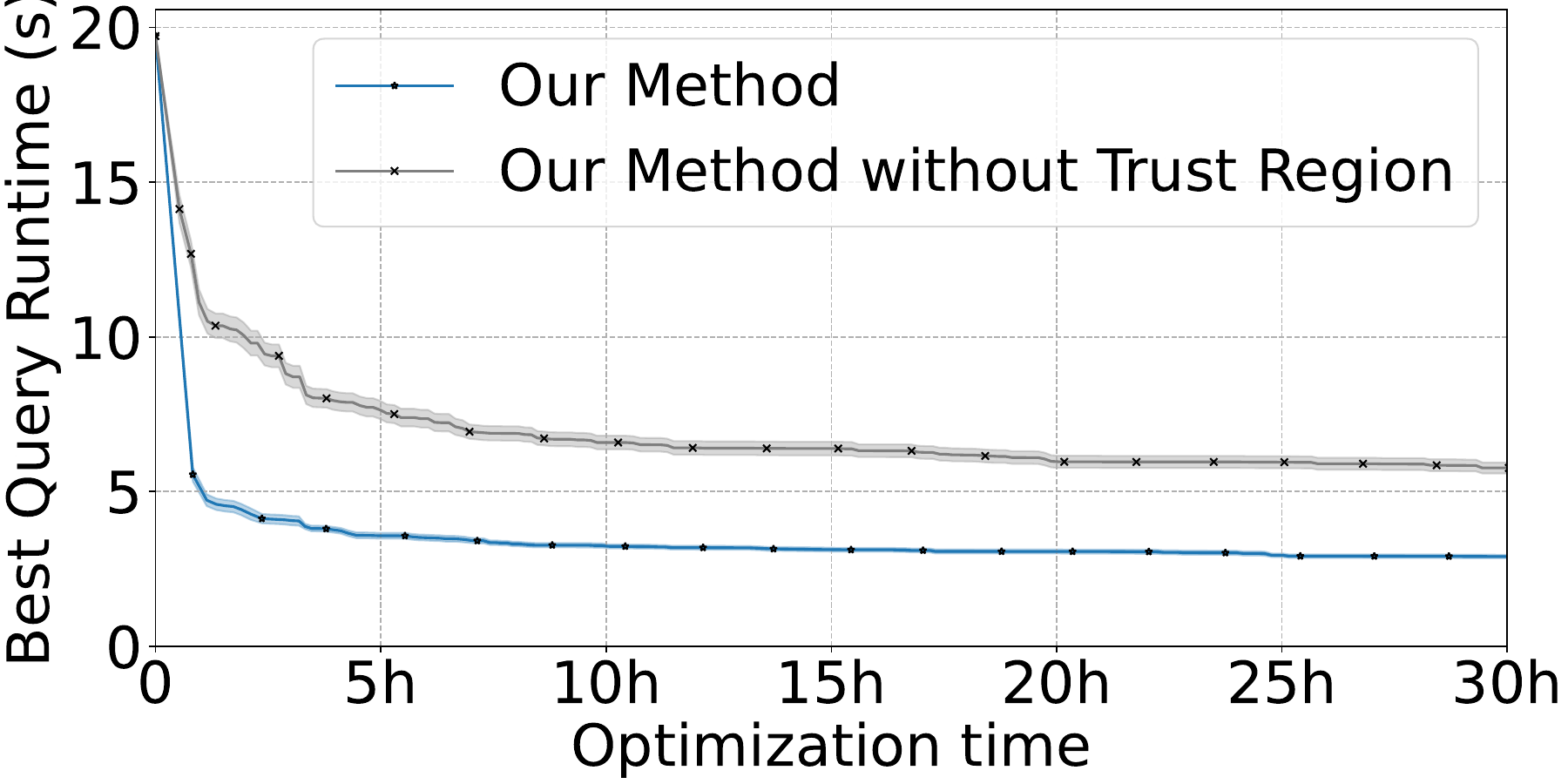}
        \caption{Optimization time vs. best plan runtime with and without the local ``Trust Region'' optimization.}
        \label{fig:bo-ablation-trust-region}
    \end{subfigure}
    \caption{Ablation study of our novel BO scheme.}
    \label{fig:bo-ablation}
\end{figure}

To answer \labelcref{rq:bo-ablation}, we justify our usage of a novel timeout strategy, as well as the choice to perform local BO based on trust regions, \footnote{\edit{Which, despite the name, is a \emph{global} optimization scheme: see \Cref{sec:bo-background}.}} via an ablation study visualized in \Cref{fig:bo-ablation}. As described in \Cref{sec:bo-background}, we utilize timeouts in order to manage the impact of executing terrible plans that take many orders of magnitude longer to execute than the optimal plan, wasting optimization budget for little gain.

In \Cref{fig:bo-ablation-timeout}, we show the results of using different timeout schemes when optimizing a single JOB query. Intuitively, using timeouts longer than the runtime of the current best-seen plan (i.e. the 0th percentile timeout) allows the surrogate model to learn more about regions of the space of plans that it is less certain about. Since we use censored observations, two plans (and their surrounding regions of plan space) will look exactly the same if they both timed out after 1 second, even if one plan would have executed for 1.2 seconds and the other for 2 hours. By using longer timeouts, the surrogate model gains greater confidence in whether a particular region of the space is still promising to explore or if it is clearly terrible. As shown in the plot, our uncertainty-based method for determining the timeout threshold results in BO finding faster plans while consuming less optimization budget. In fact, using the best-seen runtime as the timeout causes BO to find the worst plan by the end of optimization, perhaps due to the artificially low timeout uniformly discouraging BO from exploring the space of plans. 

We also justify the choice to use trust region-based local BO instead of global BO by performing another ablation, shown in \Cref{fig:bo-ablation-trust-region}. Our choice to represent the space of plans via the latent space of a VAE comes at the cost of high dimensionality (64). Though it is possible in principle that local BO will miss globally optimal plans, in our experiments we find that local BO initialized with plans derived from the PostgreSQL optimizer can find highly-optimized plans for many queries. In the ablation plots, we can see that even after many hours of optimization, global BO does not catch up to the quality of plans found by local BO due to the exponentially larger space of plans that it must explore.

\subsection{Data Drift After Optimization} \label{sec:eval-drift}
\begin{figure*}
    \centering
    \includegraphics[width=\textwidth]{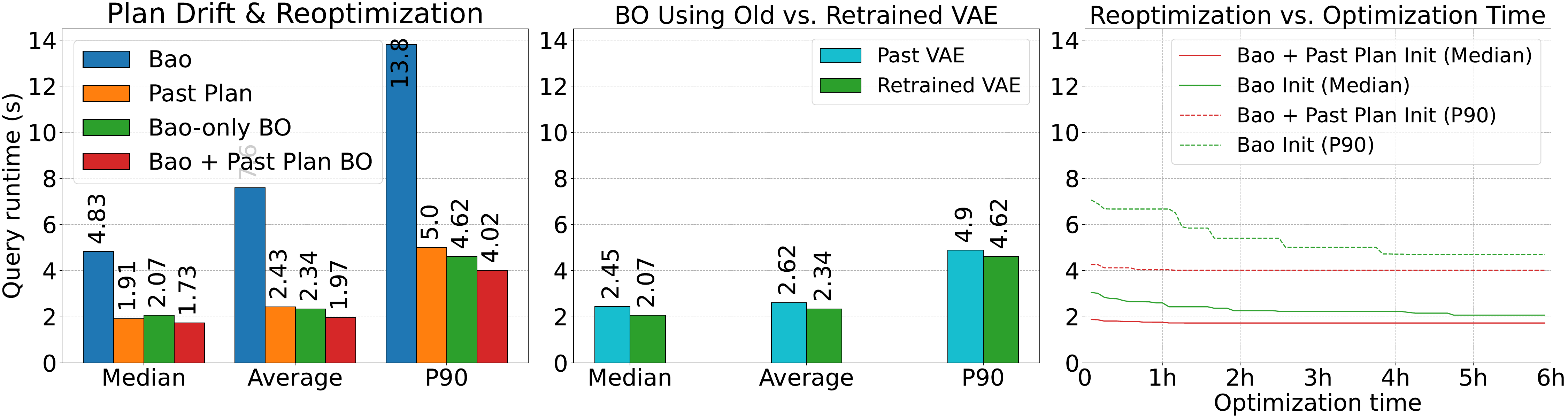}
    \caption{Left: Plans from the past vs. plans optimized in the future. Middle: BO run results when using the VAE from the past vs. the VAE retrained in the future. Right: Optimization speed of BO initialized with Bao vs. those including the past plan.}
    \label{fig:drift}
\end{figure*}

\begin{figure}
    \centering
    \includegraphics[width=\linewidth]{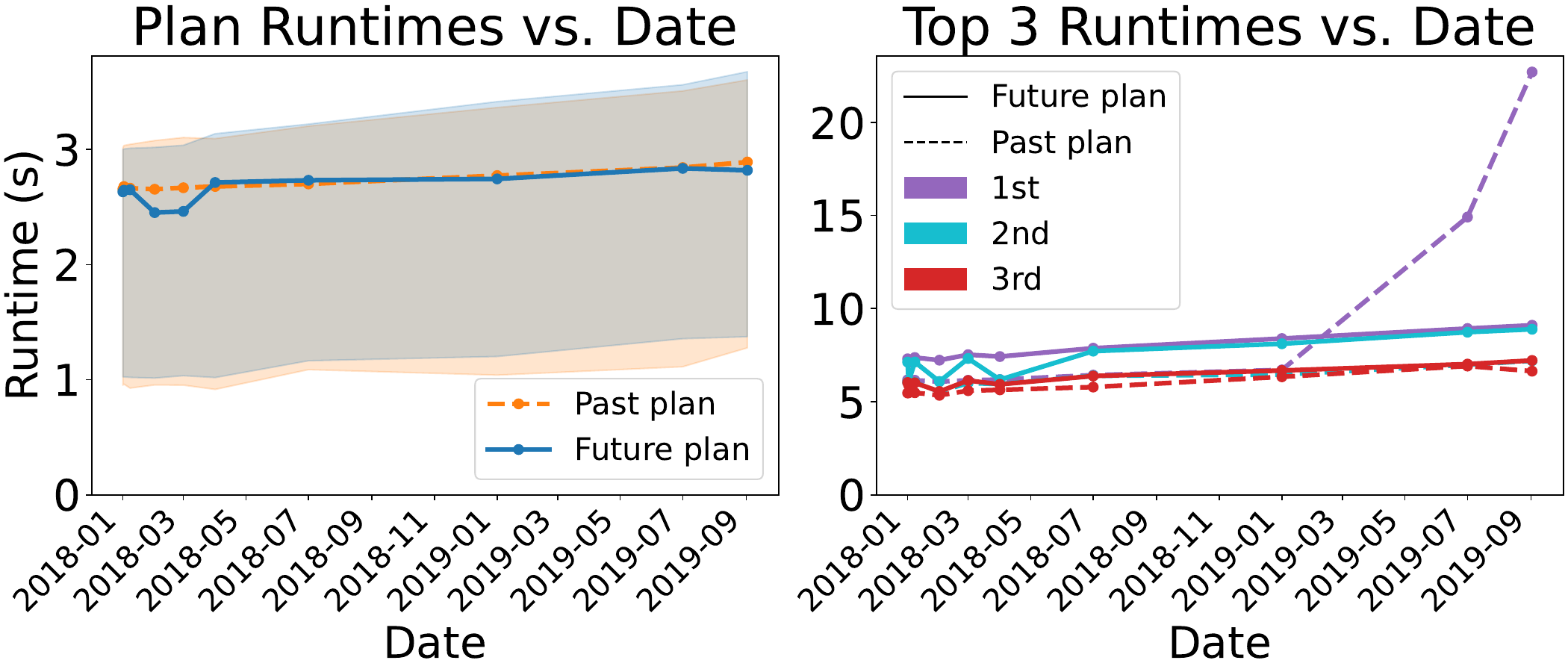}
    \caption{Left: Performance of plans optimized in the past vs. plans optimized in the future executed on dates in between. Shaded regions show the 25th to 75th percentile runtimes. Right: The top 3 longest plan runtimes per date.}
    \label{fig:drift_over_dates}
\end{figure}

In order to model data drift, we modified the StackOverflow dataset by deleting all rows in all tables with timestamps after 2017, as well as the transitive closure of all rows whose foreign keys became invalidated as a result of deleting those rows. This reduced the overall dataset size by roughly 20\%, with individual tables decreasing in row count between 0\% and 28\%. This deletion effectively restored the database to a snapshot from the end of 2017, while the original StackOverflow dataset snapshot was taken in late 2019. \edit{We present this two year shift as a worst-case scenario for data drift, expecting that if plan performance were to degrade due to data drift, it would degrade more over a longer period of time.} For the rest of this section, we will refer to this 2017 snapshot as the ``past'' and the original 2019 snapshot as the ``future''.

We sought to answer three questions about the impact of data drift:

\begin{enumerate}[label=\textbf{RQ3.\arabic*.}]
    \item How do the past plans perform in comparison to plans produced by reoptimizing from scratch on the future dataset? \label{rq:drift-degradation}
    \item Is it important to retrain the VAE for the future dataset before performing BO? \label{rq:drift-vae}
    \item Does including the past plan as an initialization point in the future BO process speed up optimization? \label{rq:drift-reoptimize}
\end{enumerate}

To answer \labelcref{rq:drift-degradation}, we trained a VAE for the past dataset using the procedure described in \Cref{sec:technique-vae}, planning the sampled queries using PostgreSQL with the past dataset, akin to conducting the full \sysname offline optimization process in the past. We then performed BO against the past dataset using this past version of the VAE and a reduced set of 50 (out of our original 200) queries. We then took the past query plans and executed them against the future dataset, effectively simulating the real-world use case of continuing to use previously optimized query plans well past when data drift may have rendered those plans suboptimal.

The results of executing these past plans against the future dataset are visualized in the left plot of \Cref{fig:drift}. We compare past plans against the Bao optimal plans for the future dataset and the future plans produced by BO from our initial experiment. We also compare them to the results of performing BO for a short time (about 1 hour) initialized with the past plan in addition to the Bao plans, discussed further below. Despite the substantial data drift, past plans perform about as well as if we had performed BO for the first time on the future dataset and continue to perform much better than the best Bao hint. This suggests that the optimality of most of the plans found by BO is not affected much data drift.

\edit{We also executed these past and future plans on dates in between the ``past'' and ``future'' endpoints, shown in \Cref{fig:drift_over_dates}. For the vast majority of queries, there is not a significant difference in runtime between the past and future plans, but as shown by the dramatic increase in runtime of the longest-running past plan, it is indeed possible for data drift to render a plan suboptimal.}


To answer \labelcref{rq:drift-vae}, we performed a set of BO runs against the future dataset using the VAE trained on the past dataset, simulating the real-world use case of attempting to perform offline optimization in the future without retraining the VAE. The results are visualized in the middle plot of \Cref{fig:drift}. We observe that keeping the VAE up-to-date has a non-negligible effect on the optimality of plans found by BO. Given the small cost of training the VAE compared to the rest of the BO process, it seems worthwhile to periodically retrain the VAE to account for data drift.


To answer \labelcref{rq:drift-reoptimize}, we performed a set of BO runs against the future dataset initialized with both the Bao initialization and the optimized past plan, simulating reoptimization when a query had previously been optimized in the past. For these BO runs, we used the VAE trained on the future dataset. The aggregate results in red for reoptimization in the left plot of \Cref{fig:drift} show not only that reoptimization to account for data drift is viable, but that it tends to produce the best plans across the entire workload. In the right side of the same figure, we observe that the BO runs converge to optimized plans much more quickly than a run started from scratch. Reoptimizations ran for an average of 1.5 hours compared to from-scratch optimization runs, which ran for an average of 8.2 hours. This suggests that plans generated by \sysname can be brought up-to-date to account for data drift while consuming much less optimization budget than the initial offline optimization run.

\subsection{Few-Shot LLM from BO Results} \label{sec:eval-llm}
\begin{figure}
    \includegraphics[width=\linewidth]{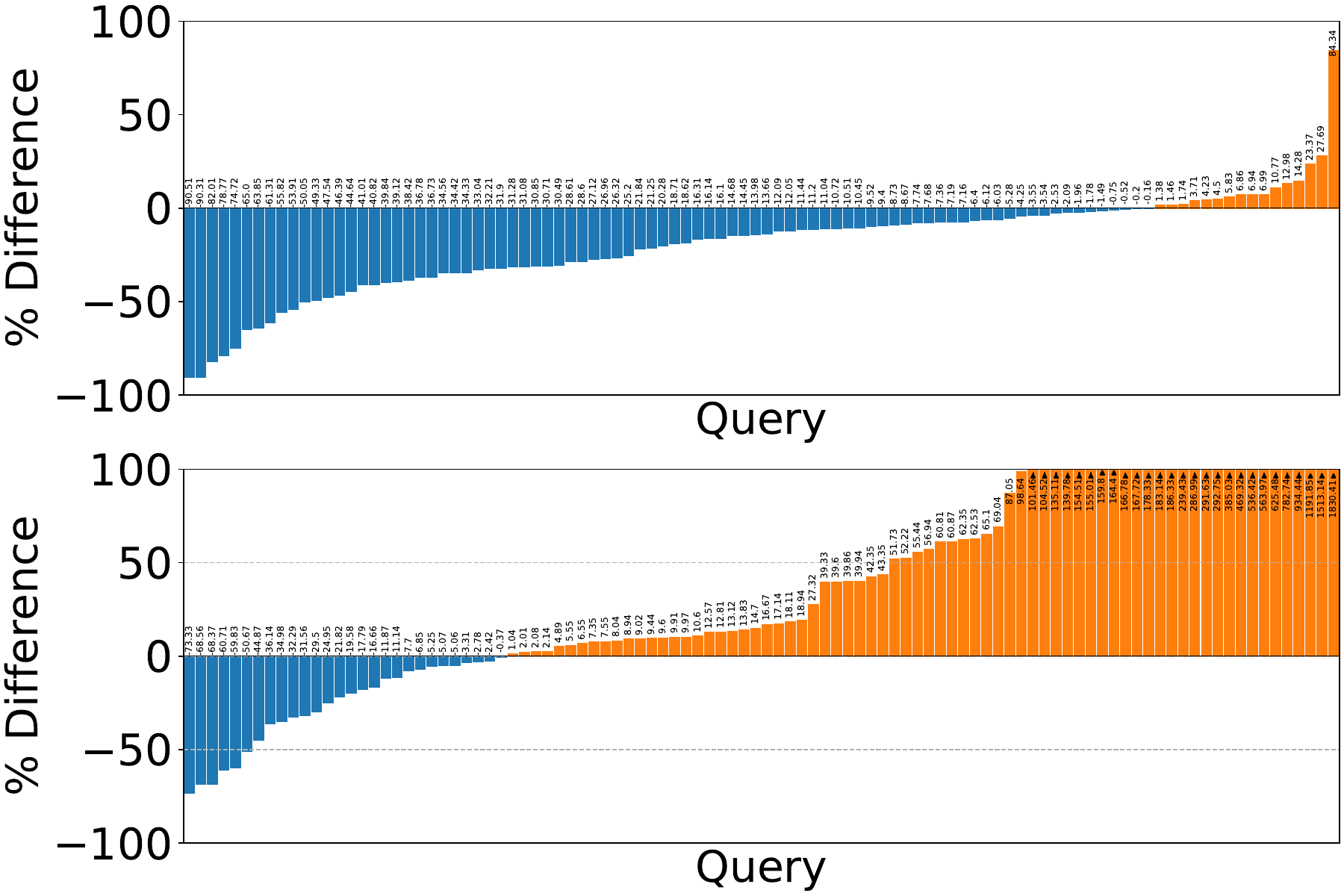}
    \caption{Top: LLM trained on plans of the same template. Bottom: LLM trained on plans not of the same template. Lower (more blue) is better.}
    \label{fig:llm-experiments}
\end{figure}

As a byproduct of performing this experimental evaluation, we generated plans that, to our knowledge, are the best plans that can currently be found for the queries in our evaluation workloads. We hypothesized that a large language model fine-tuned using these high-quality plans could potentially be a better few-shot plan optimizer than existing techniques. Here, we evaluate the LLM's ability to generate initialization points for \sysname and defer offline optimization initialized using the LLM outputs to future work.

In order to answer \labelcref{rq:llm}, we performed two experiments. In the first experiment, visualized in the top plot of \Cref{fig:llm-experiments}, we fine-tuned GPT-4o mini on the fastest 10 optimized plans for each query from running BO on the CEB workload as described in \Cref{sec:initialization_strategies}. This training set included queries from all query templates present in the workload. We then compared the best runtime out of 50 plans (giving it as many plans as Bao) per-query generated by the LLM for a particular query against the runtime of the optimal Bao plan.

In the second experiment, we performed the same process, but withheld BO results from two query templates from the fine-tuning process. We then performed the same test, comparing the best runtime out of 50 plans per-query from the LLM against the optimal Bao plans. The results are visualized in the bottom plot of \Cref{fig:llm-experiments}. The LLM clearly does not perform as well when it has not been fine-tuned with results from the same query template.

\subsection{\sysname Overhead} \label{sec:overhead}
\begin{figure}
    \centering
    \includegraphics[width=0.9\linewidth]{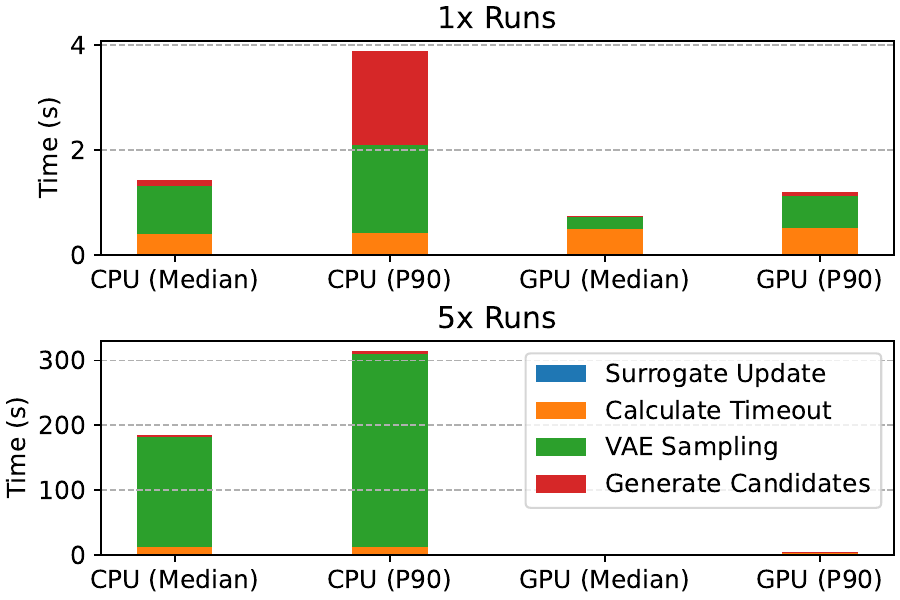}
    \caption{Overheads per iteration of BO. Top: 1x simultaneous BO run. Bottom: 5x simultaneous BO runs.}
    \label{fig:overheads}
\end{figure}

We measured the overhead of running \sysname under multiple conditions in order to better understand its resource requirements and to determine if a GPU is necessary to perform optimization. We recorded time spent in each part of the optimization loop when executing the BO components on both CPU and GPU while varying the number of simultaneous BO runs. The results are visualized in \Cref{fig:overheads}. We find that with only one BO run, while overhead on a CPU is worse than on a GPU, the absolute time spent on overhead (i.e., everything but query execution) is in the single-digit seconds range. For sufficiently long-running queries, this CPU overhead may be tolerable. However, even with just 5 simultaneous runs, \edit{we note that VAE sampling scales significantly better on GPUs than CPUs -- this is attributable to the GPU's hardware support for scaled dot product attention~\cite{dao_flashattention}}.

\subsection{Comparison to LimeQO} \label{sec:limeqo-comparison}
\begin{figure}
    \centering
    \includegraphics[width=0.9\linewidth]{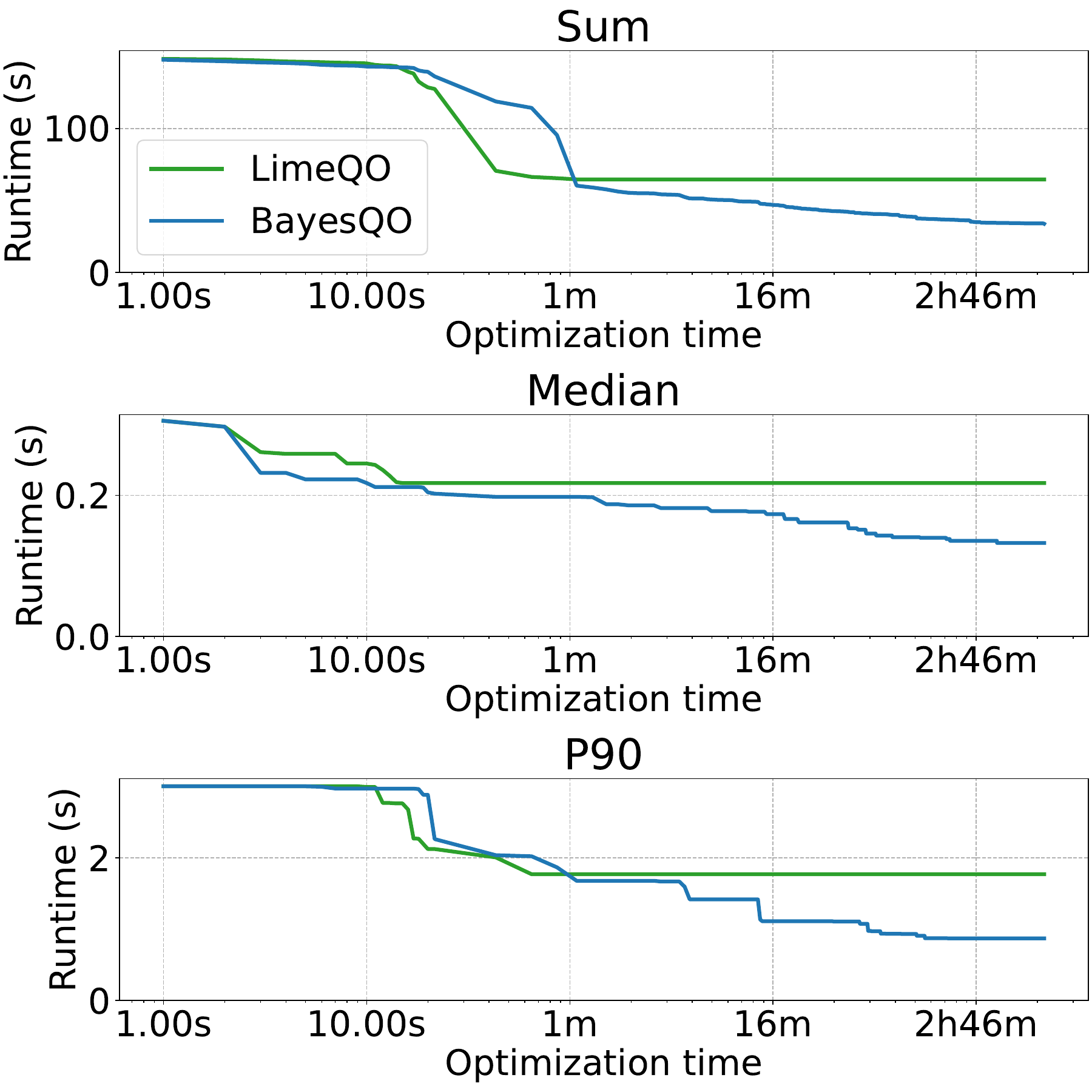}
    \caption{Optimization using \sysname vs. LimeQO across the entire JOB. Lower is better. The x-axis is log scale so both final performance and the initial improvement is visible.}
    \label{fig:limeqo-comparison}
\end{figure}

\edit{
LimeQO ~\cite{limeqo} is another work that performs offline optimization for repetitive workloads, but LimeQO and chiefly differs from \sysname in that its potential optimizations are limited to finding optimal hints from a small set, as opposed to \sysname which constructs entire join orders. LimeQO uses the hint sets from Bao~\cite{bao}, which \sysname also uses to initiate the Bayesian optimization process. As shown in \Cref{fig:limeqo-comparison}, both techniques explore all of the Bao hints: once LimeQO has exhausted all of the hints, there are no remaining avenues for further optimization, whereas BayesQO continues exploring and finds better plans.
}
\section{Conclusion}
In this work, we presented \sysname, an \emph{offline} learned query optimizer. \sysname uses modern Bayesian optimization techniques to search for fast query plans for important repeated queries. Experimentally, we show that \sysname was able to optimize nearly every query in a set of common, non-synthetic benchmarks, sometimes achieving multiple orders-of-magnitude improvements.

In the future, we plan to integrate \sysname into a wider variety of database systems. Given the promising results from our LLM experiment, we plan to investigate how we can ``close the loop,'' using the LLM to generate initialization data for future optimization runs. We are also excited to examine how Bayesian optimization can be applied to other databases problems such as automatic index selection, data layout, and benchmark curation.


\bibliographystyle{ACM-Reference-Format}
\bibliography{references,ryan-cites-long,s3}


\begin{thebibliography}{106}


\ifx \showCODEN    \undefined \def \showCODEN     #1{\unskip}     \fi
\ifx \showDOI      \undefined \def \showDOI       #1{#1}\fi
\ifx \showISBNx    \undefined \def \showISBNx     #1{\unskip}     \fi
\ifx \showISBNxiii \undefined \def \showISBNxiii  #1{\unskip}     \fi
\ifx \showISSN     \undefined \def \showISSN      #1{\unskip}     \fi
\ifx \showLCCN     \undefined \def \showLCCN      #1{\unskip}     \fi
\ifx \shownote     \undefined \def \shownote      #1{#1}          \fi
\ifx \showarticletitle \undefined \def \showarticletitle #1{#1}   \fi
\ifx \showURL      \undefined \def \showURL       {\relax}        \fi
\providecommand\bibfield[2]{#2}
\providecommand\bibinfo[2]{#2}
\providecommand\natexlab[1]{#1}
\providecommand\showeprint[2][]{arXiv:#2}

\bibitem[Anneser et~al\mbox{.}(2023)]%
        {autosteer}
\bibfield{author}{\bibinfo{person}{Christoph Anneser}, \bibinfo{person}{Nesime
  Tatbul}, \bibinfo{person}{David Cohen}, \bibinfo{person}{Zhenggang Xu},
  \bibinfo{person}{Prithvi Pandian}, \bibinfo{person}{Nikolay Leptev}, {and}
  \bibinfo{person}{Ryan Marcus}.} \bibinfo{year}{2023}\natexlab{}.
\newblock \showarticletitle{{AutoSteer}: {Learned} {Query} {Optimization} for
  {Any} {SQL} {Database}}.
\newblock \bibinfo{journal}{\emph{PVLDB}} \bibinfo{volume}{14},
  \bibinfo{number}{1} (\bibinfo{date}{Aug.} \bibinfo{year}{2023}).
\newblock
\showISSN{2150-8097}
\urldef\tempurl%
\url{https://doi.org/10.14778/3611540.3611544}
\showDOI{\tempurl}


\bibitem[Astrahan et~al\mbox{.}(1976)]%
        {system_r}
\bibfield{author}{\bibinfo{person}{M.~M. Astrahan}, \bibinfo{person}{M.~W.
  Blasgen}, \bibinfo{person}{D.~D. Chamberlin}, \bibinfo{person}{K.~P.
  Eswaran}, \bibinfo{person}{J.~N. Gray}, \bibinfo{person}{P.~P. Griffiths},
  \bibinfo{person}{W.~F. King}, \bibinfo{person}{R.~A. Lorie},
  \bibinfo{person}{P.~R. McJones}, \bibinfo{person}{J.~W. Mehl},
  \bibinfo{person}{G.~R. Putzolu}, \bibinfo{person}{I.~L. Traiger},
  \bibinfo{person}{B.~W. Wade}, {and} \bibinfo{person}{V. Watson}.}
  \bibinfo{year}{1976}\natexlab{}.
\newblock \showarticletitle{System R: relational approach to database
  management}.
\newblock \bibinfo{journal}{\emph{ACM Trans. Database Syst.}}
  \bibinfo{volume}{1}, \bibinfo{number}{2} (\bibinfo{date}{jun}
  \bibinfo{year}{1976}), \bibinfo{pages}{97–137}.
\newblock
\showISSN{0362-5915}
\urldef\tempurl%
\url{https://doi.org/10.1145/320455.320457}
\showDOI{\tempurl}


\bibitem[Babu et~al\mbox{.}(2005)]%
        {rio_reopt}
\bibfield{author}{\bibinfo{person}{Shivnath Babu}, \bibinfo{person}{Pedro
  Bizarro}, {and} \bibinfo{person}{David DeWitt}.}
  \bibinfo{year}{2005}\natexlab{}.
\newblock \showarticletitle{Proactive re-optimization with {Rio}}. In
  \bibinfo{booktitle}{\emph{Proceedings of the 2005 {ACM} {SIGMOD}
  international conference on {Management} of data}}
  \emph{(\bibinfo{series}{{SIGMOD} '05})}. \bibinfo{publisher}{Association for
  Computing Machinery}, \bibinfo{address}{New York, NY, USA},
  \bibinfo{pages}{936--938}.
\newblock
\showISBNx{978-1-59593-060-6}
\urldef\tempurl%
\url{https://doi.org/10.1145/1066157.1066294}
\showDOI{\tempurl}


\bibitem[Behr et~al\mbox{.}(2023)]%
        {qo_rank}
\bibfield{author}{\bibinfo{person}{Henriette Behr}, \bibinfo{person}{Volker
  Markl}, {and} \bibinfo{person}{Zoi Kaoudi}.} \bibinfo{year}{2023}\natexlab{}.
\newblock \showarticletitle{Learn {What} {Really} {Matters}: {A}
  {Learning}-to-{Rank} {Approach} for {ML}-based {Query} {Optimization}}. In
  \bibinfo{booktitle}{\emph{Database {Systems} for {Business}, {Technology},
  and the {Web} 2023}} \emph{(\bibinfo{series}{{BTW} '23})},
  \bibfield{editor}{\bibinfo{person}{Birgitta König-Ries},
  \bibinfo{person}{Stefanie Scherzinger}, \bibinfo{person}{Wolfgang Lehner},
  {and} \bibinfo{person}{Gottfried Vossen}} (Eds.).
  \bibinfo{publisher}{Gesellschaft für Informatik e.V.}
\newblock
\urldef\tempurl%
\url{https://doi.org/10.18420/BTW2023-25}
\showDOI{\tempurl}


\bibitem[Cereda et~al\mbox{.}(2021)]%
        {cgptuner}
\bibfield{author}{\bibinfo{person}{Stefano Cereda}, \bibinfo{person}{Stefano
  Valladares}, \bibinfo{person}{Paolo Cremonesi}, {and}
  \bibinfo{person}{Stefano Doni}.} \bibinfo{year}{2021}\natexlab{}.
\newblock \showarticletitle{CGPTuner: a contextual gaussian process bandit
  approach for the automatic tuning of IT configurations under varying workload
  conditions}.
\newblock \bibinfo{journal}{\emph{Proc. VLDB Endow.}} \bibinfo{volume}{14},
  \bibinfo{number}{8} (\bibinfo{date}{apr} \bibinfo{year}{2021}),
  \bibinfo{pages}{1401–1413}.
\newblock
\showISSN{2150-8097}
\urldef\tempurl%
\url{https://doi.org/10.14778/3457390.3457404}
\showDOI{\tempurl}


\bibitem[Chaudhuri(2009)]%
        {chaudhuri2009_rethinking_qo}
\bibfield{author}{\bibinfo{person}{Surajit Chaudhuri}.}
  \bibinfo{year}{2009}\natexlab{}.
\newblock \showarticletitle{Query optimizers: time to rethink the contract?}.
  In \bibinfo{booktitle}{\emph{Proceedings of the 2009 ACM SIGMOD International
  Conference on Management of Data}} (Providence, Rhode Island, USA)
  \emph{(\bibinfo{series}{SIGMOD '09})}. \bibinfo{publisher}{Association for
  Computing Machinery}, \bibinfo{address}{New York, NY, USA},
  \bibinfo{pages}{961–968}.
\newblock
\showISBNx{9781605585512}
\urldef\tempurl%
\url{https://doi.org/10.1145/1559845.1559955}
\showDOI{\tempurl}


\bibitem[Chen et~al\mbox{.}(2023b)]%
        {loger}
\bibfield{author}{\bibinfo{person}{Tianyi Chen}, \bibinfo{person}{Jun Gao},
  \bibinfo{person}{Hedui Chen}, {and} \bibinfo{person}{Yaofeng Tu}.}
  \bibinfo{year}{2023}\natexlab{b}.
\newblock \showarticletitle{{LOGER}: {A} {Learned} {Optimizer} {Towards}
  {Generating} {Efficient} and {Robust} {Query} {Execution} {Plans}}.
\newblock \bibinfo{journal}{\emph{Proceedings of the VLDB Endowment}}
  \bibinfo{volume}{16}, \bibinfo{number}{7} (\bibinfo{date}{March}
  \bibinfo{year}{2023}), \bibinfo{pages}{1777--1789}.
\newblock
\showISSN{2150-8097}
\urldef\tempurl%
\url{https://doi.org/10.14778/3587136.3587150}
\showDOI{\tempurl}


\bibitem[Chen et~al\mbox{.}(2023a)]%
        {leon_qo}
\bibfield{author}{\bibinfo{person}{Xu Chen}, \bibinfo{person}{Haitian Chen},
  \bibinfo{person}{Zibo Liang}, \bibinfo{person}{Shuncheng Liu},
  \bibinfo{person}{Jinghong Wang}, \bibinfo{person}{Kai Zeng},
  \bibinfo{person}{Han Su}, {and} \bibinfo{person}{Kai Zheng}.}
  \bibinfo{year}{2023}\natexlab{a}.
\newblock \showarticletitle{{LEON}: {A} {New} {Framework} for {ML}-{Aided}
  {Query} {Optimization}}.
\newblock \bibinfo{journal}{\emph{Proc. VLDB Endow.}} \bibinfo{volume}{16},
  \bibinfo{number}{9} (\bibinfo{date}{May} \bibinfo{year}{2023}),
  \bibinfo{pages}{2261--2273}.
\newblock
\showISSN{2150-8097}
\urldef\tempurl%
\url{https://doi.org/10.14778/3598581.3598597}
\showDOI{\tempurl}


\bibitem[Damasio et~al\mbox{.}(2019)]%
        {workload_reopt}
\bibfield{author}{\bibinfo{person}{Guilherme Damasio}, \bibinfo{person}{Vincent
  Corvinelli}, \bibinfo{person}{Parke Godfrey}, \bibinfo{person}{Piotr
  Mierzejewski}, \bibinfo{person}{Alex Mihaylov}, \bibinfo{person}{Jaroslaw
  Szlichta}, {and} \bibinfo{person}{Calisto Zuzarte}.}
  \bibinfo{year}{2019}\natexlab{}.
\newblock \showarticletitle{Guided automated learning for query workload
  re-optimization}.
\newblock \bibinfo{journal}{\emph{Proceedings of the VLDB Endowment}}
  \bibinfo{volume}{12}, \bibinfo{number}{12} (\bibinfo{date}{Aug.}
  \bibinfo{year}{2019}), \bibinfo{pages}{2010--2021}.
\newblock
\showISSN{2150-8097}
\urldef\tempurl%
\url{https://doi.org/10.14778/3352063.3352120}
\showDOI{\tempurl}


\bibitem[Dao et~al\mbox{.}(2022)]%
        {dao_flashattention}
\bibfield{author}{\bibinfo{person}{Tri Dao}, \bibinfo{person}{Daniel~Y. Fu},
  \bibinfo{person}{Stefano Ermon}, \bibinfo{person}{Atri Rudra}, {and}
  \bibinfo{person}{Christopher Ré}.} \bibinfo{year}{2022}\natexlab{}.
\newblock \bibinfo{title}{FlashAttention: Fast and Memory-Efficient Exact
  Attention with IO-Awareness}.
\newblock
\newblock
\showeprint[arxiv]{2205.14135}~[cs.LG]
\urldef\tempurl%
\url{https://arxiv.org/abs/2205.14135}
\showURL{%
\tempurl}


\bibitem[Deshwal and Doppa(2021)]%
        {ladder}
\bibfield{author}{\bibinfo{person}{Aryan Deshwal} {and}
  \bibinfo{person}{Janardhan~Rao Doppa}.} \bibinfo{year}{2021}\natexlab{}.
\newblock \showarticletitle{Combining Latent Space and Structured Kernels for
  {Bayesian} Optimization over Combinatorial Spaces}.
\newblock \bibinfo{journal}{\emph{CoRR}}  \bibinfo{volume}{abs/2111.01186}
  (\bibinfo{year}{2021}).
\newblock
\showeprint[arXiv]{2111.01186}
\urldef\tempurl%
\url{https://arxiv.org/abs/2111.01186}
\showURL{%
\tempurl}


\bibitem[Ding et~al\mbox{.}(2021)]%
        {ding_dsb}
\bibfield{author}{\bibinfo{person}{Bailu Ding}, \bibinfo{person}{Surajit
  Chaudhuri}, \bibinfo{person}{Johannes Gehrke}, {and} \bibinfo{person}{Vivek
  Narasayya}.} \bibinfo{year}{2021}\natexlab{}.
\newblock \showarticletitle{DSB: a decision support benchmark for
  workload-driven and traditional database systems}.
\newblock \bibinfo{journal}{\emph{Proc. VLDB Endow.}} \bibinfo{volume}{14},
  \bibinfo{number}{13} (\bibinfo{date}{Sept.} \bibinfo{year}{2021}),
  \bibinfo{pages}{3376–3388}.
\newblock
\showISSN{2150-8097}
\urldef\tempurl%
\url{https://doi.org/10.14778/3484224.3484234}
\showDOI{\tempurl}


\bibitem[Dong et~al\mbox{.}(2023)]%
        {slabcity}
\bibfield{author}{\bibinfo{person}{Rui Dong}, \bibinfo{person}{Jie Liu},
  \bibinfo{person}{Yuxuan Zhu}, \bibinfo{person}{Cong Yan},
  \bibinfo{person}{Barzan Mozafari}, {and} \bibinfo{person}{Xinyu Wang}.}
  \bibinfo{year}{2023}\natexlab{}.
\newblock \showarticletitle{{SlabCity}: {Whole}-{Query} {Optimization} {Using}
  {Program} {Synthesis}}.
\newblock \bibinfo{journal}{\emph{Proceedings of the VLDB Endowment}}
  \bibinfo{volume}{16}, \bibinfo{number}{11} (\bibinfo{date}{July}
  \bibinfo{year}{2023}), \bibinfo{pages}{3151--3164}.
\newblock
\showISSN{2150-8097}
\urldef\tempurl%
\url{https://doi.org/10.14778/3611479.3611515}
\showDOI{\tempurl}


\bibitem[Doshi et~al\mbox{.}(2023)]%
        {kepler}
\bibfield{author}{\bibinfo{person}{Lyric Doshi}, \bibinfo{person}{Vincent
  Zhuang}, \bibinfo{person}{Gaurav Jain}, \bibinfo{person}{Ryan Marcus},
  \bibinfo{person}{Haoyu Huang}, \bibinfo{person}{Deniz Altinbuken},
  \bibinfo{person}{Eugene Brevdo}, {and} \bibinfo{person}{Campbell Fraser}.}
  \bibinfo{year}{2023}\natexlab{}.
\newblock \showarticletitle{Kepler: {Robust} {Learning} for {Faster}
  {Parametric} {Query} {Optimization}}.
\newblock \bibinfo{journal}{\emph{Proceedings of the 2023 ACM SIGMOD
  Conference}} \bibinfo{volume}{1}, \bibinfo{number}{1} (\bibinfo{date}{May}
  \bibinfo{year}{2023}), \bibinfo{pages}{109}.
\newblock
\urldef\tempurl%
\url{https://doi.org/10.1145/3588963}
\showDOI{\tempurl}


\bibitem[Duggan et~al\mbox{.}(2011)]%
        {jennie_sigmod11}
\bibfield{author}{\bibinfo{person}{Jennie Duggan}, \bibinfo{person}{Ugur
  Cetintemel}, \bibinfo{person}{Olga Papaemmanouil}, {and} \bibinfo{person}{Eli
  Upfal}.} \bibinfo{year}{2011}\natexlab{}.
\newblock \showarticletitle{Performance {Prediction} for {Concurrent}
  {Database} {Workloads}}. In \bibinfo{booktitle}{\emph{Proceedings of the 2011
  {ACM} {SIGMOD} {International} {Conference} on {Management} of {Data}}}
  \emph{(\bibinfo{series}{{SIGMOD} '11})}. \bibinfo{publisher}{ACM},
  \bibinfo{address}{Athens, Greece}, \bibinfo{pages}{337--348}.
\newblock
\showISBNx{978-1-4503-0661-4}
\urldef\tempurl%
\url{https://doi.org/10.1145/1989323.1989359}
\showDOI{\tempurl}
\newblock
\shownote{tex.acmid= 1989359 tex.numpages= 12}.


\bibitem[Duggan et~al\mbox{.}(2014)]%
        {contender}
\bibfield{author}{\bibinfo{person}{Jennie Duggan}, \bibinfo{person}{Olga
  Papaemmanouil}, \bibinfo{person}{Ugur Cetintemel}, {and} \bibinfo{person}{Eli
  Upfal}.} \bibinfo{year}{2014}\natexlab{}.
\newblock \showarticletitle{Contender: {A} {Resource} {Modeling} {Approach} for
  {Concurrent} {Query} {Performance} {Prediction}}. In
  \bibinfo{booktitle}{\emph{Proceedings of the 14th {International}
  {Conference} on {Extending} {Database} {Technology}}}
  \emph{(\bibinfo{series}{{EDBT} '14})}. \bibinfo{pages}{109--120}.
\newblock


\bibitem[Eggensperger et~al\mbox{.}(2020a)]%
        {eggensperger2020neural}
\bibfield{author}{\bibinfo{person}{Katharina Eggensperger},
  \bibinfo{person}{Kai Haase}, \bibinfo{person}{Philipp Müller},
  \bibinfo{person}{Marius Lindauer}, {and} \bibinfo{person}{Frank Hutter}.}
  \bibinfo{year}{2020}\natexlab{a}.
\newblock \bibinfo{title}{Neural Model-based Optimization with Right-Censored
  Observations}.
\newblock
\newblock
\showeprint[arxiv]{2009.13828}~[cs.AI]


\bibitem[Eggensperger et~al\mbox{.}(2020b)]%
        {eggensperger2020_censored}
\bibfield{author}{\bibinfo{person}{Katharina Eggensperger},
  \bibinfo{person}{Kai Haase}, \bibinfo{person}{Philipp Müller},
  \bibinfo{person}{Marius Lindauer}, {and} \bibinfo{person}{Frank Hutter}.}
  \bibinfo{year}{2020}\natexlab{b}.
\newblock \bibinfo{title}{Neural Model-based Optimization with Right-Censored
  Observations}.
\newblock
\newblock
\showeprint[arxiv]{2009.13828}~[cs.AI]
\urldef\tempurl%
\url{https://arxiv.org/abs/2009.13828}
\showURL{%
\tempurl}


\bibitem[Eissman et~al\mbox{.}(2018)]%
        {eissman2018bayesian}
\bibfield{author}{\bibinfo{person}{Stephan Eissman}, \bibinfo{person}{Daniel
  Levy}, \bibinfo{person}{Rui Shu}, \bibinfo{person}{Stefan Bartzsch}, {and}
  \bibinfo{person}{Stefano Ermon}.} \bibinfo{year}{2018}\natexlab{}.
\newblock \showarticletitle{Bayesian optimization and attribute adjustment}. In
  \bibinfo{booktitle}{\emph{Proc. 34th Conference on Uncertainty in Artificial
  Intelligence}}.
\newblock


\bibitem[Eriksson and Jankowiak(2021)]%
        {high_dim_bo}
\bibfield{author}{\bibinfo{person}{David Eriksson} {and}
  \bibinfo{person}{Martin Jankowiak}.} \bibinfo{year}{2021}\natexlab{}.
\newblock \showarticletitle{High-dimensional {Bayesian} optimization with
  sparse axis-aligned subspaces}. In \bibinfo{booktitle}{\emph{Proceedings of
  the {Thirty}-{Seventh} {Conference} on {Uncertainty} in {Artificial}
  {Intelligence}}}. \bibinfo{publisher}{PMLR}, \bibinfo{pages}{493--503}.
\newblock
\urldef\tempurl%
\url{https://proceedings.mlr.press/v161/eriksson21a.html}
\showURL{%
\tempurl}
\newblock
\shownote{ISSN: 2640-3498}.


\bibitem[Eriksson et~al\mbox{.}(2019a)]%
        {eriksson2019scalable}
\bibfield{author}{\bibinfo{person}{David Eriksson}, \bibinfo{person}{Michael
  Pearce}, \bibinfo{person}{Jacob Gardner}, \bibinfo{person}{Ryan~D Turner},
  {and} \bibinfo{person}{Matthias Poloczek}.} \bibinfo{year}{2019}\natexlab{a}.
\newblock \showarticletitle{Scalable Global Optimization via Local {Bayesian}
  Optimization}. In \bibinfo{booktitle}{\emph{Advances in Neural Information
  Processing Systems}}. \bibinfo{pages}{5496--5507}.
\newblock
\urldef\tempurl%
\url{http://papers.nips.cc/paper/8788-scalable-global-optimization-via-local-bayesian-optimization.pdf}
\showURL{%
\tempurl}


\bibitem[Eriksson et~al\mbox{.}(2019b)]%
        {bayes_local}
\bibfield{author}{\bibinfo{person}{David Eriksson}, \bibinfo{person}{Michael
  Pearce}, \bibinfo{person}{Jacob~R Gardner}, \bibinfo{person}{Ryan Turner},
  {and} \bibinfo{person}{Matthias Poloczek}.} \bibinfo{year}{2019}\natexlab{b}.
\newblock \showarticletitle{Scalable global optimization via local {Bayesian}
  optimization}. In \bibinfo{booktitle}{\emph{Proceedings of the 33rd
  {International} {Conference} on {Neural} {Information} {Processing}
  {Systems}}} \emph{(\bibinfo{series}{{NeurIPS} '19})}.
  \bibinfo{publisher}{Curran Associates Inc.}, \bibinfo{address}{Red Hook, NY,
  USA}, \bibinfo{pages}{5496--5507}.
\newblock


\bibitem[Frazier et~al\mbox{.}(2009)]%
        {frazier2009knowledge}
\bibfield{author}{\bibinfo{person}{Peter Frazier}, \bibinfo{person}{Warren
  Powell}, {and} \bibinfo{person}{Savas Dayanik}.}
  \bibinfo{year}{2009}\natexlab{}.
\newblock \showarticletitle{The knowledge-gradient policy for correlated normal
  beliefs}.
\newblock \bibinfo{journal}{\emph{INFORMS journal on Computing}}
  \bibinfo{volume}{21}, \bibinfo{number}{4} (\bibinfo{year}{2009}),
  \bibinfo{pages}{599--613}.
\newblock


\bibitem[Garnett(2023)]%
        {garnett_bayesoptbook_2023}
\bibfield{author}{\bibinfo{person}{Roman Garnett}.}
  \bibinfo{year}{2023}\natexlab{}.
\newblock \bibinfo{booktitle}{\emph{{Bayesian Optimization}}}.
\newblock \bibinfo{publisher}{Cambridge University Press}.
\newblock


\bibitem[Gjurovski et~al\mbox{.}(2024)]%
        {gridar_lce}
\bibfield{author}{\bibinfo{person}{Damjan Gjurovski}, \bibinfo{person}{Angjela
  Davitkova}, {and} \bibinfo{person}{Sebastian Michel}.}
  \bibinfo{year}{2024}\natexlab{}.
\newblock \bibinfo{title}{Grid-{AR}: {A} {Grid}-based {Booster} for {Learned}
  {Cardinality} {Estimation} and {Range} {Joins}}.
\newblock
\newblock
\urldef\tempurl%
\url{https://doi.org/10.48550/arXiv.2410.07895}
\showDOI{\tempurl}
\newblock
\shownote{arXiv:2410.07895 [cs]}.


\bibitem[G{\'o}mez-Bombarelli et~al\mbox{.}(2018)]%
        {gomez2018automatic}
\bibfield{author}{\bibinfo{person}{Rafael G{\'o}mez-Bombarelli},
  \bibinfo{person}{Jennifer~N Wei}, \bibinfo{person}{David Duvenaud},
  \bibinfo{person}{Jos{\'e}~Miguel Hern{\'a}ndez-Lobato},
  \bibinfo{person}{Benjam{\'\i}n S{\'a}nchez-Lengeling},
  \bibinfo{person}{Dennis Sheberla}, \bibinfo{person}{Jorge
  Aguilera-Iparraguirre}, \bibinfo{person}{Timothy~D Hirzel},
  \bibinfo{person}{Ryan~P Adams}, {and} \bibinfo{person}{Al{\'a}n
  Aspuru-Guzik}.} \bibinfo{year}{2018}\natexlab{}.
\newblock \showarticletitle{Automatic chemical design using a data-driven
  continuous representation of molecules}.
\newblock \bibinfo{journal}{\emph{ACS central science}} \bibinfo{volume}{4},
  \bibinfo{number}{2} (\bibinfo{year}{2018}), \bibinfo{pages}{268--276}.
\newblock


\bibitem[Graefe(1995)]%
        {cascades}
\bibfield{author}{\bibinfo{person}{Goetz Graefe}.}
  \bibinfo{year}{1995}\natexlab{}.
\newblock \showarticletitle{The Cascades Framework for Query Optimization}.
\newblock \bibinfo{journal}{\emph{IEEE Data Eng. Bull.}} \bibinfo{volume}{18},
  \bibinfo{number}{3} (\bibinfo{year}{1995}), \bibinfo{pages}{19--29}.
\newblock


\bibitem[Grosnit et~al\mbox{.}(2021)]%
        {Huawei}
\bibfield{author}{\bibinfo{person}{Antoine Grosnit}, \bibinfo{person}{Rasul
  Tutunov}, \bibinfo{person}{Alexandre~Max Maraval},
  \bibinfo{person}{Ryan{-}Rhys Griffiths}, \bibinfo{person}{Alexander~Imani
  Cowen{-}Rivers}, \bibinfo{person}{Lin Yang}, \bibinfo{person}{Lin Zhu},
  \bibinfo{person}{Wenlong Lyu}, \bibinfo{person}{Zhitang Chen},
  \bibinfo{person}{Jun Wang}, \bibinfo{person}{Jan Peters}, {and}
  \bibinfo{person}{Haitham Bou{-}Ammar}.} \bibinfo{year}{2021}\natexlab{}.
\newblock \showarticletitle{High-Dimensional {Bayesian} Optimisation with
  Variational Autoencoders and Deep Metric Learning}.
\newblock \bibinfo{journal}{\emph{CoRR}}  \bibinfo{volume}{abs/2106.03609}
  (\bibinfo{year}{2021}).
\newblock
\showeprint[arXiv]{2106.03609}


\bibitem[Hensman et~al\mbox{.}(2013)]%
        {hensman2013gaussian}
\bibfield{author}{\bibinfo{person}{James Hensman}, \bibinfo{person}{Nicol\`{o}
  Fusi}, {and} \bibinfo{person}{Neil~D. Lawrence}.}
  \bibinfo{year}{2013}\natexlab{}.
\newblock \showarticletitle{Gaussian processes for Big data}. In
  \bibinfo{booktitle}{\emph{Proceedings of the Twenty-Ninth Conference on
  Uncertainty in Artificial Intelligence}} (Bellevue, WA)
  \emph{(\bibinfo{series}{UAI'13})}. \bibinfo{publisher}{AUAI Press},
  \bibinfo{address}{Arlington, Virginia, USA}, \bibinfo{pages}{282–290}.
\newblock


\bibitem[Hensman et~al\mbox{.}(2014)]%
        {svgp}
\bibfield{author}{\bibinfo{person}{James Hensman}, \bibinfo{person}{Alex
  Matthews}, {and} \bibinfo{person}{Zoubin Ghahramani}.}
  \bibinfo{year}{2014}\natexlab{}.
\newblock \bibinfo{title}{Scalable Variational Gaussian Process
  Classification}.
\newblock
\newblock
\showeprint[arxiv]{1411.2005}~[stat.ML]


\bibitem[Hilprecht and Binnig(2022)]%
        {zeroshot_latency_model}
\bibfield{author}{\bibinfo{person}{Benjamin Hilprecht} {and}
  \bibinfo{person}{Carsten Binnig}.} \bibinfo{year}{2022}\natexlab{}.
\newblock \showarticletitle{Zero-shot cost models for out-of-the-box learned
  cost prediction}.
\newblock \bibinfo{journal}{\emph{Proceedings of the VLDB Endowment}}
  \bibinfo{volume}{15}, \bibinfo{number}{11} (\bibinfo{date}{July}
  \bibinfo{year}{2022}), \bibinfo{pages}{2361--2374}.
\newblock
\showISSN{2150-8097}
\urldef\tempurl%
\url{https://doi.org/10.14778/3551793.3551799}
\showDOI{\tempurl}


\bibitem[Hutter et~al\mbox{.}(2013a)]%
        {hutter2013_bocensored}
\bibfield{author}{\bibinfo{person}{Frank Hutter}, \bibinfo{person}{Holger
  Hoos}, {and} \bibinfo{person}{Kevin Leyton-Brown}.}
  \bibinfo{year}{2013}\natexlab{a}.
\newblock \bibinfo{title}{Bayesian Optimization With Censored Response Data}.
\newblock
\newblock
\showeprint[arxiv]{1310.1947}~[cs.AI]
\urldef\tempurl%
\url{https://arxiv.org/abs/1310.1947}
\showURL{%
\tempurl}


\bibitem[Hutter et~al\mbox{.}(2013b)]%
        {DBLP:journals/corr/HutterHL13}
\bibfield{author}{\bibinfo{person}{Frank Hutter}, \bibinfo{person}{Holger~H.
  Hoos}, {and} \bibinfo{person}{Kevin Leyton{-}Brown}.}
  \bibinfo{year}{2013}\natexlab{b}.
\newblock \showarticletitle{Bayesian Optimization With Censored Response Data}.
\newblock \bibinfo{journal}{\emph{CoRR}}  \bibinfo{volume}{abs/1310.1947}
  (\bibinfo{year}{2013}).
\newblock
\showeprint[arXiv]{1310.1947}
\urldef\tempurl%
\url{http://arxiv.org/abs/1310.1947}
\showURL{%
\tempurl}


\bibitem[{Immanuel Trummer}(2022)]%
        {genesisdb}
\bibfield{author}{\bibinfo{person}{{Immanuel Trummer}}.}
  \bibinfo{year}{2022}\natexlab{}.
\newblock \bibinfo{booktitle}{\emph{{GenesisDB}: {Synthesizing} {Customized}
  {SQL} {Execution} {Engines} from {Natural} {Language} {Instructions} {Using}
  {GPT}-3 {Codex}}}.
\newblock \bibinfo{type}{Technical {Report}}. \bibinfo{institution}{Cornell},
  \bibinfo{address}{Ithaca. NY}.
\newblock
\urldef\tempurl%
\url{https://rm.cab/genesisdb}
\showURL{%
\tempurl}


\bibitem[Jin et~al\mbox{.}(2018)]%
        {JTVAE}
\bibfield{author}{\bibinfo{person}{Wengong Jin}, \bibinfo{person}{Regina
  Barzilay}, {and} \bibinfo{person}{Tommi~S. Jaakkola}.}
  \bibinfo{year}{2018}\natexlab{}.
\newblock \showarticletitle{Junction Tree Variational Autoencoder for Molecular
  Graph Generation}. In \bibinfo{booktitle}{\emph{Proceedings of the 35th
  International Conference on Machine Learning, {ICML} 2018,
  Stockholmsm{\"{a}}ssan, Stockholm, Sweden, July 10-15, 2018}}
  \emph{(\bibinfo{series}{Proceedings of Machine Learning Research},
  Vol.~\bibinfo{volume}{80})}, \bibfield{editor}{\bibinfo{person}{Jennifer~G.
  Dy} {and} \bibinfo{person}{Andreas Krause}} (Eds.).
  \bibinfo{publisher}{{PMLR}}, \bibinfo{pages}{2328--2337}.
\newblock
\urldef\tempurl%
\url{http://proceedings.mlr.press/v80/jin18a.html}
\showURL{%
\tempurl}


\bibitem[Kajino(2019)]%
        {kajino2019molecular}
\bibfield{author}{\bibinfo{person}{Hiroshi Kajino}.}
  \bibinfo{year}{2019}\natexlab{}.
\newblock \showarticletitle{Molecular hypergraph grammar with its application
  to molecular optimization}. In \bibinfo{booktitle}{\emph{International
  Conference on Machine Learning}}. PMLR, \bibinfo{pages}{3183--3191}.
\newblock


\bibitem[Kamali et~al\mbox{.}(2024a)]%
        {robopt}
\bibfield{author}{\bibinfo{person}{Amin Kamali}, \bibinfo{person}{Verena
  Kantere}, \bibinfo{person}{Calisto Zuzarte}, {and} \bibinfo{person}{Vincent
  Corvinelli}.} \bibinfo{year}{2024}\natexlab{a}.
\newblock \showarticletitle{{RobOpt}: {A} {Tool} for {Robust} {Workload}
  {Optimization} {Based} on {Uncertainty}-{Aware} {Machine} {Learning}}. In
  \bibinfo{booktitle}{\emph{Companion of the 2024 {International} {Conference}
  on {Management} of {Data}}} \emph{(\bibinfo{series}{{SIGMOD} '24})}.
  \bibinfo{publisher}{Association for Computing Machinery},
  \bibinfo{address}{New York, NY, USA}, \bibinfo{pages}{468--471}.
\newblock
\showISBNx{979-8-4007-0422-2}
\urldef\tempurl%
\url{https://doi.org/10.1145/3626246.3654755}
\showDOI{\tempurl}


\bibitem[Kamali et~al\mbox{.}(2024b)]%
        {roq}
\bibfield{author}{\bibinfo{person}{Amin Kamali}, \bibinfo{person}{Verena
  Kantere}, \bibinfo{person}{Calisto Zuzarte}, {and} \bibinfo{person}{Vincent
  Corvinelli}.} \bibinfo{year}{2024}\natexlab{b}.
\newblock \showarticletitle{Roq: {Robust} {Query} {Optimization} {Based} on a
  {Risk}-aware {Learned} {Cost} {Model}}.
\newblock  (\bibinfo{year}{2024}).
\newblock
\urldef\tempurl%
\url{https://doi.org/10.48550/ARXIV.2401.15210}
\showDOI{\tempurl}


\bibitem[Kanellis et~al\mbox{.}(2022)]%
        {kanellis_llamatune}
\bibfield{author}{\bibinfo{person}{Konstantinos Kanellis},
  \bibinfo{person}{Cong Ding}, \bibinfo{person}{Brian Kroth},
  \bibinfo{person}{Andreas M\"{u}ller}, \bibinfo{person}{Carlo Curino}, {and}
  \bibinfo{person}{Shivaram Venkataraman}.} \bibinfo{year}{2022}\natexlab{}.
\newblock \showarticletitle{LlamaTune: sample-efficient DBMS configuration
  tuning}.
\newblock \bibinfo{journal}{\emph{Proc. VLDB Endow.}} \bibinfo{volume}{15},
  \bibinfo{number}{11} (\bibinfo{date}{July} \bibinfo{year}{2022}),
  \bibinfo{pages}{2953–2965}.
\newblock
\showISSN{2150-8097}
\urldef\tempurl%
\url{https://doi.org/10.14778/3551793.3551844}
\showDOI{\tempurl}


\bibitem[Kim et~al\mbox{.}(2024)]%
        {asm_lce}
\bibfield{author}{\bibinfo{person}{Kyoungmin Kim}, \bibinfo{person}{Sangoh
  Lee}, \bibinfo{person}{Injung Kim}, {and} \bibinfo{person}{Wook-Shin Han}.}
  \bibinfo{year}{2024}\natexlab{}.
\newblock \showarticletitle{{ASM}: {Harmonizing} {Autoregressive} {Model},
  {Sampling}, and {Multi}-dimensional {Statistics} {Merging} for {Cardinality}
  {Estimation}}.
\newblock \bibinfo{journal}{\emph{Proceedings of the ACM on Management of
  Data}} \bibinfo{volume}{2}, \bibinfo{number}{1} (\bibinfo{date}{March}
  \bibinfo{year}{2024}), \bibinfo{pages}{1--27}.
\newblock
\showISSN{2836-6573}
\urldef\tempurl%
\url{https://doi.org/10.1145/3639300}
\showDOI{\tempurl}


\bibitem[Kingma and Welling(2014)]%
        {KingmaW13}
\bibfield{author}{\bibinfo{person}{Diederik~P. Kingma} {and}
  \bibinfo{person}{Max Welling}.} \bibinfo{year}{2014}\natexlab{}.
\newblock \showarticletitle{Auto-Encoding Variational Bayes}. In
  \bibinfo{booktitle}{\emph{2nd International Conference on Learning
  Representations, {ICLR} 2014, Banff, AB, Canada, April 14-16, 2014,
  Conference Track Proceedings}}, \bibfield{editor}{\bibinfo{person}{Yoshua
  Bengio} {and} \bibinfo{person}{Yann LeCun}} (Eds.).
\newblock
\urldef\tempurl%
\url{http://arxiv.org/abs/1312.6114}
\showURL{%
\tempurl}


\bibitem[Kipf et~al\mbox{.}(2019)]%
        {deep_card_est2}
\bibfield{author}{\bibinfo{person}{Andreas Kipf}, \bibinfo{person}{Thomas
  Kipf}, \bibinfo{person}{Bernhard Radke}, \bibinfo{person}{Viktor Leis},
  \bibinfo{person}{Peter Boncz}, {and} \bibinfo{person}{Alfons Kemper}.}
  \bibinfo{year}{2019}\natexlab{}.
\newblock \showarticletitle{Learned {Cardinalities}: {Estimating} {Correlated}
  {Joins} with {Deep} {Learning}}. In \bibinfo{booktitle}{\emph{9th {Biennial}
  {Conference} on {Innovative} {Data} {Systems} {Research}}}
  \emph{(\bibinfo{series}{{CIDR} '19})}.
\newblock
\urldef\tempurl%
\url{http://arxiv.org/abs/1809.00677}
\showURL{%
\tempurl}


\bibitem[Krenn et~al\mbox{.}(2020)]%
        {selfies}
\bibfield{author}{\bibinfo{person}{Mario Krenn}, \bibinfo{person}{Florian
  Häse}, \bibinfo{person}{AkshatKumar Nigam}, \bibinfo{person}{Pascal
  Friederich}, {and} \bibinfo{person}{Alán Aspuru-Guzik}.}
  \bibinfo{year}{2020}\natexlab{}.
\newblock \showarticletitle{Self-{Referencing} {Embedded} {Strings}
  ({SELFIES}): {A} 100\% robust molecular string representation}.
\newblock \bibinfo{journal}{\emph{Machine Learning: Science and Technology}}
  \bibinfo{volume}{1}, \bibinfo{number}{4} (\bibinfo{date}{Dec.}
  \bibinfo{year}{2020}), \bibinfo{pages}{045024}.
\newblock
\showISSN{2632-2153}
\urldef\tempurl%
\url{https://doi.org/10.1088/2632-2153/aba947}
\showDOI{\tempurl}
\newblock
\shownote{arXiv:1905.13741 [physics, physics:quant-ph, stat]}.


\bibitem[Lao et~al\mbox{.}(2024)]%
        {gptuner}
\bibfield{author}{\bibinfo{person}{Jiale Lao}, \bibinfo{person}{Yibo Wang},
  \bibinfo{person}{Yufei Li}, \bibinfo{person}{Jianping Wang},
  \bibinfo{person}{Yunjia Zhang}, \bibinfo{person}{Zhiyuan Cheng},
  \bibinfo{person}{Wanghu Chen}, \bibinfo{person}{Mingjie Tang}, {and}
  \bibinfo{person}{Jianguo Wang}.} \bibinfo{year}{2024}\natexlab{}.
\newblock \showarticletitle{GPTuner: A Manual-Reading Database Tuning System
  via GPT-Guided Bayesian Optimization}.
\newblock \bibinfo{journal}{\emph{Proc. VLDB Endow.}} \bibinfo{volume}{17},
  \bibinfo{number}{8} (\bibinfo{date}{may} \bibinfo{year}{2024}),
  \bibinfo{pages}{1939–1952}.
\newblock
\showISSN{2150-8097}
\urldef\tempurl%
\url{https://doi.org/10.14778/3659437.3659449}
\showDOI{\tempurl}


\bibitem[Leis et~al\mbox{.}(2015a)]%
        {howgood}
\bibfield{author}{\bibinfo{person}{Viktor Leis}, \bibinfo{person}{Andrey
  Gubichev}, \bibinfo{person}{Atanas Mirchev}, \bibinfo{person}{Peter Boncz},
  \bibinfo{person}{Alfons Kemper}, {and} \bibinfo{person}{Thomas Neumann}.}
  \bibinfo{year}{2015}\natexlab{a}.
\newblock \showarticletitle{How {Good} {Are} {Query} {Optimizers}, {Really}?}
\newblock \bibinfo{journal}{\emph{PVLDB}} \bibinfo{volume}{9},
  \bibinfo{number}{3} (\bibinfo{year}{2015}), \bibinfo{pages}{204--215}.
\newblock
\showISSN{2150-8097}
\urldef\tempurl%
\url{https://doi.org/10.14778/2850583.2850594}
\showDOI{\tempurl}


\bibitem[Leis et~al\mbox{.}(2015b)]%
        {job}
\bibfield{author}{\bibinfo{person}{Viktor Leis}, \bibinfo{person}{Andrey
  Gubichev}, \bibinfo{person}{Atanas Mirchev}, \bibinfo{person}{Peter Boncz},
  \bibinfo{person}{Alfons Kemper}, {and} \bibinfo{person}{Thomas Neumann}.}
  \bibinfo{year}{2015}\natexlab{b}.
\newblock \showarticletitle{How good are query optimizers, really?}
\newblock \bibinfo{journal}{\emph{Proc. VLDB Endow.}} \bibinfo{volume}{9},
  \bibinfo{number}{3} (\bibinfo{date}{nov} \bibinfo{year}{2015}),
  \bibinfo{pages}{204–215}.
\newblock
\showISSN{2150-8097}
\urldef\tempurl%
\url{https://doi.org/10.14778/2850583.2850594}
\showDOI{\tempurl}


\bibitem[Li et~al\mbox{.}(2021)]%
        {opengauss}
\bibfield{author}{\bibinfo{person}{Guoliang Li}, \bibinfo{person}{Xuanhe Zhou},
  \bibinfo{person}{Ji Sun}, \bibinfo{person}{Xiang Yu}, \bibinfo{person}{Yue
  Han}, \bibinfo{person}{Lianyuan Jin}, \bibinfo{person}{Wenbo Li},
  \bibinfo{person}{Tianqing Wang}, {and} \bibinfo{person}{Shifu Li}.}
  \bibinfo{year}{2021}\natexlab{}.
\newblock \showarticletitle{{openGauss}: an autonomous database system}.
\newblock \bibinfo{journal}{\emph{Proceedings of the VLDB Endowment}}
  \bibinfo{volume}{14}, \bibinfo{number}{12} (\bibinfo{date}{July}
  \bibinfo{year}{2021}), \bibinfo{pages}{3028--3042}.
\newblock
\showISSN{2150-8097}
\urldef\tempurl%
\url{https://doi.org/10.14778/3476311.3476380}
\showDOI{\tempurl}


\bibitem[Li et~al\mbox{.}(2023)]%
        {alece_lce}
\bibfield{author}{\bibinfo{person}{Pengfei Li}, \bibinfo{person}{Wenqing Wei},
  \bibinfo{person}{Rong Zhu}, \bibinfo{person}{Bolin Ding},
  \bibinfo{person}{Jingren Zhou}, {and} \bibinfo{person}{Hua Lu}.}
  \bibinfo{year}{2023}\natexlab{}.
\newblock \showarticletitle{{ALECE}: {An} {Attention}-based {Learned}
  {Cardinality} {Estimator} for {SPJ} {Queries} on {Dynamic} {Workloads}}.
\newblock \bibinfo{journal}{\emph{Proc. VLDB Endow.}} \bibinfo{volume}{17},
  \bibinfo{number}{2} (\bibinfo{date}{Oct.} \bibinfo{year}{2023}),
  \bibinfo{pages}{197--210}.
\newblock
\showISSN{2150-8097}
\urldef\tempurl%
\url{https://doi.org/10.14778/3626292.3626302}
\showDOI{\tempurl}


\bibitem[Lim et~al\mbox{.}(2024)]%
        {hitthegym}
\bibfield{author}{\bibinfo{person}{Wan~Shen Lim}, \bibinfo{person}{Lin Ma},
  \bibinfo{person}{William Zhang}, \bibinfo{person}{Matthew Butrovich},
  \bibinfo{person}{Samuel Arch}, {and} \bibinfo{person}{Andrew Pavlo}.}
  \bibinfo{year}{2024}\natexlab{}.
\newblock \showarticletitle{Hit the {Gym}: {Accelerating} {Query} {Execution}
  to {Efficiently} {Bootstrap} {Behavior} {Models} for {Self}-{Driving}
  {Database} {Management} {Systems}}.
\newblock \bibinfo{journal}{\emph{Proceedings of the VLDB Endowment}}
  \bibinfo{volume}{17}, \bibinfo{number}{11} (\bibinfo{date}{July}
  \bibinfo{year}{2024}), \bibinfo{pages}{3680--3693}.
\newblock
\showISSN{2150-8097}
\urldef\tempurl%
\url{https://doi.org/10.14778/3681954.3682030}
\showDOI{\tempurl}


\bibitem[Liu et~al\mbox{.}(2016)]%
        {inc_reopt}
\bibfield{author}{\bibinfo{person}{Mengmeng Liu}, \bibinfo{person}{Zachary~G.
  Ives}, {and} \bibinfo{person}{Boon~Thau Loo}.}
  \bibinfo{year}{2016}\natexlab{}.
\newblock \showarticletitle{Enabling {Incremental} {Query}
  {Re}-{Optimization}}. In \bibinfo{booktitle}{\emph{Proceedings of the 2016
  {International} {Conference} on {Management} of {Data}}}
  \emph{(\bibinfo{series}{{SIGMOD} '16})}. \bibinfo{publisher}{Association for
  Computing Machinery}, \bibinfo{address}{New York, NY, USA},
  \bibinfo{pages}{1705--1720}.
\newblock
\showISBNx{978-1-4503-3531-7}
\urldef\tempurl%
\url{https://doi.org/10.1145/2882903.2915212}
\showDOI{\tempurl}


\bibitem[Lohman(2014)]%
        {qo_unsolved}
\bibfield{author}{\bibinfo{person}{Guy Lohman}.}
  \bibinfo{year}{2014}\natexlab{}.
\newblock \showarticletitle{Is {Query} {Optimization} a "{Solved}" {Problem}?}.
  In \bibinfo{booktitle}{\emph{{ACM} {SIGMOD} {Blog}}}
  \emph{(\bibinfo{series}{{ACM} {Blog} '14})}.
\newblock
\urldef\tempurl%
\url{https://wp.sigmod.org/?p=1075}
\showURL{%
\tempurl}


\bibitem[Lu et~al\mbox{.}(1995)]%
        {fittest_qo}
\bibfield{author}{\bibinfo{person}{Hongjun Lu}, \bibinfo{person}{K. Tan}, {and}
  \bibinfo{person}{S. Dao}.} \bibinfo{year}{1995}\natexlab{}.
\newblock \showarticletitle{The {Fittest} {Survives}: {An} {Adaptive}
  {Approach} to {Query} {Optimization}} \emph{(\bibinfo{series}{{VLDB} '95})}.
  \bibinfo{address}{Zurich, Switzerland}.
\newblock
\urldef\tempurl%
\url{https://www.semanticscholar.org/paper/The-Fittest-Survives%3A-An-Adaptive-Approach-to-Query-Lu-Tan/3d2625f15445adc9dd23324d4839c5dd364630fa}
\showURL{%
\tempurl}


\bibitem[Maddox et~al\mbox{.}(2021)]%
        {maddox2021conditioning}
\bibfield{author}{\bibinfo{person}{Wesley~J Maddox}, \bibinfo{person}{Samuel
  Stanton}, {and} \bibinfo{person}{Andrew~G Wilson}.}
  \bibinfo{year}{2021}\natexlab{}.
\newblock \showarticletitle{Conditioning sparse variational gaussian processes
  for online decision-making}.
\newblock \bibinfo{journal}{\emph{Advances in Neural Information Processing
  Systems}}  \bibinfo{volume}{34} (\bibinfo{year}{2021}),
  \bibinfo{pages}{6365--6379}.
\newblock


\bibitem[Marcus(2023)]%
        {superopt_vision}
\bibfield{author}{\bibinfo{person}{Ryan Marcus}.}
  \bibinfo{year}{2023}\natexlab{}.
\newblock \showarticletitle{Learned {Query} {Superoptimization}}. In
  \bibinfo{booktitle}{\emph{Joint {Workshops} at 49th {International}
  {Conference} on {Very} {Large} {Data} {Bases}}}
  \emph{(\bibinfo{series}{{AIDB}@{VLDB} '23})}. \bibinfo{publisher}{CEUR
  Workshop Proceedings}, \bibinfo{address}{Vancouver, BC, Canada}.
\newblock


\bibitem[Marcus et~al\mbox{.}(2021)]%
        {bao}
\bibfield{author}{\bibinfo{person}{Ryan Marcus}, \bibinfo{person}{Parimarjan
  Negi}, \bibinfo{person}{Hongzi Mao}, \bibinfo{person}{Nesime Tatbul},
  \bibinfo{person}{Mohammad Alizadeh}, {and} \bibinfo{person}{Tim Kraska}.}
  \bibinfo{year}{2021}\natexlab{}.
\newblock \showarticletitle{Bao: Making Learned Query Optimization Practical}.
  In \bibinfo{booktitle}{\emph{Proceedings of the 2021 International Conference
  on Management of Data}} (Virtual Event, China) \emph{(\bibinfo{series}{SIGMOD
  '21})}. \bibinfo{publisher}{Association for Computing Machinery},
  \bibinfo{address}{New York, NY, USA}, \bibinfo{pages}{1275–1288}.
\newblock
\showISBNx{9781450383431}
\urldef\tempurl%
\url{https://doi.org/10.1145/3448016.3452838}
\showDOI{\tempurl}


\bibitem[Marcus et~al\mbox{.}(2019)]%
        {neo}
\bibfield{author}{\bibinfo{person}{Ryan Marcus}, \bibinfo{person}{Parimarjan
  Negi}, \bibinfo{person}{Hongzi Mao}, \bibinfo{person}{Chi Zhang},
  \bibinfo{person}{Mohammad Alizadeh}, \bibinfo{person}{Tim Kraska},
  \bibinfo{person}{Olga Papaemmanouil}, {and} \bibinfo{person}{Nesime Tatbul}.}
  \bibinfo{year}{2019}\natexlab{}.
\newblock \showarticletitle{Neo: a learned query optimizer}.
\newblock \bibinfo{journal}{\emph{Proc. VLDB Endow.}} \bibinfo{volume}{12},
  \bibinfo{number}{11} (\bibinfo{date}{jul} \bibinfo{year}{2019}),
  \bibinfo{pages}{1705–1718}.
\newblock
\showISSN{2150-8097}
\urldef\tempurl%
\url{https://doi.org/10.14778/3342263.3342644}
\showDOI{\tempurl}


\bibitem[Marcus and Papaemmanouil(2018)]%
        {rejoin}
\bibfield{author}{\bibinfo{person}{Ryan Marcus} {and} \bibinfo{person}{Olga
  Papaemmanouil}.} \bibinfo{year}{2018}\natexlab{}.
\newblock \showarticletitle{Deep {Reinforcement} {Learning} for {Join} {Order}
  {Enumeration}}. In \bibinfo{booktitle}{\emph{First {International} {Workshop}
  on {Exploiting} {Artificial} {Intelligence} {Techniques} for {Data}
  {Management}}} \emph{(\bibinfo{series}{{aiDM} @ {SIGMOD} '18})}.
  \bibinfo{address}{Houston, TX}.
\newblock


\bibitem[Marcus and Papaemmanouil(2019)]%
        {qppnet}
\bibfield{author}{\bibinfo{person}{Ryan Marcus} {and} \bibinfo{person}{Olga
  Papaemmanouil}.} \bibinfo{year}{2019}\natexlab{}.
\newblock \showarticletitle{Plan-{Structured} {Deep} {Neural} {Network}
  {Models} for {Query} {Performance} {Prediction}}.
\newblock \bibinfo{journal}{\emph{PVLDB}} \bibinfo{volume}{12},
  \bibinfo{number}{11} (\bibinfo{year}{2019}), \bibinfo{pages}{1733--1746}.
\newblock
\urldef\tempurl%
\url{https://doi.org/10.14778/3342263.3342646}
\showDOI{\tempurl}


\bibitem[Markl et~al\mbox{.}(2007)]%
        {fleeing_from_knowledge}
\bibfield{author}{\bibinfo{person}{V. Markl}, \bibinfo{person}{P.~J. Haas},
  \bibinfo{person}{M. Kutsch}, \bibinfo{person}{N. Megiddo},
  \bibinfo{person}{U. Srivastava}, {and} \bibinfo{person}{T.~M. Tran}.}
  \bibinfo{year}{2007}\natexlab{}.
\newblock \showarticletitle{Consistent selectivity estimation via maximum
  entropy}.
\newblock \bibinfo{journal}{\emph{The VLDB Journal}} \bibinfo{volume}{16},
  \bibinfo{number}{1} (\bibinfo{date}{Jan.} \bibinfo{year}{2007}),
  \bibinfo{pages}{55--76}.
\newblock
\showISSN{1066-8888}
\urldef\tempurl%
\url{https://doi.org/10.1007/s00778-006-0030-1}
\showDOI{\tempurl}


\bibitem[Massalin(1987)]%
        {superopt_coined}
\bibfield{author}{\bibinfo{person}{Alexia Massalin}.}
  \bibinfo{year}{1987}\natexlab{}.
\newblock \showarticletitle{Superoptimizer: a look at the smallest program}.
\newblock \bibinfo{journal}{\emph{ACM SIGARCH Computer Architecture News}}
  \bibinfo{volume}{15}, \bibinfo{number}{5} (\bibinfo{date}{Oct.}
  \bibinfo{year}{1987}), \bibinfo{pages}{122--126}.
\newblock
\showISSN{0163-5964}
\urldef\tempurl%
\url{https://doi.org/10.1145/36177.36194}
\showDOI{\tempurl}


\bibitem[Maus et~al\mbox{.}(2022a)]%
        {maus2022local}
\bibfield{author}{\bibinfo{person}{Natalie Maus}, \bibinfo{person}{Haydn
  Jones}, \bibinfo{person}{Juston Moore}, \bibinfo{person}{Matt~J. Kusner},
  \bibinfo{person}{John Bradshaw}, {and} \bibinfo{person}{Jacob~R. Gardner}.}
  \bibinfo{year}{2022}\natexlab{a}.
\newblock \showarticletitle{Local Latent Space Bayesian Optimization over
  Structured Inputs}. In \bibinfo{booktitle}{\emph{Advances in Neural
  Information Processing Systems 35: Annual Conference on Neural Information
  Processing Systems 2022, NeurIPS 2022, New Orleans, LA, USA, November 28 -
  December 9, 2022}}, \bibfield{editor}{\bibinfo{person}{Sanmi Koyejo},
  \bibinfo{person}{S.~Mohamed}, \bibinfo{person}{A.~Agarwal},
  \bibinfo{person}{Danielle Belgrave}, \bibinfo{person}{K.~Cho}, {and}
  \bibinfo{person}{A.~Oh}} (Eds.).
\newblock
\urldef\tempurl%
\url{http://papers.nips.cc/paper\_files/paper/2022/hash/ded98d28f82342a39f371c013dfb3058-Abstract-Conference.html}
\showURL{%
\tempurl}


\bibitem[Maus et~al\mbox{.}(2022b)]%
        {bayes_latent}
\bibfield{author}{\bibinfo{person}{Natalie Maus}, \bibinfo{person}{Haydn~Thomas
  Jones}, \bibinfo{person}{Juston Moore}, \bibinfo{person}{Matt Kusner},
  \bibinfo{person}{John Bradshaw}, {and} \bibinfo{person}{Jacob~R. Gardner}.}
  \bibinfo{year}{2022}\natexlab{b}.
\newblock \showarticletitle{Local {Latent} {Space} {Bayesian} {Optimization}
  over {Structured} {Inputs}} \emph{(\bibinfo{series}{{NeurIPS} '22})}.
\newblock
\urldef\tempurl%
\url{https://openreview.net/forum?id=nZRTRevUO-}
\showURL{%
\tempurl}


\bibitem[Maus et~al\mbox{.}(2024)]%
        {lolbo}
\bibfield{author}{\bibinfo{person}{Natalie~T. Maus}, \bibinfo{person}{Haydn~T.
  Jones}, \bibinfo{person}{Juston~S. Moore}, \bibinfo{person}{Matt~J. Kusner},
  \bibinfo{person}{John Bradshaw}, {and} \bibinfo{person}{Jacob~R. Gardner}.}
  \bibinfo{year}{2024}\natexlab{}.
\newblock \showarticletitle{Local latent space Bayesian optimization over
  structured inputs}. In \bibinfo{booktitle}{\emph{Proceedings of the 36th
  International Conference on Neural Information Processing Systems}} (New
  Orleans, LA, USA) \emph{(\bibinfo{series}{NIPS '22})}.
  \bibinfo{publisher}{Curran Associates Inc.}, \bibinfo{address}{Red Hook, NY,
  USA}, Article \bibinfo{articleno}{2500}, \bibinfo{numpages}{14}~pages.
\newblock
\showISBNx{9781713871088}


\bibitem[{Mert Akdere} and {Ugur Cetintemel}(2012)]%
        {learning_latency}
\bibfield{author}{\bibinfo{person}{{Mert Akdere}} {and} \bibinfo{person}{{Ugur
  Cetintemel}}.} \bibinfo{year}{2012}\natexlab{}.
\newblock \showarticletitle{Learning-based query performance modeling and
  prediction}. In \bibinfo{booktitle}{\emph{2012 {IEEE} 28th {International}
  {Conference} on {Data} {Engineering}}} \emph{(\bibinfo{series}{{ICDE} '12})}.
  \bibinfo{publisher}{IEEE}, \bibinfo{pages}{390--401}.
\newblock


\bibitem[Mo et~al\mbox{.}(2023)]%
        {concurrent_lqo}
\bibfield{author}{\bibinfo{person}{Songsong Mo}, \bibinfo{person}{Yile Chen},
  \bibinfo{person}{Hao Wang}, \bibinfo{person}{Gao Cong}, {and}
  \bibinfo{person}{Zhifeng Bao}.} \bibinfo{year}{2023}\natexlab{}.
\newblock \showarticletitle{Lemo: {A} {Cache}-{Enhanced} {Learned} {Optimizer}
  for {Concurrent} {Queries}}.
\newblock \bibinfo{journal}{\emph{Proc. ACM Manag. Data}} \bibinfo{volume}{1},
  \bibinfo{number}{4} (\bibinfo{date}{Dec.} \bibinfo{year}{2023}),
  \bibinfo{pages}{247:1--247:26}.
\newblock
\urldef\tempurl%
\url{https://doi.org/10.1145/3626734}
\showDOI{\tempurl}


\bibitem[Negi et~al\mbox{.}(2021a)]%
        {bao_scope}
\bibfield{author}{\bibinfo{person}{Parimarjan Negi}, \bibinfo{person}{Matteo
  Interlandi}, \bibinfo{person}{Ryan Marcus}, \bibinfo{person}{Mohammad
  Alizadeh}, \bibinfo{person}{Tim Kraska}, \bibinfo{person}{Marc Friedman},
  {and} \bibinfo{person}{Alekh Jindal}.} \bibinfo{year}{2021}\natexlab{a}.
\newblock \showarticletitle{Steering {Query} {Optimizers}: {A} {Practical}
  {Take} on {Big} {Data} {Workloads}}. In \bibinfo{booktitle}{\emph{Proceedings
  of the 2021 {International} {Conference} on {Management} of {Data}}}
  \emph{(\bibinfo{series}{{SIGMOD} '21})}. \bibinfo{publisher}{ACM},
  \bibinfo{address}{Virtual Event China}, \bibinfo{pages}{2557--2569}.
\newblock
\showISBNx{978-1-4503-8343-1}
\urldef\tempurl%
\url{https://doi.org/10.1145/3448016.3457568}
\showDOI{\tempurl}
\newblock
\shownote{Award: 'best paper honorable mention'}.


\bibitem[Negi et~al\mbox{.}(2021b)]%
        {flowloss}
\bibfield{author}{\bibinfo{person}{Parimarjan Negi}, \bibinfo{person}{Ryan
  Marcus}, \bibinfo{person}{Andreas Kipf}, \bibinfo{person}{Hongzi Mao},
  \bibinfo{person}{Nesime Tatbul}, \bibinfo{person}{Tim Kraska}, {and}
  \bibinfo{person}{Mohammad Alizadeh}.} \bibinfo{year}{2021}\natexlab{b}.
\newblock \showarticletitle{Flow-Loss: Learning Cardinality Estimates That
  Matter}.
\newblock \bibinfo{journal}{\emph{Proc. VLDB Endow.}} \bibinfo{volume}{14},
  \bibinfo{number}{11} (\bibinfo{date}{jul} \bibinfo{year}{2021}),
  \bibinfo{pages}{2019–2032}.
\newblock
\showISSN{2150-8097}
\urldef\tempurl%
\url{https://doi.org/10.14778/3476249.3476259}
\showDOI{\tempurl}


\bibitem[Ono and Lohman(1990)]%
        {joe_complexity}
\bibfield{author}{\bibinfo{person}{Kiyoshi Ono} {and} \bibinfo{person}{Guy~M.
  Lohman}.} \bibinfo{year}{1990}\natexlab{}.
\newblock \showarticletitle{Measuring the {Complexity} of {Join} {Enumeration}
  in {Query} {Optimization}}. In \bibinfo{booktitle}{\emph{{VLDB}}}
  \emph{(\bibinfo{series}{{VLDB} '90})}. \bibinfo{pages}{314--325}.
\newblock
\showISBNx{978-1-55860-149-9}
\urldef\tempurl%
\url{http://dl.acm.org/citation.cfm?id=645916.671976}
\showURL{%
\tempurl}


\bibitem[Ortiz et~al\mbox{.}(2018)]%
        {qo_state_rep}
\bibfield{author}{\bibinfo{person}{Jennifer Ortiz}, \bibinfo{person}{Magdalena
  Balazinska}, \bibinfo{person}{Johannes Gehrke}, {and}
  \bibinfo{person}{S.~Sathiya Keerthi}.} \bibinfo{year}{2018}\natexlab{}.
\newblock \showarticletitle{Learning {State} {Representations} for {Query}
  {Optimization} with {Deep} {Reinforcement} {Learning}}. In
  \bibinfo{booktitle}{\emph{2nd {Workshop} on {Data} {Managmeent} for
  {End}-to-{End} {Machine} {Learning}}} \emph{(\bibinfo{series}{{DEEM} '18})}.
\newblock
\urldef\tempurl%
\url{https://arxiv.org/abs/1803.08604}
\showURL{%
\tempurl}


\bibitem[Park et~al\mbox{.}(2018)]%
        {quicksel}
\bibfield{author}{\bibinfo{person}{Yongjoo Park}, \bibinfo{person}{Shucheng
  Zhong}, {and} \bibinfo{person}{Barzan Mozafari}.}
  \bibinfo{year}{2018}\natexlab{}.
\newblock \showarticletitle{{QuickSel}: {Quick} {Selectivity} {Learning} with
  {Mixture} {Models}}.
\newblock \bibinfo{journal}{\emph{arXiv:1812.10568 [cs]}} (\bibinfo{date}{Dec.}
  \bibinfo{year}{2018}).
\newblock
\urldef\tempurl%
\url{http://arxiv.org/abs/1812.10568}
\showURL{%
\tempurl}


\bibitem[Perron et~al\mbox{.}(2019)]%
        {reopt}
\bibfield{author}{\bibinfo{person}{Matthew Perron}, \bibinfo{person}{Zeyuan
  Shang}, \bibinfo{person}{Tim Kraska}, {and} \bibinfo{person}{Michael
  Stonebraker}.} \bibinfo{year}{2019}\natexlab{}.
\newblock \showarticletitle{How {I} {Learned} to {Stop} {Worrying} and {Love}
  {Re}-optimization}.
\newblock \bibinfo{journal}{\emph{2019 IEEE 35th International Conference on
  Data Engineering (ICDE)}} (\bibinfo{date}{April} \bibinfo{year}{2019}),
  \bibinfo{pages}{1758--1761}.
\newblock
\urldef\tempurl%
\url{https://doi.org/10.1109/ICDE.2019.00191}
\showDOI{\tempurl}
\newblock
\shownote{Conference Name: 2019 IEEE 35th International Conference on Data
  Engineering (ICDE) ISBN: 9781538674741 Place: Macao, Macao Publisher: IEEE}.


\bibitem[{PostgreSQL Developers}(2024)]%
        {url-pg_hints}
\bibfield{author}{\bibinfo{person}{{PostgreSQL Developers}}.}
  \bibinfo{year}{2024}\natexlab{}.
\newblock \bibinfo{title}{{PostgreSQL} hints,
  https://www.postgresql.org/docs/current/runtime-config-query.html}.
\newblock
\newblock
\urldef\tempurl%
\url{https://www.postgresql.org/docs/current/runtime-config-query.html}
\showURL{%
\tempurl}
\newblock
\shownote{tex.key= 1}.


\bibitem[Reiner and Grossniklaus(2024)]%
        {geom_card_est}
\bibfield{author}{\bibinfo{person}{Silvan Reiner} {and}
  \bibinfo{person}{Michael Grossniklaus}.} \bibinfo{year}{2024}\natexlab{}.
\newblock \showarticletitle{Sample-{Efficient} {Cardinality} {Estimation}
  {Using} {Geometric} {Deep} {Learning}}.
\newblock \bibinfo{journal}{\emph{Proc. VLDB Endow.}} \bibinfo{volume}{17},
  \bibinfo{number}{4} (\bibinfo{date}{March} \bibinfo{year}{2024}),
  \bibinfo{pages}{740--752}.
\newblock
\showISSN{2150-8097}
\urldef\tempurl%
\url{https://doi.org/10.14778/3636218.3636229}
\showDOI{\tempurl}


\bibitem[Schmidt et~al\mbox{.}(2024)]%
        {redshift_pred_cache}
\bibfield{author}{\bibinfo{person}{Tobias Schmidt}, \bibinfo{person}{Andreas
  Kipf}, \bibinfo{person}{Dominik Horn}, \bibinfo{person}{Gaurav Saxena}, {and}
  \bibinfo{person}{Tim Kraska}.} \bibinfo{year}{2024}\natexlab{}.
\newblock \showarticletitle{Predicate {Caching}: {Query}-{Driven} {Secondary}
  {Indexing} for {Cloud} {Data} {Warehouses}}. In
  \bibinfo{booktitle}{\emph{Companion of the 2024 {International} {Conference}
  on {Management} of {Data}}} \emph{(\bibinfo{series}{{SIGMOD} '24})}.
  \bibinfo{publisher}{Association for Computing Machinery},
  \bibinfo{address}{New York, NY, USA}, \bibinfo{pages}{347--359}.
\newblock
\showISBNx{979-8-4007-0422-2}
\urldef\tempurl%
\url{https://doi.org/10.1145/3626246.3653395}
\showDOI{\tempurl}


\bibitem[Shahriari et~al\mbox{.}(2016)]%
        {bayes_survey}
\bibfield{author}{\bibinfo{person}{Bobak Shahriari}, \bibinfo{person}{Kevin
  Swersky}, \bibinfo{person}{Ziyu Wang}, \bibinfo{person}{Ryan~P. Adams}, {and}
  \bibinfo{person}{Nando de Freitas}.} \bibinfo{year}{2016}\natexlab{}.
\newblock \showarticletitle{Taking the {Human} {Out} of the {Loop}: {A}
  {Review} of {Bayesian} {Optimization}}.
\newblock \bibinfo{journal}{\emph{Proc. IEEE}} \bibinfo{volume}{104},
  \bibinfo{number}{1} (\bibinfo{date}{Jan.} \bibinfo{year}{2016}),
  \bibinfo{pages}{148--175}.
\newblock
\showISSN{1558-2256}
\urldef\tempurl%
\url{https://doi.org/10.1109/JPROC.2015.2494218}
\showDOI{\tempurl}


\bibitem[Shamir et~al\mbox{.}(2010)]%
        {bottleneck}
\bibfield{author}{\bibinfo{person}{Ohad Shamir}, \bibinfo{person}{Sivan
  Sabato}, {and} \bibinfo{person}{Naftali Tishby}.}
  \bibinfo{year}{2010}\natexlab{}.
\newblock \showarticletitle{Learning and generalization with the information
  bottleneck}.
\newblock \bibinfo{journal}{\emph{Theoretical Computer Science}}
  \bibinfo{volume}{411}, \bibinfo{number}{29} (\bibinfo{date}{June}
  \bibinfo{year}{2010}), \bibinfo{pages}{2696--2711}.
\newblock
\showISSN{0304-3975}
\urldef\tempurl%
\url{https://doi.org/10.1016/j.tcs.2010.04.006}
\showDOI{\tempurl}


\bibitem[Shypula et~al\mbox{.}(2024)]%
        {pie}
\bibfield{author}{\bibinfo{person}{Alexander Shypula}, \bibinfo{person}{Aman
  Madaan}, \bibinfo{person}{Yimeng Zeng}, \bibinfo{person}{Uri Alon},
  \bibinfo{person}{Jacob Gardner}, \bibinfo{person}{Milad Hashemi},
  \bibinfo{person}{Graham Neubig}, \bibinfo{person}{Parthasarathy Ranganathan},
  \bibinfo{person}{Osbert Bastani}, {and} \bibinfo{person}{Amir Yazdanbakhsh}.}
  \bibinfo{year}{2024}\natexlab{}.
\newblock
  \showarticletitle{\href{https://openreview.net/pdf?id=ix7rLVHXyY}{Learning
  Performance-Improving Code Edits}}. In \bibinfo{booktitle}{\emph{The Twelfth
  International Conference on Learning Representations (ICLR)}}.
\newblock
\urldef\tempurl%
\url{https://openreview.net/pdf?id=ix7rLVHXyY}
\showURL{%
\tempurl}


\bibitem[Sun and Li(2019)]%
        {learn_cost}
\bibfield{author}{\bibinfo{person}{Ji Sun} {and} \bibinfo{person}{Guoliang
  Li}.} \bibinfo{year}{2019}\natexlab{}.
\newblock \showarticletitle{An end-to-end learning-based cost estimator}.
\newblock \bibinfo{journal}{\emph{Proceedings of the VLDB Endowment}}
  \bibinfo{volume}{13}, \bibinfo{number}{3} (\bibinfo{date}{Nov.}
  \bibinfo{year}{2019}), \bibinfo{pages}{307--319}.
\newblock
\showISSN{2150-8097}
\urldef\tempurl%
\url{https://doi.org/10.14778/3368289.3368296}
\showDOI{\tempurl}


\bibitem[Sutton and Barto(1998)]%
        {rl_book}
\bibfield{author}{\bibinfo{person}{Richard~S. Sutton} {and}
  \bibinfo{person}{Andrew~G. Barto}.} \bibinfo{year}{1998}\natexlab{}.
\newblock \bibinfo{booktitle}{\emph{Introduction to {Reinforcement} {Learning}}
  (\bibinfo{edition}{1st} ed.)}.
\newblock \bibinfo{publisher}{MIT Press}, \bibinfo{address}{Cambridge, MA,
  USA}.
\newblock
\showISBNx{978-0-262-19398-6}


\bibitem[Tai et~al\mbox{.}(2015)]%
        {tree_lstm}
\bibfield{author}{\bibinfo{person}{Kai~Sheng Tai}, \bibinfo{person}{Richard
  Socher}, {and} \bibinfo{person}{Christopher~D. Manning}.}
  \bibinfo{year}{2015}\natexlab{}.
\newblock \showarticletitle{Improved {Semantic} {Representations} {From}
  {Tree}-{Structured} {Long} {Short}-{Term} {Memory} {Networks}}.
\newblock \bibinfo{journal}{\emph{arXiv:1503.00075 [cs]}} (\bibinfo{date}{Feb.}
  \bibinfo{year}{2015}).
\newblock
\urldef\tempurl%
\url{http://arxiv.org/abs/1503.00075}
\showURL{%
\tempurl}


\bibitem[Thompson(1933)]%
        {thompson}
\bibfield{author}{\bibinfo{person}{William~R. Thompson}.}
  \bibinfo{year}{1933}\natexlab{}.
\newblock \showarticletitle{On the {Likelihood} that {One} {Unknown}
  {Probability} {Exceeds} {Another} in {View} of the {Evidence} of {Two}
  {Samples}}.
\newblock \bibinfo{journal}{\emph{Biometrika}} (\bibinfo{year}{1933}).
\newblock


\bibitem[Titsias(2009)]%
        {titsias2009variational}
\bibfield{author}{\bibinfo{person}{Michalis~K. Titsias}.}
  \bibinfo{year}{2009}\natexlab{}.
\newblock \showarticletitle{Variational Learning of Inducing Variables in
  Sparse Gaussian Processes}. In \bibinfo{booktitle}{\emph{Proceedings of the
  Twelfth International Conference on Artificial Intelligence and Statistics,
  {AISTATS} 2009, Clearwater Beach, Florida, USA, April 16-18, 2009}}
  \emph{(\bibinfo{series}{{JMLR} Proceedings}, Vol.~\bibinfo{volume}{5})},
  \bibfield{editor}{\bibinfo{person}{David A.~Van Dyk} {and}
  \bibinfo{person}{Max Welling}} (Eds.). \bibinfo{publisher}{JMLR.org},
  \bibinfo{pages}{567--574}.
\newblock
\urldef\tempurl%
\url{http://proceedings.mlr.press/v5/titsias09a.html}
\showURL{%
\tempurl}


\bibitem[Tripp et~al\mbox{.}(2020)]%
        {Weighted_Retraining}
\bibfield{author}{\bibinfo{person}{Austin Tripp}, \bibinfo{person}{Erik~A.
  Daxberger}, {and} \bibinfo{person}{Jos{\'{e}}~Miguel
  Hern{\'{a}}ndez{-}Lobato}.} \bibinfo{year}{2020}\natexlab{}.
\newblock \showarticletitle{Sample-Efficient Optimization in the Latent Space
  of Deep Generative Models via Weighted Retraining}. In
  \bibinfo{booktitle}{\emph{Advances in Neural Information Processing Systems
  33}}.
\newblock


\bibitem[Trummer et~al\mbox{.}(2018)]%
        {skinnerdb}
\bibfield{author}{\bibinfo{person}{Immanuel Trummer}, \bibinfo{person}{Samuel
  Moseley}, \bibinfo{person}{Deepak Maram}, \bibinfo{person}{Saehan Jo}, {and}
  \bibinfo{person}{Joseph Antonakakis}.} \bibinfo{year}{2018}\natexlab{}.
\newblock \showarticletitle{{SkinnerDB}: {Regret}-bounded {Query} {Evaluation}
  via {Reinforcement} {Learning}}.
\newblock \bibinfo{journal}{\emph{PVLDB}} \bibinfo{volume}{11},
  \bibinfo{number}{12} (\bibinfo{year}{2018}), \bibinfo{pages}{2074--2077}.
\newblock
\showISSN{2150-8097}
\urldef\tempurl%
\url{https://doi.org/10.14778/3229863.3236263}
\showDOI{\tempurl}


\bibitem[Van De~Water et~al\mbox{.}(2022)]%
        {datafarm}
\bibfield{author}{\bibinfo{person}{Robin Van De~Water},
  \bibinfo{person}{Francesco Ventura}, \bibinfo{person}{Zoi Kaoudi},
  \bibinfo{person}{Jorge-Arnulfo Quiané-Ruiz}, {and} \bibinfo{person}{Volker
  Markl}.} \bibinfo{year}{2022}\natexlab{}.
\newblock \showarticletitle{Farming {Your} {ML}-based {Query} {Optimizer}'s
  {Food}}. In \bibinfo{booktitle}{\emph{2022 {IEEE} 38th {International}
  {Conference} on {Data} {Engineering} ({ICDE})}}
  \emph{(\bibinfo{series}{{ICDE} '22})}. \bibinfo{pages}{3186--3189}.
\newblock
\urldef\tempurl%
\url{https://doi.org/10.1109/ICDE53745.2022.00294}
\showDOI{\tempurl}
\newblock
\shownote{ISSN: 2375-026X}.


\bibitem[van Renen et~al\mbox{.}(2024)]%
        {redshift_workload}
\bibfield{author}{\bibinfo{person}{Alexander van Renen},
  \bibinfo{person}{Dominik Horn}, \bibinfo{person}{Pascal Pfeil},
  \bibinfo{person}{Kapil~Eknath Vaidya}, \bibinfo{person}{Wenjian Dong},
  \bibinfo{person}{Murali Narayanaswamy}, \bibinfo{person}{Zhengchun Liu},
  \bibinfo{person}{Gaurav Saxena}, \bibinfo{person}{Andreas Kipf}, {and}
  \bibinfo{person}{Tim Kraska}.} \bibinfo{year}{2024}\natexlab{}.
\newblock \showarticletitle{Why {TPC} is not enough: {An} analysis of the
  {Amazon} {Redshift} fleet}.
\newblock \bibinfo{journal}{\emph{Proceedings of the VLDB Endowment}}
  (\bibinfo{year}{2024}).
\newblock
\urldef\tempurl%
\url{https://www.amazon.science/publications/why-tpc-is-not-enough-an-analysis-of-the-amazon-redshift-fleet}
\showURL{%
\tempurl}


\bibitem[Vaswani et~al\mbox{.}(2017)]%
        {attention}
\bibfield{author}{\bibinfo{person}{Ashish Vaswani}, \bibinfo{person}{Noam
  Shazeer}, \bibinfo{person}{Niki Parmar}, \bibinfo{person}{Jakob Uszkoreit},
  \bibinfo{person}{Llion Jones}, \bibinfo{person}{Aidan~N Gomez},
  \bibinfo{person}{Łukasz Kaiser}, {and} \bibinfo{person}{Illia Polosukhin}.}
  \bibinfo{year}{2017}\natexlab{}.
\newblock \showarticletitle{Attention is {All} you {Need}}. In
  \bibinfo{booktitle}{\emph{Advances in {Neural} {Information} {Processing}
  {Systems}}} \emph{(\bibinfo{series}{{NeurIPS} '17},
  Vol.~\bibinfo{volume}{30})}. \bibinfo{publisher}{Curran Associates, Inc.}
\newblock
\urldef\tempurl%
\url{https://papers.nips.cc/paper/2017/hash/3f5ee243547dee91fbd053c1c4a845aa-Abstract.html}
\showURL{%
\tempurl}


\bibitem[Waas and Pellenkoft(2000)]%
        {quickpick}
\bibfield{author}{\bibinfo{person}{F.~Michael Waas} {and}
  \bibinfo{person}{Arjan Pellenkoft}.} \bibinfo{year}{2000}\natexlab{}.
\newblock \showarticletitle{Join Order Selection - Good Enough Is Easy}. In
  \bibinfo{booktitle}{\emph{British National Conference on Databases}}.
\newblock
\urldef\tempurl%
\url{https://api.semanticscholar.org/CorpusID:15111246}
\showURL{%
\tempurl}


\bibitem[Weininger(1988)]%
        {weininger_smiles}
\bibfield{author}{\bibinfo{person}{David Weininger}.}
  \bibinfo{year}{1988}\natexlab{}.
\newblock \showarticletitle{SMILES, a chemical language and information system.
  1. Introduction to methodology and encoding rules}.
\newblock \bibinfo{journal}{\emph{Journal of Chemical Information and Computer
  Sciences}} \bibinfo{volume}{28}, \bibinfo{number}{1} (\bibinfo{year}{1988}),
  \bibinfo{pages}{31--36}.
\newblock
\urldef\tempurl%
\url{https://doi.org/10.1021/ci00057a005}
\showDOI{\tempurl}
\showeprint{https://doi.org/10.1021/ci00057a005}


\bibitem[Weng et~al\mbox{.}(2024)]%
        {eraser_lqo}
\bibfield{author}{\bibinfo{person}{Lianggui Weng}, \bibinfo{person}{Rong Zhu},
  \bibinfo{person}{Di Wu}, \bibinfo{person}{Bolin Ding},
  \bibinfo{person}{Bolong Zheng}, {and} \bibinfo{person}{Jingren Zhou}.}
  \bibinfo{year}{2024}\natexlab{}.
\newblock \showarticletitle{Eraser: {Eliminating} {Performance} {Regression} on
  {Learned} {Query} {Optimizer}}.
\newblock \bibinfo{journal}{\emph{PVLDB}} \bibinfo{volume}{17},
  \bibinfo{number}{5} (\bibinfo{year}{2024}), \bibinfo{pages}{926--938}.
\newblock
\urldef\tempurl%
\url{https://doi.org/10.14778/3641204.3641205}
\showDOI{\tempurl}


\bibitem[Woltmann et~al\mbox{.}(2023)]%
        {fastgres}
\bibfield{author}{\bibinfo{person}{Lucas Woltmann}, \bibinfo{person}{Jerome
  Thiessat}, \bibinfo{person}{Claudio Hartmann}, \bibinfo{person}{Dirk Habich},
  {and} \bibinfo{person}{Wolfgang Lehner}.} \bibinfo{year}{2023}\natexlab{}.
\newblock \showarticletitle{{FASTgres}: {Making} {Learned} {Query} {Optimizer}
  {Hinting} {Effective}}.
\newblock \bibinfo{journal}{\emph{Proceedings of the VLDB Endowment}}
  \bibinfo{volume}{16}, \bibinfo{number}{11} (\bibinfo{date}{Aug.}
  \bibinfo{year}{2023}), \bibinfo{pages}{3310--3322}.
\newblock
\showISSN{2150-8097}
\urldef\tempurl%
\url{https://doi.org/10.14778/3611479.3611528}
\showDOI{\tempurl}


\bibitem[Wu and Ives(2024)]%
        {learned_shift}
\bibfield{author}{\bibinfo{person}{Peizhi Wu} {and} \bibinfo{person}{Zachary~G.
  Ives}.} \bibinfo{year}{2024}\natexlab{}.
\newblock \showarticletitle{Modeling {Shifting} {Workloads} for {Learned}
  {Database} {Systems}}.
\newblock \bibinfo{journal}{\emph{Proceedings of the ACM on Management of
  Data}} \bibinfo{volume}{2}, \bibinfo{number}{1} (\bibinfo{date}{March}
  \bibinfo{year}{2024}), \bibinfo{pages}{1--27}.
\newblock
\showISSN{2836-6573}
\urldef\tempurl%
\url{https://doi.org/10.1145/3639293}
\showDOI{\tempurl}


\bibitem[Wu et~al\mbox{.}(2024a)]%
        {stage}
\bibfield{author}{\bibinfo{person}{Ziniu Wu}, \bibinfo{person}{Ryan Marcus},
  \bibinfo{person}{Zhengchun Liu}, \bibinfo{person}{Parimarjan Negi},
  \bibinfo{person}{Vikram Nathan}, \bibinfo{person}{Pascal Pfeil},
  \bibinfo{person}{Gaurav Saxena}, \bibinfo{person}{Mohammad Rahman},
  \bibinfo{person}{Balakrishnan Narayanaswamy}, {and} \bibinfo{person}{Tim
  Kraska}.} \bibinfo{year}{2024}\natexlab{a}.
\newblock \showarticletitle{Stage: {Query} {Execution} {Time} {Prediction} in
  {Amazon} {Redshift}}. In \bibinfo{booktitle}{\emph{Proceedings of the 2024
  {International} {Conference} on {Management} of {Data} ({SIGMOD} ’24)}}
  \emph{(\bibinfo{series}{{SIGMOD} '24})}. \bibinfo{address}{Santiago, Chile}.
\newblock
\urldef\tempurl%
\url{https://doi.org/10.48550/arXiv.2403.02286}
\showDOI{\tempurl}


\bibitem[Wu et~al\mbox{.}(2024b)]%
        {wu_stage_redshift}
\bibfield{author}{\bibinfo{person}{Ziniu Wu}, \bibinfo{person}{Ryan Marcus},
  \bibinfo{person}{Zhengchun Liu}, \bibinfo{person}{Parimarjan Negi},
  \bibinfo{person}{Vikram Nathan}, \bibinfo{person}{Pascal Pfeil},
  \bibinfo{person}{Gaurav Saxena}, \bibinfo{person}{Mohammad Rahman},
  \bibinfo{person}{Balakrishnan Narayanaswamy}, {and} \bibinfo{person}{Tim
  Kraska}.} \bibinfo{year}{2024}\natexlab{b}.
\newblock \showarticletitle{Stage: Query Execution Time Prediction in Amazon
  Redshift}. In \bibinfo{booktitle}{\emph{Companion of the 2024 International
  Conference on Management of Data}} (Santiago AA, Chile)
  \emph{(\bibinfo{series}{SIGMOD/PODS '24})}. \bibinfo{publisher}{Association
  for Computing Machinery}, \bibinfo{address}{New York, NY, USA},
  \bibinfo{pages}{280–294}.
\newblock
\showISBNx{9798400704222}
\urldef\tempurl%
\url{https://doi.org/10.1145/3626246.3653391}
\showDOI{\tempurl}


\bibitem[Yang et~al\mbox{.}(2022)]%
        {balsa}
\bibfield{author}{\bibinfo{person}{Zongheng Yang}, \bibinfo{person}{Wei-Lin
  Chiang}, \bibinfo{person}{Sifei Luan}, \bibinfo{person}{Gautam Mittal},
  \bibinfo{person}{Michael Luo}, {and} \bibinfo{person}{Ion Stoica}.}
  \bibinfo{year}{2022}\natexlab{}.
\newblock \showarticletitle{Balsa: {Learning} a {Query} {Optimizer} {Without}
  {Expert} {Demonstrations}}. In \bibinfo{booktitle}{\emph{Proceedings of the
  2022 {International} {Conference} on {Management} of {Data}}}.
  \bibinfo{publisher}{ACM}, \bibinfo{address}{Philadelphia PA USA},
  \bibinfo{pages}{931--944}.
\newblock
\showISBNx{978-1-4503-9249-5}
\urldef\tempurl%
\url{https://doi.org/10.1145/3514221.3517885}
\showDOI{\tempurl}


\bibitem[Yang et~al\mbox{.}(2020)]%
        {neurocard}
\bibfield{author}{\bibinfo{person}{Zongheng Yang}, \bibinfo{person}{Amog
  Kamsetty}, \bibinfo{person}{Sifei Luan}, \bibinfo{person}{Eric Liang},
  \bibinfo{person}{Yan Duan}, \bibinfo{person}{Xi Chen}, {and}
  \bibinfo{person}{Ion Stoica}.} \bibinfo{year}{2020}\natexlab{}.
\newblock \showarticletitle{{NeuroCard}: {One} {Cardinality} {Estimator} for
  {All} {Tables}}.
\newblock \bibinfo{journal}{\emph{arXiv:2006.08109 [cs]}} (\bibinfo{date}{June}
  \bibinfo{year}{2020}).
\newblock
\urldef\tempurl%
\url{http://arxiv.org/abs/2006.08109}
\showURL{%
\tempurl}
\newblock
\shownote{arXiv: 2006.08109}.


\bibitem[Yi et~al\mbox{.}(2024)]%
        {limeqo}
\bibfield{author}{\bibinfo{person}{Zixuan Yi}, \bibinfo{person}{Yao Tian},
  \bibinfo{person}{Zachary~G. Ives}, {and} \bibinfo{person}{Ryan Marcus}.}
  \bibinfo{year}{2024}\natexlab{}.
\newblock \showarticletitle{Low {Rank} {Approximation} for {Learned} {Query}
  {Optimization}}. In \bibinfo{booktitle}{\emph{International {Workshop} on
  {Exploiting} {Artificial} {Intelligence} {Techniques} for {Data}
  {Management}}} \emph{(\bibinfo{series}{{aiDM} @ {SIGMOD} '24})}.
  \bibinfo{publisher}{ACM}, \bibinfo{address}{Santiago, Chile}.
\newblock
\urldef\tempurl%
\url{https://doi.org/10.1145/3663742.3663974}
\showDOI{\tempurl}


\bibitem[Yu et~al\mbox{.}(2022)]%
        {hybrid_lqo}
\bibfield{author}{\bibinfo{person}{Xiang Yu}, \bibinfo{person}{Chengliang
  Chai}, \bibinfo{person}{Guoliang Li}, {and} \bibinfo{person}{Jiabin Liu}.}
  \bibinfo{year}{2022}\natexlab{}.
\newblock \showarticletitle{Cost-{Based} or {Learning}-{Based}? {A} {Hybrid}
  {Query} {Optimizer} for {Query} {Plan} {Selection}}.
\newblock \bibinfo{journal}{\emph{Proceedings of the VLDB Endowment}}
  \bibinfo{volume}{15}, \bibinfo{number}{13} (\bibinfo{date}{Sept.}
  \bibinfo{year}{2022}), \bibinfo{pages}{3924--3936}.
\newblock
\showISSN{2150-8097}
\urldef\tempurl%
\url{https://doi.org/10.14778/3565838.3565846}
\showDOI{\tempurl}


\bibitem[Yu et~al\mbox{.}(2020)]%
        {lstm_jo}
\bibfield{author}{\bibinfo{person}{Xiang Yu}, \bibinfo{person}{Guoliang Li},
  \bibinfo{person}{Chengliang Chai}, {and} \bibinfo{person}{Nan Tang}.}
  \bibinfo{year}{2020}\natexlab{}.
\newblock \showarticletitle{Reinforcement {Learning} with {Tree}-{LSTM} for
  {Join} {Order} {Selection}}. In \bibinfo{booktitle}{\emph{2020 {IEEE} 36th
  {International} {Conference} on {Data} {Engineering}}}
  \emph{(\bibinfo{series}{{ICDE} '20})}. \bibinfo{pages}{1297--1308}.
\newblock
\urldef\tempurl%
\url{https://doi.org/10.1109/ICDE48307.2020.00116}
\showDOI{\tempurl}
\newblock
\shownote{ISSN: 2375-026X}.


\bibitem[Zhang et~al\mbox{.}(2022b)]%
        {bao_scope2}
\bibfield{author}{\bibinfo{person}{Wangda Zhang}, \bibinfo{person}{Matteo
  Interlandi}, \bibinfo{person}{Paul Mineiro}, \bibinfo{person}{Shi Qiao},
  \bibinfo{person}{Nasim Ghazanfari}, \bibinfo{person}{Karlen Lie},
  \bibinfo{person}{Marc Friedman}, \bibinfo{person}{Rafah Hosn},
  \bibinfo{person}{Hiren Patel}, {and} \bibinfo{person}{Alekh Jindal}.}
  \bibinfo{year}{2022}\natexlab{b}.
\newblock \showarticletitle{Deploying a {Steered} {Query} {Optimizer} in
  {Production} at {Microsoft}}. In \bibinfo{booktitle}{\emph{Proceedings of the
  2022 {International} {Conference} on {Management} of {Data}}}
  \emph{(\bibinfo{series}{{SIGMOD} '22})}. \bibinfo{publisher}{ACM},
  \bibinfo{address}{Philadelphia PA USA}, \bibinfo{pages}{2299--2311}.
\newblock
\showISBNx{978-1-4503-9249-5}
\urldef\tempurl%
\url{https://doi.org/10.1145/3514221.3526052}
\showDOI{\tempurl}


\bibitem[Zhang et~al\mbox{.}(2022a)]%
        {database_hyperparameter_optimization}
\bibfield{author}{\bibinfo{person}{Xinyi Zhang}, \bibinfo{person}{Zhuo Chang},
  \bibinfo{person}{Yang Li}, \bibinfo{person}{Hong Wu}, \bibinfo{person}{Jian
  Tan}, \bibinfo{person}{Feifei Li}, {and} \bibinfo{person}{Bin Cui}.}
  \bibinfo{year}{2022}\natexlab{a}.
\newblock \showarticletitle{Facilitating database tuning with hyper-parameter
  optimization: a comprehensive experimental evaluation}.
\newblock \bibinfo{journal}{\emph{Proc. VLDB Endow.}} \bibinfo{volume}{15},
  \bibinfo{number}{9} (\bibinfo{date}{may} \bibinfo{year}{2022}),
  \bibinfo{pages}{1808–1821}.
\newblock
\showISSN{2150-8097}
\urldef\tempurl%
\url{https://doi.org/10.14778/3538598.3538604}
\showDOI{\tempurl}


\bibitem[Zhao et~al\mbox{.}(2022)]%
        {q_transformer}
\bibfield{author}{\bibinfo{person}{Yue Zhao}, \bibinfo{person}{Gao Cong},
  \bibinfo{person}{Jiachen Shi}, {and} \bibinfo{person}{Chunyan Miao}.}
  \bibinfo{year}{2022}\natexlab{}.
\newblock \showarticletitle{{QueryFormer}: a tree transformer model for query
  plan representation}.
\newblock \bibinfo{journal}{\emph{Proceedings of the VLDB Endowment}}
  \bibinfo{volume}{15}, \bibinfo{number}{8} (\bibinfo{date}{April}
  \bibinfo{year}{2022}), \bibinfo{pages}{1658--1670}.
\newblock
\showISSN{2150-8097}
\urldef\tempurl%
\url{https://doi.org/10.14778/3529337.3529349}
\showDOI{\tempurl}


\bibitem[Zhou et~al\mbox{.}(2021)]%
        {mcts_qo}
\bibfield{author}{\bibinfo{person}{Xuanhe Zhou}, \bibinfo{person}{Guoliang Li},
  \bibinfo{person}{Chengliang Chai}, {and} \bibinfo{person}{Jianhua Feng}.}
  \bibinfo{year}{2021}\natexlab{}.
\newblock \showarticletitle{A learned query rewrite system using {Monte}
  {Carlo} tree search}.
\newblock \bibinfo{journal}{\emph{Proceedings of the VLDB Endowment}}
  \bibinfo{volume}{15}, \bibinfo{number}{1} (\bibinfo{date}{Sept.}
  \bibinfo{year}{2021}), \bibinfo{pages}{46--58}.
\newblock
\showISSN{2150-8097}
\urldef\tempurl%
\url{https://doi.org/10.14778/3485450.3485456}
\showDOI{\tempurl}


\bibitem[Zhu et~al\mbox{.}(2023)]%
        {lero}
\bibfield{author}{\bibinfo{person}{Rong Zhu}, \bibinfo{person}{Wei Chen},
  \bibinfo{person}{Bolin Ding}, \bibinfo{person}{Xingguang Chen},
  \bibinfo{person}{Andreas Pfadler}, \bibinfo{person}{Ziniu Wu}, {and}
  \bibinfo{person}{Jingren Zhou}.} \bibinfo{year}{2023}\natexlab{}.
\newblock \showarticletitle{Lero: A Learning-to-Rank Query Optimizer}.
\newblock \bibinfo{journal}{\emph{Proc. VLDB Endow.}} \bibinfo{volume}{16},
  \bibinfo{number}{6} (\bibinfo{date}{feb} \bibinfo{year}{2023}),
  \bibinfo{pages}{1466–1479}.
\newblock
\showISSN{2150-8097}
\urldef\tempurl%
\url{https://doi.org/10.14778/3583140.3583160}
\showDOI{\tempurl}


\bibitem[Zhu et~al\mbox{.}(2024)]%
        {pilotscope}
\bibfield{author}{\bibinfo{person}{Rong Zhu}, \bibinfo{person}{Lianggui Weng},
  \bibinfo{person}{Wenqing Wei}, \bibinfo{person}{Di Wu},
  \bibinfo{person}{Jiazhen Peng}, \bibinfo{person}{Yifan Wang},
  \bibinfo{person}{Bolin Ding}, \bibinfo{person}{Defu Lian},
  \bibinfo{person}{Bolong Zheng}, {and} \bibinfo{person}{Jingren Zhou}.}
  \bibinfo{year}{2024}\natexlab{}.
\newblock \showarticletitle{{PilotScope}: {Steering} {Databases} with {Machine}
  {Learning} {Drivers}}.
\newblock \bibinfo{journal}{\emph{PVLDB}} \bibinfo{volume}{17},
  \bibinfo{number}{5} (\bibinfo{year}{2024}), \bibinfo{pages}{980--993}.
\newblock
\urldef\tempurl%
\url{https://doi.org/10.14778/3641204.3641209}
\showDOI{\tempurl}


\bibitem[Zinchenko and Iazov(2024)]%
        {hero_qo}
\bibfield{author}{\bibinfo{person}{Sergey Zinchenko} {and}
  \bibinfo{person}{Sergey Iazov}.} \bibinfo{year}{2024}\natexlab{}.
\newblock \bibinfo{title}{{HERO}: {Hint}-{Based} {Efficient} and {Reliable}
  {Query} {Optimizer}}.
\newblock
\newblock
\urldef\tempurl%
\url{https://doi.org/10.48550/arXiv.2412.02372}
\showDOI{\tempurl}
\newblock
\shownote{arXiv:2412.02372 [cs]}.


\end{thebibliography}



\end{document}